\documentclass[fleqn,usenatbib]{mnras}

\usepackage[T1]{fontenc}
\usepackage{ae,aecompl}
\usepackage[dvipsnames]{xcolor}

\usepackage{graphicx}	% Including figure files
\usepackage{amsmath}	% Advanced maths commands
\usepackage{amssymb}	% Extra maths symbols
\usepackage{url,hyperref}

\title[Asteroids in the 1:2 resonance with Mars]{Chaotic diffusion of asteroids in the exterior 1:2 mean motion resonance with Mars}

\author[A. A. Christou et al.]{
Apostolos A. Christou,$^{1,2}$\thanks{E-mail: apostolos.christou@armagh.ac.uk (AAC)}
Stanley F. Dermott,$^{2}$
and Dan Li$^{3}$
\\
$^{1}$Armagh Observatory and Planetarium, College Hill, Armagh BT61 9DG, Northern Ireland, UK\\
$^{2}$Department of Astronomy, University of Florida at Gainsville, USA\\
$^{3}$NSF's National Optical-Infrared Astronomy Research Laboratory, Tucson, USA
}
\date{Accepted  2022 August 02. Received  2022 August 02; in original form 2022 July 12}

\pubyear{2022}

\begin{document}
\label{firstpage}
\pagerange{\pageref{firstpage}--\pageref{lastpage}}
\maketitle

% Abstract of the paper
\begin{abstract}
The inner asteroid belt between 2.1 and 2.5 au is of particular dynamical significance because it is the dominant source of both chondritic meteorites and near-Earth asteroids. This inner belt is bounded by an eccentricity-type secular resonance and by the 1:3 mean motion resonance with Jupiter. Unless asteroid perihelia are low enough to allow scattering by Mars, escape requires transport to one of the bounding resonances. In addition Yarkovsky forces are generally ineffective in changing either the eccentricity and/or inclination for asteroids with diameter $\gtrsim$30 km. Thus, large asteroids with pericentres far from Mars may only escape from the inner belt through large changes in their eccentricities. In this paper we study chaotic diffusion of orbits near the 1:2 mean motion resonance with Mars in a systematic way. We show that, while chaotic orbital evolution in both resonant and non-resonant orbits increase the dispersion of the inclinations and eccentricities, it does not significantly change their mean values. We show further that, while the dispersive growth is greatest for resonant orbits, at high $e$ the resonance acts to mitigate asteroid scattering by Mars - making the asteroid lifetime in the belt longer than it would have been for a non-resonant orbit. For asteroids of all sizes in both resonant and non-resonant orbits, the changes in eccentricity needed to account for the observations cannot be achieved by gravitational forces alone. The role of resonant trapping in protecting asteroids from encounters with Mars is also analysed.  
\end{abstract}

% Select between one and six entries from the list of approved keywords.
% Don't make up new ones.
% 15 pages, 16 figures, 3 tables
% Submitted for publication in MNRAS 2021 July 12
%%%%%%%%%%%%%%%%%%%%%%%%%%%%%%%%%%%%%%%%%%%%%%%%%%

%%%%%%%%%%%%%%%%% BODY OF PAPER %%%%%%%%%%%%%%%%%%

\section{Introduction}
% Most Near-Earth Asteroids (NEAs) and meteorites originate from the Main Belt (MB hereafter) as fragments of larger asteroids \citep{Michel.et.al2001,Burbine.et.al2002, Jacobson.et.al2014} mobilised by chaotic orbital evolution \citep{MorbidelliNesvorny1999, MintonMalhotra2010} and Yarkovsky radiation forces \citep{FarinellaVokhroulicky1999,Bottke.et.al2002a}. The process of orbit mobilisation while still in the MB is a key link in this chain \cite[eg][]{MorbidelliNesvorny1999}, with implications for the past evolution and genealogy of bodies in the asteroid belt \citep{MintonMalhotra2010,Dermott.et.al2018}.
Most Near-Earth Asteroids (NEAs) and meteorites originate from the Main asteroid Belt (or MB for short) as fragments of larger asteroids mobilised by chaotic orbital evolution and Yarkovsky radiation forces \citep{FarinellaVokhroulicky1999,Bottke.et.al2002a}. The process of orbit mobilisation while still in the MB is a key link in this chain \cite[eg][]{MorbidelliNesvorny1999}, with implications for the past evolution and genealogy of bodies in the asteroid belt \citep{MintonMalhotra2010,Dermott.et.al2021}.
The inner region of the Main belt (Inner Main Belt or IMB for short), containing asteroids with orbital semimajor axis between $2.1$ and $2.5$ au, is bracketed by the $\nu_{6}$ secular resonance on its sunward border and on the other by the 3:1 mean motion resonance with Jupiter. It is of particular significance to solar system scientists as the source of most near-Earth asteroids and of chondritic meteorites \citep{GranvikBrown2018,Granvik.et.al2018}.

% Mean motion resonances with Mars have a destabilising effect on Main Belt asteroid orbits interior to 2.5 au \citep{MorbidelliNesvorny1999}, meaning that asteroids eventually become Mars- and Earth-crossers. Asteroids$\geq$10 km suffer a radial drift  no higher than 0.02 au over 1 Gyr \citep{NesvornyRoig2018} and do not evolve sufficiently to escape the MB but may still migrate to become interlopers within neighbouring families or enter the non-family background. This has implications for the origin of non-family asteroids \citep{Dermott.et.al2018}.
% but also efforts to debias the observed NEA population \citep{Bottke.et.al2002b,Greenstreet.et.al2012}.

% There is merit in re-examining the mobility.
% The introduction will reference the work by 
 To remove asteroids from the IMB, \citet{MorbidelliNesvorny1999} invoked the action of weak Martian resonances. While we agree with their conclusions, initial conditions in their simulations were chosen from a very wide range of $e$ and $I$ - also the case in \citet{MintonMalhotra2010}. To improve our understanding of asteroid mobilisation and escape from the IMB, we must examine the orbital evolution of a tightly specified distribution of $a$, $e$ and $I$ optimised to distinguish between resonant and non-resonant orbits.
% excerpt from  NM99: The initial eccentricity was set equal to 0.1, while the initial inclination and phase angles were all set equal to zero with respect to the ecliptic and the Vernal point 

As a first step in this direction, we study here the efficiency of a mean motion resonance with Mars in mobilising IMB asteroids. We do this by making use of intensive numerical simulations to characterise orbital mobility as a random process. In particular, we calculate the timescales for the growth of the overall dispersion in $e$ and $I$ for a precisely controlled range of initial conditions and for both resonant and non-resonant orbits.

In this work we consider only point-mass gravitational forces, since these are the primary mobility drivers for asteroids larger than about 30 km but would affect all asteroids independent of size. Pinning down the long-term gravitational diffusion would also allow to separate it out in later work that will include non-gravitational forces. We focus on the exterior 1:2 mean motion resonance with Mars (1M-2A hereafter) as belonging to a finite number of IMB resonances shown to contribute to orbital diffusion \citep{Migliorini.et.al1998,MorbidelliNesvorny1999}. The dynamics of asteroids in this resonance were investigated by \citet{Gallardo2007,Gallardo.et.al2011,Dermott.et.al2022} showing a long lifetime ($> 10$ Myr) of asteroids in the resonance as well as size-dependent evolutionary outcomes under the Yarkovsky effect, the latter being consistent with the observed statistical distinctiveness of resonant asteroids compared to the non-resonant background population. 
% This resonance shows a $\sim 20$\% excess in the number density of resonant asteroids as well as a higher mean eccentricity and a relative lack of larger asteroids compared to the non-resonant background, pointing to size- and orbit-dependent evolution of asteroids driven by the Yarkovsky effect \citet{Gallardo2007,Gallardo.et.al2011}.

% We complement, and expand on, earlier work \citep{MorbidelliNesvorny1999,Gallardo2007,Gallardo.et.al2011} by quantifying the mobility as a random process

Our ultimate aim is to systematically map out the mobility of asteroids both within and outside resonances, including all relevant forces and over a wide region of phase space so that the flow of asteroids both through and out of the belt can be precisely quantified.
\section{Methods}
\label{sec:methods}
\subsection{Clone generation}
\label{sec:clones}
 % Picked four asteroids: 17791, 32766, 121193, 337668
 Osculating elements represent the instantaneous keplerian orbit of the asteroid and vary periodically over time because of the gravitational influence of the planets. Proper elements, on the other hand, are immune to periodic variations and only change by the slow process of chaotic diffusion \cite[e.\,g.][]{MilaniFarinella1994}. To investigate how the orbits diffuse over long periods, we therefore need to follow the proper, rather than the osculating, elements. For this purpose we generated two distinct sets of test particles both within and outside the resonance. We refer to these as the ``resonant'' and the ``non-resonant'' or ``control'' sets respectively. 
 
 Our clone generation method is essentially identical to that used in \citet{Dermott.et.al2018}, however for ease of reference we describe it also here with some additional information provided in Appendix~\ref{sec:appendixa}. In short, this is a two-step process where we first average the osculating element vectors (eg $\left( e \cos{\varpi},e \sin{\varpi} \right)$ for the eccentricity vector) for a large number of real asteroids to estimate the - assumed common - forced vector. The initial osculating vector for the test particles is then formed as the vector sum of the user-defined proper vector with magnitude $e_{P}(t=0)$ for the eccentricity ($I_{P}(t=0)$ for the inclination) and the estimated forced vector.  
 % For the 1M:2A resonance, we used 14653 numbered asteroids \cite{NUMB} with proper mean motion $n_{p}$ between 94.5 and 96.5 deg/yr.
 The starting values of $e_{P}$ and $I_{P}$ for the particles are shown in Table~\ref{tab:sims} where we have generally sampled $e_{P}(0)$ from 0.11 to 0.27 with a step of $0.04$ and for $I_{P}(0)=2.5^{\circ}$, $6.5^{\circ}$ and $10^{\circ}$. For the control particle set we have additionally included a group with $e_{P}(0)=0.20$, near the boundary with the chaotic region \cite[$e_{P}(0)$$\gtrsim$$ 0.2$;][]{MorbidelliNesvorny1999,Gallardo.et.al2011}. 
 
 The proper longitudes of perihelion \& the ascending node were assigned uniformly random values and the initial osculating values formed by again adding the chosen proper elements vectorially to the forced component. 
 %(see Appendix~\ref{app_a}). 
 The mean anomaly $M$ was randomised and the osculating semimajor axis $a$ held fixed to the values of $2.4183$ au and $2.4230$ au for resonant and non-resonant particles respectively.
 % The proper orbital elements of the particles near the beginning of our runs, shown in Fig.~\ref{fig:ic_100m}, are good proxies for the "starting" proper orbits. Here we observe some scatter in proper $a$, $e$ and $I$. Scatter in $e_{P}$ and $I_{P}$ is generally low, but higher for the control particles, $\pm$0.01 and $\pm$$1^{\circ}$ respectively. The scatter in $a_{P}$ is likely a consequence of fixing the osculating $a$ and randomising $M$ at $t=t_{0}$, due to the coupling between osculating and proper elements through the Lagrange equations \citep{MurrayDermott1999}. Additionally, we expect that errors introduced through the estimation of the forced vectors and in our averaging over a finite time interval to obtain the proper elements contribute to the scatter about the target values of $e_{P}$ and $I_{P}$.
\subsection{Numerical simulations}
\label{sec:numint}
Test particles were integrated for $T=10^{8}$ yr using the {\sl orbit9} code \citep{MilaniNobili1988,KnezevicMilani2000} available within the {\sl orbfit} package \citep{ORBFIT} with a maximum step size of $0.2$ yr and an output step of $10^{4}$ yr. The code uses a multistep predictor scheme with automatic step size control combined with a symplectic fixed-step scheme as starter. The solar system model included the seven major planets from Venus to Neptune where the mass of Mercury was added to that of the Sun. The initial planetary state vectors were taken from the JPL DE405 ephemeris \citep{Giorgini.et.al1996}\footnote{url: \url{https://ssd.jpl.nasa.gov/horizons/\#telnet}} at $t_{0}=$\,JD2457000.5 and referred to the Sun-Mercury barycentre. {\sl Orbit9} also carries out on-the-fly digital filtering to suppress periodic variations in the orbital elements shorter than a few hundred years \citep{KnezevicMilani2000} and furnishes both filtered and unfiltered numerical output. Here, we have chosen to use the unfiltered output for post-processing and analysis. Note also that our model does not include large asteroids such as (1) Ceres, (2) Pallas or (4) Vesta, as done in \citet{Nesvorny.et.al2002b,Carruba.et.al2003,Gallardo.et.al2011}.  Gravitational scattering by these objects has been shown to contribute to orbit diffusion for IMB asteroids by mobilising $a$, therefore causing asteroids to jump in and out of resonances. However, based on these previous works we expect the scattering effect to be small compared to the chaotic diffusion and we shall show evidence that this is actually the case.

To identify each run, we utilise the designation {\tt XxxmMnAIIiTTT} where ``Xxx'' refers to either Res(onant) or Con(trol) asteroids, ``mMnA'' to the mM-nA martian resonance, ``IIi'' to the starting value of the proper inclination (II.i degrees) and ``TTT'' to the run  duration in Myr. To extract the time history of the proper elements, we average the integration output every 32 samples or $3.2 \times 10^{5}$ yr. The integration of the control particle group with $e_{P}(0)$$=$$0.2$ and $I_{P}(0)$$=$$2.5^{\circ}$ was stopped at $0.74\times10^{8}$ yr due to a technical issue with the computing server.
\section{Results}
\label{sec:results}
The calculated proper orbital elements of the particles near the beginning of our runs are shown in Fig.~\ref{fig:ic_100m}, where we observe some scatter both in proper $a$ and in proper $e$. The scatter in $a_{P}$ is likely a consequence of fixing the osculating $a$ and randomising $M$ at $t=t_{0}$, due to the coupling between osculating and proper elements through the Lagrange equations \citep{MurrayDermott1999}. Because of this scatter, a few particles - between 10 and 20 from each group of 400 - are inadvertently placed just outside and to the left of the resonance and do not participate in the subsequent chaotic evolution induced by the resonant dynamics, however as their number in all cases is small they do not affect our computations. Errors in the estimation of the forced vectors and through our averaging over a finite time interval to obtain the proper elements, contribute to the scatter about the target values of $e_{P}$ and $I_{P}$. The scatter is generally low, but higher for the control particles, $\pm$0.01 and $\pm$$1^{\circ}$ respectively. Note also that the effective width $\Delta a$ of the resonance for our chosen values of $e_{P}(0)$ is $\gtrsim$$3 \times 10^{-3}$ au and larger than the expected diffusion from large asteroids ($\sim$$10^{-3}$ au), justifying our choice to omit those bodies from our model.

During our runs, many particles beginning in non-Mars-crossing orbits within the resonance have their eccentricity diffuse to higher values and eventually become Mars-crossing. To isolate the effect we want to measure, namely the chaotic diffusion within the resonance, in our analysis we only consider particles approaching the orbit of Mars no closer than twice the martian Hill sphere radius $R_{H}$, in other words 
\begin{equation}
 q>Q_{M}+n R_{H}
 \label{eq:crit}
 \end{equation}
 with $n=2$, where $q$ is the osculating perihelion distance of the asteroid and $Q_{M}$ the osculating aphelion distance of Mars. We note here that {\sl orbit9} has a feature that records Mars-particle encounters to within a specified {\it cartesian} distance $d_{enc}$. This distance was left to its default value of $0.1$ au for these runs and criterion (\ref{eq:crit}) was instead used in the analysis. By preferentially removing data from our sample, this filtering procedure potentially skews the statistical analysis, but we will see later in the Section that this only affects a limited number of particles and even then it is readily recognised and taken into account so that it does not affect our conclusions.

Fig.~\ref{fig:itime_100m} shows the proper eccentricity history of two groups of test particles, the group {\tt Res1M2A065100} with $e_{P}(0)=0.19$ (top panel) and the corresponding control group {\tt Con1M2A065100} (bottom panel). The plotted quantity is the difference between $e_{P}(t)$ and the starting value $e_{P}(0)$ for each particle. The eccentricity diffusion of the resonant clones clearly increases over time, with individual particles reaching values of $|e_{P} - e_{P}(0)|$ in excess of $0.03$ after $10^{8}$ yr, compared to $\lesssim$0.008 for the control particles. 
% Moreover, high values for the control set are reached only for a few particles, with the majority of particles reaching values a few times $10^{-3}$ towards the end of the simulation. 
For the control group, we note also the modulation of the output with a $\sim$$1 \mbox{\,Myr}$ frequency and amplitude $\sim$5$\times$$10^{-3}$, several times smaller than the dispersion magnitude in the resonant group, but comparable to the more modest diffusion within the control group. This motivated us to apply more aggressive numerical averaging for control particles to isolate the chaotic orbital variations, hence we chose to average with a bin width of 128 points ($1.28\mbox{\,Myr}$, black lines) instead of the 32 points used for resonant particles. Applying this coarser averaging window revealed the presence of several particles undergoing longer-period ($5-20$ Myr) oscillations, probably caused by proximity to secular resonances.

\subsection{Orbital mobility: Dispersion and mean}
\label{sec:diff100}
The cumulative chaotic evolution of the orbits may be quantified through the function \citep{Carruba.et.al2003}:
\begin{equation}
\label{eq:sigma}
V_{c}(t)=\frac{1}{N-1}\sum_{i=0}^{N} {[c_{i}(t) - c_{i}(0)]}^{2}
\end{equation}
where $i$ runs through the particles for each group, $c$ can be $a_{P}$, $e_{P}$ or $I_{P}$ and $N$ is the sample size. The sample standard deviation is then $\sigma_{c}(t)=\sqrt{V_{c}(t)}$. Using the averaged numerical output, we can compute $\sigma(t)$ for each group of particles and for the elements of interest. Plots of $\sigma_{e}(t)$ and $\sigma_{I}(t)$ are shown in Fig.~\ref{fig:sigma_100m_in2to1m} for the control particles (black and grey lines) and for the resonant particles, where a warmer colour indicates a higher starting value of $e_{P}$. The time dependence of $\sigma$ was modelled as a power law
\begin{equation}
 \sigma(t)=\sigma_{T} \left(\frac{t}{T}\right)^{b}
 \label{eq:power_law}
 \end{equation}
and the exponent $b$ and normalising constant $\sigma_{T}$ estimated from the numerical data by fitting to the log-linear relationship $\log_{10}{\sigma(t)} = c + b \log_{10}{t}$ where $c=\log_{10}(\sigma_{T}/T^{b})$ and omitting the first $10^{7}$ yr of averaged numerical output from the fit. Since all our runs have $T=10^{8}$ yr we find it convenient to adopt the notation $\sigma_{8}$ to refer to the normalising constant in Eq~\ref{eq:power_law} for the remainder of the paper. 
The fit results are reported in Table~\ref{tab:results100in} where the values in brackets represent our estimate of $\sigma_{8}$.
The dashed straight lines in Fig.~\ref{fig:sigma_100m_in2to1m} represent future extrapolations of these relationships beyond the timespan of our numerical integrations and for up to 2 Gyr.

The time evolution of eccentricity diffusion for resonant particles consistently fits a power law with $b$ values between $0.306$ and $0.434$, a rather narrow range, independently of $e_{P}(0)$ or $I_{P}(0)$. We find this is also the case for the inclination diffusion, with $b_{I}$ varying between $0.350$ and $0.518$. These slopes suggest a chaotic diffusion rate near, though somewhat lower than, that expected for a purely random walk ($b=0.5$). 

If we consider the power law data in Table~\ref{tab:results100in} as estimates of a common true slope, and omit estimates for $e_{P}(0)$$=$$0.23$ for reasons we will come to later in the Section, we obtain $b_{e}=0.419\pm 0.030$ for the eccentricity and $b_{I}=0.437\pm0.053$ for the inclination. Setting $b=0.4$ appears therefore to be a good approximation to the slope of these power laws, meaning the estimated $\sigma_{e,8}$ for each group depend only on the starting values, with the highest value of $\sigma_{e,8}$ recorded for the smallest $e_{P}(0)$ and highest $I_{P}(0)$. For $\sigma_{I,8}$ the form of the dependence is somewhat different; while we see no correlation with $e_{P}(0)$, the values of $\sigma_{I,8}$ recorded for the particle group with $I_{P}(0)$=10${}^{\circ}$ are 1.5-2.0$\times$ higher than those for groups with $I_{P}(0)$=2.5${}^{\circ}$ and $I_{P}(0)$=6.5${}^{\circ}$. 

The diffusion among control particles (Fig.~\ref{fig:sigma_100m_in2to1m}) is significantly less than for resonant particles and there also appears to be a dependence of the slope on the starting eccentricity. Specifically, while particles with $e_{P}(0)$$=$$0.20$ (grey curves) display a similar slope to the resonant particles, for smaller values of $e_{P}(0)$ (black curves) we observe shallower slopes. It is not entirely clear to us that this variation in the slope, which we do not observe for resonant particles, is a real feature of the dynamics or some numerical artifact, perhaps caused by our aggressive numerical averaging strategy. However, the similarity in slope between the $e_{P}(0)$$=$$0.19/0.20$ control particles and those of resonant particles is consistent with the action of weak resonances, which would be stronger at higher $e_{P}(0)$ and should affect those particles the most.    
% The growth of $\sigma(t)$ for the control particle sets (Fig.~\ref{fig:sigma_100m_in2to1m}, black \& grey curves) shows a weaker time dependence: $b$$=$$0$$-$$0.35$ for $\sigma_{e}$ and $b$$=$$0$$-$$0.53$ for $\sigma_{I}$ (Table~\ref{tab:results100out}). Overall, The dispersion of groups of control particles is significantly less than for resonant particles. 
% Unlike the resonant particle sets, the slope is consistently shallower for control sets with the smallest values of $e_{P}(0)$ with the corresponding values of $b_{e}$, $b_{I}$, $\sigma_{e,8}$ and $\sigma_{I,8}$ being highest for particles with $e_{P}(0)=0.20$ (grey curves in Fig.~\ref{fig:sigma_100m_in2to1m}). 

In relation to this, we note here that our $b$ estimates for the group with $e_{P}(0)$$=$$0.2$ and $I_{P}(0)$$=$$2.5^{\circ}$ are higher than those reported by \citet{Dermott.et.al2018} for the same data, namely $b$$=$$0.15$ for $\sigma_{e}$ and $b$$=$$0.29$ for $\sigma_{I}$. This is due to the 4$\times$ shorter averaging window used to calculate control particle proper elements in that work. It is shorter than the period of the $\sim$$1$ Myr modulation observed in Fig~\ref{fig:itime_100m} and skews those estimates of $b$ to artificially low values. % Fig.~\ref{fig:sigma_100m_in2to1m} and Table \ref{tab:results100out} 
Our log-linear extrapolations indicate that $e_{P}(t)-e_{P}(0)$ for the control sample would grow to $7 \times 10^{-3}$ and $I_{P}(t)-I_{P}(0)$ to $0.4^{\circ}$ at $t$$=$$2$ Gyr, compared to $\gtrsim 0.04$ (5$\times$) and $\gtrsim 1$ degree (2.5-3$\times$) respectively for the resonant set.

Returning to the resonant particles, we find that using the criterion in Eq.~\ref{eq:crit} to filter out Mars-encountering particles from our data has two consequences. Firstly, all particles in the group with $e_{P}(0)$$=$$0.27$ do not satisfy it, either at $t$$=$$0$ or later on during the simulation; these particles are therefore excluded from the analysis in this Section; Secondly, most particles in the group with $e_{P}(0)$$=$$0.23$ also fail the criterion, leaving $\sim$100 out of the original 400 particles in each group suitable for statistical analysis. Their orbital diffusion is represented by the yellow curves in Fig.~\ref{fig:sigma_100m_in2to1m}. To understand how this might affect the time dependence of the diffusion, we found it instructive to compare the growth of $\sigma_{I}$ (bottom panels of Fig.~\ref{fig:sigma_100m_in2to1m}) among the groups with different $I_{P}(0)$. We find that the data for $e_{P}(0)$=0.23 is offset relative to the other curves and, furthermore (see Table \ref{tab:results100in}), the offset decreases from a factor of $\sim$4 for particles with $I_{P}(0)$=2.5${}^{\circ}$ down to a factor of $\sim$2 for particles with $I_{P}(0)$=6.5${}^{\circ}$, becoming negligible for particles with $I_{P}(0)$=10${}^{\circ}$. 

We believe the offset is related to changes in the particle orbit during moderately close approaches to Mars when $q_{A} \gtrsim Q_{M}+2 R_{H}$, acting to modify the orbital elements in addition to chaotic diffusion in the resonance. In this scenario, the offset is smaller for the higher $I_{P}(0)$ due to the higher encounter velocity between Mars and the particle, mitigating the scattering effect of Mars on the orbit. The existence of an offset and its dependence on $I_{P}(0)$ is hinted at also in the behaviour of $\sigma_{e}$ (top panels of Fig.~\ref{fig:sigma_100m_in2to1m}), though the evidence consists solely of a slightly higher $\sigma_{e}$ than expected for particles with $e_{P}(0)$$=$$0.23$ (leftmost panel; $I_{P}(0)$=2.5${}^{\circ}$).
%, based on the dependence of this quantity on $e_{P}(0)$ observed in the other panels and discussed earlier in this Section. 
It is instructive to compare our findings on the diffusion of resonant particle groups with \citet{Gallardo.et.al2011} where, in agreement with that work, we find that $\sigma_{e}$ and $\sigma_{I}$ both increase with increasing $I_{P}(0)$. Specifically, those authors report $e$- and $I$-diffusion values of $0.05$/$1.2^{\circ}$ and $0.08$/$1.6^{\circ}$ for starting orbits typical of Massalia family ($e_{P}$$=$$0.16$, $I_{P}$$=$$1.5^{\circ}$) and for Vesta family ($e_{P}$$=$$0.10$, $I_{P}$$=$$6.6^{\circ}$) asteroids respectively.
Reading the appropriate values from Table~\ref{tab:results100in} and extrapolating to $T=1$ Gyr as in \citeauthor{Gallardo.et.al2011} but using Eq~\ref{eq:power_law}, we find $\sigma_{e}(10^{9} \mbox{yr})=0.036$, $\sigma_{I}(10^{9} \mbox{yr})=0.6^{\circ}$ for Massalia-like and $0.066$/$0.7^{\circ}$ for Vesta-like starting orbits.
% , in fairly good agreement considering the differences in choosing the respective samples.

The robust diffusive growth rates and clear dependence of $\sigma_{e}(T)$ on $e_{P}(0)$ and $I_{P}(0)$ motivated us to model this quantity as an empirical function of the proper orbit so that
\begin{equation}
\sigma_{e,8}[e_{P}(0), I_{P}(0)]=c (1-b_{e} e_{P}(0)) (1-b_{I} I_{P}(0))
\label{eq:sigma_100m_fit}
\end{equation}
and fit it to the data in Table~\ref{tab:results100in}, obtaining $c = 0.0231\pm 0.0026$, $b_e = 3.238 \pm 0.188$ and $b_{I}  = -4.518  \pm 1.012$ where $I_{P}(0)$ is in radians. Fit residuals are shown as vertical blue segments in Fig.~\ref{fig:e_diffusion_fit}. These estimates are insensitive to the measurement uncertainties; the quoted values are formally valid for errors equal to 10\% of the nominal value for each datum. For the reasons put forward earlier in this Section we do not consider data points with $e_{P}(0)=0.23$ in the fit, but do include them in Fig.~\ref{fig:e_diffusion_fit}. Expression~(\ref{eq:sigma_100m_fit}) may then be used to interpolate $\sigma_{e,8}=\sigma_{e,8}[e_{P}(0), I_{P}(0)]$ within the domain $[0.11,0.23]\times[2.5^{\circ},10^{\circ}]$ in order to evaluate the e-diffusion function $\sigma_{e}(t)\simeq\sigma_{e,8}(t / 10^{8}\mbox{yr})^{0.4}$. Unfortunately, $\sigma_{I}(T)$ does not appear to depend on $e_{P}(0)$ and there is insufficient information to carry out the same exercise, other than to state that it appears to be an increasing function of $I_{P}(0)$.

Apart from the dispersion quantified through $\sigma(t)$, the other statistical measure of orbital mobility is the population {\it mean}. We carry out a similar procedure as for $\sigma(t)$ for each group of resonant particles and find (Fig.~\ref{fig:mean_100m_in2to1m}) the evolution of the means to be consistent with a constant time function given the 1-$\sigma$ dispersion of individual test particles diluted by $\sqrt{N}$ (dotted lines). Particles with $e_{P}(0)$$=$$0.23$ (yellow colour) exhibit the largest excursion from flat time-dependence, significant at $\sim$4$\sigma$. We attribute this to moderately close approaches with Mars that also increase $\sigma_{I}$ (Fig.~\ref{fig:sigma_100m_in2to1m}). Overall, we find no strong evidence for net orbital migration of asteroids in resonance, thus further strengthening the case for a causal link between Yarkovsky-induced migration and the unique orbital characteristics of asteroids in this resonance \citep{Gallardo.et.al2011,Dermott.et.al2021,Dermott.et.al2022}. A similar conclusion regarding the symmetry of the diffusive process may be drawn for particles outside the resonance (Fig.~\ref{fig:mean_100m_out2to1m}), though we note that the final mean values for the $e_{P}(0)$$=$$0.20$ groups - though still formally consistent with zero - are the highest within particles with the same starting inclination.
% This is related to the asymmetry observed for the same particle groups (see next section) and we attribute it to moderately close encounters with Mars.
% decreases for all particle sets. Such a trend would arise if particles diffusing to higher $e_{P}$ reach Mars-overlapping orbits before the end of the simulation, allowing those particles with $e_{P}(t) < e_{P}(0)$ to dominate the calculation of the mean. The largest excursions in $e_{P}$ and $I_{P}$ are significant at $\sim$4$\sigma$ but, overall, there is no strong evidence for net orbital migration of resonant asteroids beyond that expected for stochastically diffusing orbits.
% and that this is true for particles both in and out of the resonance. For particles with $e_{P}(0)$$=$$0.23$, $<$$e_{P}$$>$ decreases for all particle sets. Such a trend would arise if particles diffusing to higher $e_{P}$ reach Mars-overlapping orbits before the end of the simulation, allowing those particles with $e_{P}(t) < e_{P}(0)$ to dominate the calculation of the mean. The largest excursions in $e_{P}$ and $I_{P}$ are significant at $\sim$4$\sigma$ but, overall, there is no strong evidence for enhanced orbital migration of resonant asteroids beyond that expected for stochastically diffusing orbits.
\subsection{Gaussianity: Skewness and kurtosis}
An important question we can answer with the dataset at hand is how well - if at all - the dispersion may be approximated by a gaussian distribution. This will affect estimates of the number of particles reaching a given value of $e_{P}$ or $I_{P}$ since the area under the curve generally depends on the shape of the distribution. For this purpose, we have computed the third and fourth moments of our sample relative to the values expected for a gaussian distribution.  

In the top panels of Fig.~\ref{fig:kurtosisskewness_100m_in2to1m} we show the fourth statistical moment (kurtosis) of the diffusion of $e_{P}$ and $I_{P}$ over time, relative to the theoretical value of $3$ for a gaussian. Bottom panels show the third moment (skewness) for the same particle groups. Only the case for $I_{P}(0)$=$6.5^{\circ}$ is shown; similar results are obtained for $I_{P}(0)$=$2.5^{\circ}$ and $I_{P}(0)$=$10^{\circ}$ and we comment below on the differences. 

For the resonant particles, the distributions appear well approximated by gaussians. 
% For the third moment, ie the skewness, the result is consistent with symmetric distributions (bottom panels of Fig.~\ref{fig:kurtosisskewness_100m_in2to1m}) leading us to conclude that the proper element dispersion has yet another feature in common with a gaussian distribution, namely that evolution towards higher or lower values appears equally likely.
For the control particles (Fig.~\ref{fig:kurtosisskewness_100m_out2to1m}) we generally find a kurtosis excess i.\,e.~the distributions are leptokurtic, also the case for particles with $I_{P}(0)$$=$$2.5^{\circ}$ and $I_{P}(0)$$=$$10^{\circ}$, not shown here. We attribute this to the relatively large orbital variations suffered by a few particles (bottom panel of Fig.~\ref{fig:itime_100m} and main text) which act to broaden the tails of the distribution. The respective plots for the skewness (Fig.~\ref{fig:kurtosisskewness_100m_out2to1m}, bottom panels) point to fairly symmetric distributions but also hint at some moderate asymmetry of $I_{P}(t) - I_{P}(0)$ for $e_{P}(0)$$=$$0.20$ particles.
% for $I_{P}(0)=6.5^{\circ}$ and $I_{P}(0)=2.5^{\circ}$ (not shown here).  
\section{Diffusive loss of inner main belt asteroids}
\label{sec:loss}
Chaotic diffusion has consequences for the escape of asteroids from the main belt. Consider a resonant asteroid with $e_{P}(0)$$=$$0.19$ and $I_{P}(0)$$=$$6.5^{\circ}$ for which the eccentricity has diffused by $\Delta e_{P}$ = + 1-$\sigma$ from its starting location after $t=4$ Gyr. From Table~\ref{tab:results100in} and Eq~\ref{eq:power_law} we find a final proper eccentricity of $\sim$0.25, sufficient for the asteroid to suffer deep encounters with Mars and be removed. At the same time, an asteroid {\it outside} the resonance with a slightly higher $e_{P}(0)$$=$$0.20$ will suffer a diffusive eccentricity increase of $\sim 0.01$, not quite sufficient to allow escape. % Furthermore, e-diffusion within the resonance is stronger for low-eccentricity orbits; for instance, $e_{P}(0)=0.11$ and $I_{P}(0)=10^{\circ}$ is $\Delta e_{P}\simeq 0.1$, meaning that asteroids are mixing between families not well separated in inclination.

To estimate the fraction of escaping asteroids, we ignore the weaker $I$-diffusion and consider objects initially at $e=e_{0}$ diffusing according to Eq.~\ref{eq:power_law} with parameter values from Tables~\ref{tab:results100in} and \ref{tab:results100out}. After an interval $\Delta T$, the number of asteroids still in the belt will have been reduced by a factor $\Phi(z)$ where $z=\left(e_{0} - e_{\rm crit}\right)/\sigma(\Delta T)$. Here $e_{\rm crit}$ represents a threshold that the asteroid is certain to escape upon crossing and $\Phi$ is the cumulative normal distribution. If the asteroid eccentricity distribution is considered initially uniform, the fractional depletion $f$ over $\Delta T$ is
\begin{equation}
\label{eq:loss}
f=\frac{\Delta n_{0}}{n_{0}}=\int_{e_{\rm LB}}^{e_{\rm crit}} \Phi(z) d e_{0}\mbox{.}
\end{equation}
By evaluating Eq.~\ref{eq:loss} we estimate the number of resonant and non-resonant asteroids lost over $\Delta T$ = 4 Gyr. We set $b=0.4$ and $e_{\rm crit}=0.23$ for both sets while the scaling factor $\sigma_{e,8}$ is set to $0.0179$ for resonant and $0.0025$ for non-resonant asteroids, representing the respective arithmetic means for entries with $e_{P}(0)$$<$$0.23$ in Table~\ref{tab:results100in} and with $e_{P}(0)$$=$$0.20$ in Table~\ref{tab:results100out}. To account for the greater mobility of the resonant asteroids, the lower bound $e_{\rm LB}$ is set to zero for the resonant and to $0.19$ for the non-resonant set. For these parameter choices, we find $f_{\rm NR}=0.0044$ (0.4\%) and $f_{\rm RES}=0.0312$ (3\%). Note these estimates are insensitive to our choice of $e_{\rm crit}$ and, particularly for the non-resonant case, $e_{\rm LB}$.

The number of IMB asteroids with $H<16.5$ is $n_{\rm IMB}(16.5)=66,494$  \citep{Dermott.et.al2018}. For this observationally complete set, we can estimate the number of non-resonant asteroids that diffused to Mars-crossing orbits over 4 Gyr as $\Delta n_{0}(H)\simeq f n(H)$, evaluating to $0.0044\times n_{\rm IMB}(16.5)$$\sim$293. At the same time, the number of resonant asteroids \cite[assuming, for simplicity, that the resonance is defined by $2.4174$<a<$2.4194$ au;][]{Gallardo.et.al2011} is $n_{\rm RES}(16.5)=1118$ and here we estimate that the number of asteroids lost over 4 Gyr is $1118\times0.0312$$\sim$35. Considering that the 1M-2A resonance is one of several so-called Main Diffusion Tracks (MDTs) that help transport IMB asteroids to Mars-crossing orbits \citep{MorbidelliNesvorny1999}, the overall number of asteroids that escaped through these tracks over 4 Gyr should be similar to those originating from outside the MDTs, so that the overall number of escaped resonant and non-resonant asteroids is $\sim$600 asteroids or a flux of 0.15 $\mbox{Myr}^{-1}$.

We now compare this number to the flux $I$ of Intermediate Mars Crossers (IMCs) to the NEO region \citep{Bottke.et.al2002a} to test the assumption that IMCs are in a steady state. From Table~3 in that paper we read $F(H$$<$$18)=65 \pm 15$ $\mbox{Myr}^{-1}$, half of which ($33 \pm 8$ $\mbox{Myr}^{-1}$) is from the IMB. Correcting for the different magnitude limits using Eq.~4 in that paper, we obtain $F_{IMB}(16.5)=9.4\pm2.3$ $\mbox{Myr}^{-1}$. Our own flux estimate is $1/60$th of this figure,
% also between one and two orders of magnitude smaller than the two-thirds depletion OF the entire MB found by \citet{MintonMalhotra2010}.
probably reflecting the dominant role of Yarkovsky drag in mobilising the smaller, more numerous asteroids towards a Mars-crossing state. We illustrate this by noting that an asteroid at our completeness limit corresponds to $D=2$ km \cite[Eq.~7 of][]{Bottke.et.al2002a}
or $\dot{a} \simeq 5 \times 10^{-5}$ au $\mbox{Myr}^{-1}$ meaning that, over 4 Gyr, this asteroid would have drifted by up to $100\times$ the width of the 1M-2A resonance.
For a meaningful comparison between the fluxes, it is therefore important to limit our analysis to only those asteroids primarily mobilised by chaotic diffusion rather than Yarkovsky drag. For this purpose, we utilise the following empirical relationship \citep{Gallardo.et.al2011} between the orbital drift rate $\dot{a}$ and the eccentricity diffusion $\Delta e_{Y}$ due to the Yarkovsky effect acting on asteroids in the 1M-2A resonance
\begin{equation}
\Delta e_{Y} \sim 6 \mbox{ }\dot{a}\mbox{ } t^{\beta}_{\rm res}
\end{equation}
where $t_{\rm res}$ is the mean lifetime of the asteroid in the resonance. 

If we assume $\Delta e_{Y}$$=$$\sigma_{e,8}$ and $\beta=1$, for $t_{\rm res}=10^{2/3}$ Myr (cf Table 4 of \citeauthor{Gallardo.et.al2011}) we obtain $\dot{a}=3\times 10^{-5/-6}$ au $\mbox{Myr}^{-1}$ or $D=3.3/33 $ km. We adopt a typical value $D=10$ km or $H=13.2$ between these two extremes and, carrying on as before, we find $F_{\rm IMB}(13.2)\simeq 0.021 F_{\rm IMB}(18) = 1.4\pm0.3$ $\mbox{Myr}^{-1}$. Therefore, chaotic diffusion appears to account for $\sim$10\% of $H<13.2$ asteroids lost from the IMB. Recall, however, that in our calculation of $F$ we ignored Yarkovsky diffusion for asteroids with $H \lesssim 13.2$ where $\Delta e_{Y}$$\lesssim$$\sigma_{e,8}$. Moreover, we assumed that all the MDTs are similarly efficient in delivering asteroids to Mars-crossing orbits. This will generally not be true and, in particular, the region interior to $a=2.17$ au that is densely populated by Martian resonances \citep{MorbidelliNesvorny1999} may be substantially more prolific in contributing IMCs than other MDTs. 
Therefore, our approach probably underestimates the true flux of IMB asteroids to the IMC region - though the underestimate is likely not severe -  leading us to expect that a figure of, say, 30\% will be closer to the true value.

In conclusion, we find that the steady-state assumption for the IMC population \citep{Bottke.et.al2002a} is reasonable, at least for the larger ($D\gtrsim10$ km) members of the population.
% $n(<13.2)/n(<16.5)\sim 0.07$ or 
%
% For an H=16.5, D=2 km (Eq 7, Bottke 2002)
%
% de = 6 adot*tres^b. eq 2 Gallardo et al 2011
%
% It supports the authors' observation that the most efficient escape pathways for MB asteroids are strong mean motion resonances with Jupiter and the $\nu_{6}$ secular resonance.
%
% The width of the resonance is eccentricity-dependent but is typically $\Delta a_{res}=2\times 10^{-3}$ au while our sets of control particles span  a range of $4 \times 10^{-3}$ au.
%If $n(H)$ represent the number of IMB asteroids brighter than H, then $n_{\rm RES}(10.5)=2$, $n_{\rm RES}(16.5)=1.1\times10^{3}$, $n_{\rm NR}(10.5)=90$, $n_{\rm NR}(16.5)=6.4 \times 10^{4}$. Because $\Delta n_{0}(H)$$\simeq$$ f n(H)$, the number of asteroids lost by non-resonant chaotic diffusion over the entire IMB is $\Delta n_{\rm 0,NR}(10.5)=0.4$ and $\Delta n_{\rm 0,NR}(16.5)=283$ while for resonant asteroids we obtain $\Delta n_{\rm 0,RES}(10.5)=0.06$ and $\Delta n_{\rm 0,RES}(16.5)=34$.

\section{Close encounters with Mars and resonant protection}
\label{sec:close}
Though initially not a goal of our investigation, we realised in the course of this study that our simulations can also shed light on the protection of resonant asteroids against encounters with Mars. 

An asteroid on a planar heliocentric orbit may undergo encounters with Mars depending on its perihelion distance, Mars's aphelion distance and the relative orientation of the orbits. In general, a mean motion resonance will act to control the asteroid's location relative to the planet so that, even if the orbits cross \citep{Malhotra1996,MurrayDermott1999} the two objects do not physically approach each other. If the Martian orbit is also assumed planar with fixed semimajor axis $a_{M}$ and eccentricity $e_{M}$, encounters between Mars and the asteroid may occur when
\begin{equation}
\label{eq:qast_adistmars}
q \leq a_{M} \left(1 + e_{M}\right)
% a_{P} (1 - e_{P}) - \underset{\rm secular}{\rm max}\{q(t) - q_{P}\} < a_{M}(1 + e_{M})
\end{equation}
where $q$ is the asteroid perihelion distance.
%where $q - q_{P}$ represents the difference between the osculating and the proper pericentre distance of the asteroid. 
Because the orbits are subject to secular perturbations, we can write Eq.~\ref{eq:qast_adistmars} as 
\begin{equation}
\label{eq:qsecular}
q_{P} + \underset{\rm secular}{\rm min}\{q - q_{P}\} \leq a_{M}(1 + e_{M})
\end{equation}
where $q_{P}=a_{P} (1 - e_{P})$ is the proper perihelion distance of the asteroid. The second term on the left-hand-side represents the mimimum value of $q - q_{P}$ over a secular cycle; its value should initially be the same for asteroids with the same $e_{P}(0)$ but may gradually change as the orbits diffuse. 

Here we have carried out some additional simulations of particles outside the resonance with $e_{P}(0)=0.23$ and $e_{P}(0)=0.27$, the aim being to directly compare with the respective runs for resonant particles (Table~\ref{tab:sims}) and, in so doing, identify any differences due to the resonance. Note that these choices of initial conditions place the particles with $e_{P}(0)=0.23$ within the chaotic region identified by \citet{MorbidelliNesvorny1999} and roughly demarcated by proper perihelion distances $q_{P}$ between $1.84$ and $1.92$ au, while particles with $e_{P}(0)$$=$$0.27$ have $q_{P}$$=$$1.77$ au. In Fig.~\ref{fig:qast_vs_adistmars} we show the location of these particle groups relative to the equality condition for Eq.~\ref{eq:qsecular} (black lines) evaluated for different values of the martian eccentricity. Both groups are therefore expected to cross the orbit of Mars over the few-Myr duration of the martian secular cycle. For these new runs, the encounter detection feature of {\sl orbit9} was triggered for the same $d_{enc}=0.1$ au as for the resonant groups, to maintain consistency with those simulations. In the following analysis, we use the time of the 1st Mars encounter for each particle recorded during the simulation. Since particles typically encounter Mars repeatedly over short time periods, this quantity efficiently separates Mars-encountering from non-Mars-encountering particles.

We begin by highlighting several features of the time distribution of encounters. Fig.~\ref{fig:ec2327_mars} shows the cumulative distribution of the data with a bin size of 5 Myr. % The fraction of resonant particles that encounter Mars during the simulation is much less than that for the respective control particles.
Loss rates of resonant particles from the different eccentricity bins (red and red dashed lines) show a similar slope, pointing to a common loss mechanism. The loss rate of particles in the $e_{P}(0)=0.23$ control group (black line) follows a shallower slope, while all 400 particles in the $e_{P}(0)=0.27$ control group (black dashed line) become Mars-encountering from the outset. A fraction of particles from each group encounter Mars during the first few million yr of the runs. Notably, the $e_{P}(0)=0.23$ control group loses $\gtrsim 50$\% of particles in this way compared to $<2$\% of the corresponding resonant group. The fraction of particles lost at the end of these runs per group is: 61/400 (resonant, $e_{P}(0)=0.23$ group); 93/400 (control, $e_{P}(0)=0.23$ group); and 145/400 (resonant, $e_{P}(0)=0.27$ group).
% For the $e_{P}(0)=0.23$ particles, 
% initially $\sim$2\% of resonant particles are Mars-encountering compared to $\sim$50\% for the control set. During the simulation, 
% the fraction of Mars-encountering particles increase by similar amounts during the simulations, 69/400 or 18\% (resonant set) and 93/400 or 23\% (control set) respectively. However, 
% only 2\% of resonant particles encountered Mars within the first 2.5 myr, compared to $\gtrsim$50\% for the control particles. For the $e_{P}(0)=0.27$ group, this increase is higher, by 185/400 or 36\% for the resonant set perhaos reflecting the higher fraction of time spent in Mars-encountering orbits. Note that all non-resonant particles encounter Mars from the outset.
% Incremental time distribution of the close encounters 130 particles encounter Mars in the period 50-70 kyr and 430-450 kyr 31 further particles. During the same period only 2 respnant particle ecounte rMars.
% This can be attributed to the evolution of Mars's orbit, 

% Control particles have undergone Mars encounters by the end of the simulation compared to $\sim$15\% of resonant particles. Because the pericentre distances for the resonant and particles are identical (Fig.~\ref{fig:qast_vs_adistmars}), the difference must be caused by the 1M-2A resonance. 
The different loss rate for resonant and control particles with $e_{P}(0)$$=$$0.23$ prompted us to focus our attention on these two groups. The top panel of Fig.~\ref{fig:ec23_adistmars} shows the incremental distribution of Mars-encountering particles during the first 20 Myr of the simulation, 
%with a bin size of $0.1$ Myr
plotted against the osculating aphelion distance of Mars (grey line). The latter is modulated with a fast and a slow frequency, with periods of $\sim$$0.1$ Myr and $\sim$$2$ Myr respectively, and changing from $<$$1.55$ au ($e_{M}$$\sim$$0$) up to $1.7$ au ($e_{M}= 0.12-0.13$). We observe that encounters between Mars and the particles occur only when Mars's aphelion distance is near maximum during the 2 myr cycle. 
%, meaning that the rate of creation of Mars-Crossers from asteroids with $e_{P}\sim 0.23$ and $a \sim2.42$ au is modulated with a period of $\sim$2 myr. 
However, a large number of particles from the control group - 160 or $\sim$40\% of the total - have already encountered Mars by $t= 10^{6}$ yr. Rebinning the data with a still finer time resolution of $5 \times 10^{3}$ yr (bottom panel), we find that most particles encounter Mars during two episodes, one at 50$-$70 kyr and the other at 300$-$600 kyr, both occurring near local {\it minima}, rather than maxima, of $Q_{M}$.

To progress further, we introduce Mars-relative orbital elements, defined through the relationships
\begin{equation}
\label{eq:erel}
\setlength\arraycolsep{0pt}
\left( \begin{array}{r}  e_{\rm rel} \cos{\varpi_{\rm rel}} \\ e_{\rm rel} \sin{\varpi_{\rm rel}} \end{array} \right) = \mbox{ }\left( \begin{array}{r}  e \cos{\varpi} \\ e \sin{\varpi}  \end{array}  \right)\mbox{}-\mbox{ }\left( \begin{array}{r} e_{\rm M} \cos{\varpi_{\rm M}} \\ e_{\rm M} \sin{\varpi_{\rm M}} \end{array}  \right)
\end{equation}
and
\begin{equation}
\label{eq:irel}
\setlength\arraycolsep{0pt}
\left( \begin{array}{r} I_{\rm rel} \cos{\Omega_{\rm rel}} \\ I_{\rm rel} \sin{\Omega_{\rm rel}} \end{array} \right) = \mbox{ }\left( \begin{array}{r}  I \cos{\Omega} \\ I \sin{\Omega}  \end{array}  \right)\mbox{}-\mbox{ }\left( \begin{array}{r} I_{\rm M} \cos{\Omega_{\rm M}} \\ I_{\rm M} \sin{\Omega_{\rm M}} \end{array}  \right)\mbox{}
\end{equation}
where we refer to the respective vectors as $\mathbf{e_{\rm rel}\mbox{, }e\mbox{ and }e_{\rm M}}$ for the eccentricity and similarly for the inclination.
% We also see that all control particles with $e_{P}(0)=0.27$ encounter Mars immediately following the start of the runs but that about 25\% of control particles with $e_{P}(0)=0.23$ have not had a close encounter after 100 Myr. This prompted us to look at the time distribution of Mars encounters for the latter particles in more detail. In Fig.~\ref{fig:ec23_adistmars} we plot the number of $e_{P}(0)=0.23$ particles on a linear time scale and a smaller bin size of $0.2$ Myr, against the osculating aphelion distance of Mars (grey line). The latter is modulated with a frequency of 2-3 Myr and changes from $<1.55$ au (i.e.~an almost circular orbit) up to $1.7$ au. During each cycle, close encounters occur when Mars's aphelion is near maximum. Therefore, unlike particles with $e_{P}(0)$$=$$0.27$ that readily cross Mars's orbit at $t=0$, those with $e_{P}(0)$$=$$0.23$ do so only as their eccentricities gradually diffuse to higher values and when the planetary osculating eccentricity is high enough for the orbits to cross. A corollary of this result is that the rate of creation of new Mars-Crossers from asteroids with $e_{P}\sim 0.23$ and $a \sim2.42$ au is modulated with a frequency of a few million years. 
% Over longer periods, the rate will slowly decay with time but may increase again if a high enough number of new asteroids are created eg by a collision.

Relative elements at Mars encounter are shown in Fig.~\ref{fig:ereleireli_6p5d_100m_inout2to1m}. While the particle relative eccentricity $e_{\rm rel}$ may vary between $0.1$ and $0.35$, its value at the moment of Mars encounter is $\geq 0.3$. The distribution of $I_{\rm rel}$, on the other hand, both at encounter and away from encounter appear similar (top right panel). To illustrate how these features arise, we replot $e_{\rm rel}$ and $I_{\rm rel}$ in the bottom panels as functions of the longitudes of the Mars-relative pericentre $\varpi- \varpi_{M}$ and ascending node $\Omega - \Omega_{M}$ in polar coordinates. We find the apses of Mars-encountering particles and the planet to be anti-aligned (i.\,e.~$\Delta \varpi = \varpi - \varpi_{M} \simeq 180^{\circ}$) and the relative inclination $I_{\rm rel}$ to be positively correlated with $\Delta \Omega = \Omega - \Omega_{M}$, with most particles
%- 44 / 67 for the resonant and 83 / 138 for the control set - 
satisfying $|\Omega - \Omega_{M}| \leq 90^{\circ}$. 
% Note that values of $\Delta \varpi$ are, on average, slightly offset from $180^{\circ}$. This is an artifact of the finite sampling rate of the simulation output for the particles and Mars where, here, we use the sample immediately following the encounter epoch. To confirm this, we used instead the sample of points {\it preceding} and obtain a mirror-image distribution of points about the y-axis. 

The similarity in the distributions of resonant and control particles suggests a common mechanism delivering particles to Mars-encountering orbits.
The observed preference for apsidal anti-correlation and nodal correlation implies that particles approach the Mars-crossing state adiabatically\footnote{In other words, the orbit shape changes slowly compared to the orbital precession timescale} and physically cross as soon as $q (t)\simeq Q_{M}(t)$.
% and further satisfy $e_{\rm rel}(t) \simeq % e(t)+e_{\rm M}(t)$. 
The observed correlation between the relative inclinations and the relative nodes can then be understood in terms of the expected enhancement of Mars's impact cross section for nodally aligned orbits, everything else being equal \citep{Opik1976}. % Therefore, if the resonant condition has an effect on has apparently no effect on the relative orientation of the orbits.

To understand the poor correlation between particle encounters and the 100 kyr modulation of the martian apocentre distance (Fig.~\ref{fig:ec23_adistmars}, bottom panel), we write Eq.~\ref{eq:erel} as
\begin{equation}
\label{eq:erel_decomp}
 \mathbf{e}_{\rm rel} = \left(\mathbf{e}  - \mathbf{e}_{\rm forced}\right) + \left(\mathbf{e}_{\rm forced} - \mathbf{e}_{M}\right)
 \end{equation}
where $e_{\rm forced}$ is the forced part of the eccentricity and
\begin{equation}
\label{eq:e_eforced}
 \left(\mathbf{e}  - \mathbf{e}_{\rm forced}\right) = \mathbf{e}_{P} + \left(\parbox{15mm}{\centering High frequency terms}\right) + \left(\parbox{15mm}{\centering Chaotic diffusion}\right)\mbox{.}
\end{equation}
In the idealised case where the orbit evolution is determined by the secular dynamics alone \cite[e.g.~][]{MurrayDermott1999}, the vector $\mathbf{e}  - \mathbf{e}_{\rm forced}$ for particles with the same initial $e_{P}$ will lie along a circle of radius $e_{P}$. 
% In Section~\ref{sec:results}, the fast harmonics were suppressed through averaging in order to estimate $e_{P}$ as a function of time whereas here these same fast harmonics will be important in determining whether the osculating orbits cross.
The presence of high frequency terms smears out the circle into an annulus (Fig.~\ref{fig:ec23_ecc_vector}), the width of which - combined with the magnitude of the second term in Eq.~\ref{eq:erel_decomp} and maximised when the vectors $\mathbf{e}_{M}$ and $\mathbf{e}_{\rm forced}$ are anti-aligned - determines if and when the orbits of Mars and the particles cross.

Because the vector $\mathbf{e}_{\rm forced}$ is common for all particles, we can estimate it as the centre location of the best-fit circle to the data. We then use the estimate to calculate $\mathbf{e}_{M}$  - $\mathbf{e}_{\rm forced}$ and compare with the Mars encounter history of the particles. We observe (Fig.~\ref{fig:ec23_eforced}) excellent agreement between episodes of Mars encounters and times when $||\mathbf{e}_{M}  - \mathbf{e}_{\rm forced}||$ is near maximum and typically $\gtrsim 0.12$. Therefore, anti-alignment between the martian eccentricity and the asteroid forced vectors is one of the necessary conditions for orbits to cross over secular timescales. 

The other condition is that the radial scatter of eccentricity vectors within the annulus in Fig.~\ref{fig:ec23_ecc_vector} is large enough to allow at least some orbits to cross Mars's. At any moment in time, the scatter is made up of both short period terms and of the chaotic diffusion, which comes into play on $\gtrsim 10^{7}$ yr timescales. In quantifying the annulus contribution, we make use of the quantity
\begin{equation}
\label{eq:ecc_dispersion}
\delta_{e}(t)=\sqrt{V\left[||\mathbf{e}_{M}  - \mathbf{e}_{\rm forced}||(t)\right]}
\end{equation}
where $V[.]$ is the statistical variance calculated over all particles in the group. The dispersion arising from chaotic diffusion and quantified through Eq.~\ref{eq:sigma} can then be thought of as one component of $\delta_{e}$. Figure~\ref{fig:ec23_dispersion} shows the evolution of $\delta_{e}(t)$ over $10^{8}$ yr separately for the control and the resonant groups. Dispersion of the resonant group (red) increases gradually and smoothly until $t=10^{7}$ yr; said behaviour is mirrored by individual particle eccentricities and semimajor axes (middle and bottom panel), pointing to the action of chaotic diffusion in increasing $e_{P}$ and therefore the resonance libration width. It it worthwhile to point out here that $a$ and $e$ oscillations due to the resonance have periods $\lesssim$10 kyr \citep{Gallardo.et.al2011}, generally shorter than the output step in our simulations.

While both groups show the same dispersion evolution initially, the control group dispersion shows a steep rise at $t$$\simeq$$10^{5}$ yr, flattening out at $t$$\simeq$$10^{6}$ yr and suffering moderate, incremental changes thereafter. The steep initial increase of $\delta_{e}$, not observed for the resonant particles, is seen also in the time history of the individual particle eccentricities (middle panel) but is absent in the semimajor axis
(bottom panel). Possible causes for this behaviour could be the onset of additional harmonics in $e$ due to proximity to the resonance or the action of chaotic diffusion due to overlap of weak resonances for high-$e_{P}$ asteroids \citep{MorbidelliNesvorny1999}. In addition, this observation is consistent with a dependence of the chaotic diffusion timescale on the fastest harmonics in the motion. For a non-resonant asteroid, these would be the synodic harmonics with period $10^{0}$$-$$10^{1}$ yr, while for a resonant asteroid the harmonics associated with the critical angle, typically of period $10^{3}$$-$$10^{4}$ yr \citep{Gallardo.et.al2011}, will dominate. Distinguishing between those mechanisms is outside the scope of the present work but, in any case, the resonance appears capable of providing a ``safe haven'' for asteroids against its tempestuous surroundings.

On timescales longer than a Myr we observe changes or ``kinks'' in the slope of the dispersion profiles, accompanied by simultaneous changes in the $e$ and/or $a$ distributions. For instance, the kink at $t$$\simeq$$10^{7}$yr for resonant particles shows simultaneous but moderate excitation of the semimajor axis, consistent with leakage of resonant particles into the chaotic region \citep{MorbidelliNesvorny1999,Gallardo.et.al2011}. Dispersion for the control particles shows two kinks, one at $t\simeq10^{7}$ yr and the other at $t\simeq 4 \times 10^{7}$ yr. We attribute both features to the continuous action of chaotic diffusion allowing control particles to reach the resonance separatrix \citep{Gallardo.et.al2011}. However, we see no clear evidence of orbit changes - beyond the offset discussed in Section 3 - attributable to encounters with Mars on the control or, for that matter, the resonant particle group which would significantly affect both $a$ and $e$. 

% to same cause as for the resonant set, since by then many control particles have been observed to reach the resonance border. The second kink is accompanied by larger changes in both $e$ and $a$ and may be caused by particles encountering Mars sufficiently close to affect the orbit.  
We conclude this Section with the synoptic observation that resonant particles, although less dispersed initially, eventually become similarly dispersed to the control particles. In this sense, the mode of resonant protection studied in this work can be effective for $\gtrsim$$10^{7}$ yr but not for significantly longer.
\section{Conclusions and Discussion}
In this paper we have studied the process of chaotic diffusion for asteroids in the vicinity of the exterior 1:2 mean motion resonance with Mars, in order to better understand and quantify long-term orbit mobility when non-gravitational forces are either weak or absent. To this end, we carried out numerical simulations of particles in precisely controlled sets of initial orbits, chosen to map out the diffusion both within and outside the resonance. Analysis of the simulation results supports the following main conclusions:
%% The last numbered section should briefly summarise what has been done, and describe the final conclusions which the authors draw from their work.
\begin{itemize}
    \item[--] 
     Chaotic diffusion of resonant asteroid orbits is a zero-mean random process where the proper $e$ and $I$ disperse over time as gaussian variates with $\sigma(t)$$\propto$$t^{0.4}$. The scaling constant in this relationship depends on the starting proper elements, so that $\sigma_{e}$ increases with decreasing $e_{P}(0)$ and increasing $I_{P}(0)$, while $\sigma_{I}$ increases with $I_{P}(0)$ but appears independent of $e_{P}(0)$. The highest $e$ dispersion - $\lesssim$$0.1$ when extrapolated over $2 \times 10^{9}$ yr - is observed for high $I_{P}(0)$ and low $e_{P}(0)$; $I$ dispersion amounts to $\sim$$1^{\circ}$ over the same period.
     \item[--]
    Chaotic diffusion is otherwise extremely limited outside the resonance so that, over the age of the solar system, we project $\sigma_{e}$$<$$0.012$ and $\sigma_{I}$$<$$0.7^{\circ}$ for non-resonant asteroids. For orbits with $e_{P}(0) \gtrsim 0.2$, the slope of the dispersion relation is similar to that for resonant asteroids while it is shallower for lower-$e_{P}(0)$ orbits. This latter observation could be the result of the overlapping of weak resonances at high eccentricity or may reflect a limitation in our ability to measure the weak chaotic diffusion of low-$e$ orbits.
\end{itemize}

In addition, the resonance acts to delay the increase of eccentricity, by $1-4 \times 10^{7}$ yr, for asteroids initially in non-Mars-orbit-crossing orbits, relative to asteroids not in resonance. That should lead to a corresponding delay for the closest, orbit-scattering, encounters although our simulations were not designed to test this.

Our findings confirm earlier conclusions, based on a more limited set of initial orbits \citep{Dermott.et.al2018}, that chaotic orbital evolution in the inner main belt can result in significant dispersion of the orbits on timescales less than the age of the Solar System. However, for those asteroids principally mobilised by chaotic diffusion rather than Yarkovsky drag, transport and mixing across regions with distinctly different $e$ and $I$ is extremely limited, supporting the argument by \citet{Dermott.et.al2021} that high-inclination asteroids in the IMB are dynamically isolated. In addition, the diffusive loss rate of large ($D\gtrsim 10$ km) asteroids from the IMB appears to be on a par with the inferred supply rate from the Mars-crossing population \citep{Bottke.et.al2002a}, suggesting that gradual chaotic evolution can account for that component of NEAs. 

Implicit in our investigation has been the assumption that the average eccentricity of Mars does not vary over timescales exceeding those in our simulations. Numerical experiments have, however, shown that changes of order a few $\times$$10^{-2}$ are possible on Gyr timescales \citep{Cuk.et.al2015,CukNesvorny2018}. For asteroids near the 1M-2A resonance, such a change would cause the locus for Mars-crossing orbits (Fig.~\ref{fig:qast_vs_adistmars}) to shift to either higher or lower eccentricity, affecting in turn the asteroid loss rate from the belt (Section~\ref{sec:loss}). A quantitative assessment of this effect is outside the scope of this work, but we cite here the relatively high phase-space density of $e_{P}$$\gtrsim$$0.2$ asteroids (see  Figs~\ref{fig:ic_100m} and \ref{fig:qast_vs_adistmars}) as an indication that the upper end of martian eccentricity values has not been significantly higher in the past than the present, at least over the typical replenishment timescale for high-$e_{P}$ asteroids.

Since our investigation focused on a single mean motion resonance and a single IMB location, a question that naturally arises is whether the character of chaotic diffusion measured in the vicinity of this resonance is also shared by other resonances and locations in the IMB. The mapping out of this fundamental dynamical property will allow us to refine projections of asteroid migration both through and out of the IMB. In the same spirit, it is important to extend the scope of these simulations to account for Yarkovsky drag affecting the more numerous smaller asteroids. This will allow construction of comprehensive statistical models to constrain asteroid evolution from both existing and new observations.
%allows mixing of resonant asteroids between neighbouring families. Taking, for example, Nysa-Polana-Eulalia (#405) and Massalia (#404) we have $e_{\rm N-P-Eu}=0.174 \pm 0.049$, $e_{\rm Mass}=0.166 \pm 0.028$ and $I_{\rm N-P-Eu}=2.85 \pm 1.03$, $I_{\rm Mass}=1.33 \pm 0.49$. Reading off the appropriate values from Table\ref{tab:results100in} we find that an asteroid with $I_{P}= 2.85$ will reach Massalia ($I_{P}<1.82^{\circ}$) after
% AN asteroid at the same location as Eulalia 
% Polana $e=0.158$ and $Ip=3.22$
% 1 out of 20 asteroids deep within the fmaily wil lhave crossed into Massalia
% According to NPEu spans the range from 0.125 to 0.222 in $e_{P}$ and $1.82^{\circ}$ to $3.88^{\circ}$ in $I_{P}$ while Massalia extends from 0.138 to 0.194 and from $0.84^{\circ}$  to $1.82^{\circ}$, therefore the respective family centers 
% NPEu: 0.1735 \pm 0.049,  2.85 \pm 1.03
% Mass: 0.166  \pm 0.028, 1.33 \pm 0.49 
\section*{Acknowledgements}
The authors wish to thank the reviewer, Jorge Correa-Otto, for his insightful comments that improved the paper. We are grateful to Matt Glover and the University of Florida Department of Astronomy for the provision of computational facilities and support. Astronomical Research at the Armagh Observatory and Planetarium is grant-aided by the Northern Ireland Department for Communities.
\section*{Data Availability}
The data underlying this paper were accessed from the Near Earth Objects Dynamic site (https://newton.spacedys.com/neodys/), the Asteroids Dynamic site (https://newton.spacedys.com/astdys/) and the JPL HORIZONS ephemeris service (https://ssd.jpl.nasa.gov/?horizons\#telnet). The derived data generated in this research are available from the corresponding author upon reasonable request.
%% The Acknowledgements section is not numbered. Here you can thank helpful colleagues, acknowledge funding agencies, telescopes and facilities used etc.
%% Try to keep it short.

%%%%%%%%%%%%%%%%%%%%%%%%%%%%%%%%%%%%%%%%%%%%%%%%%%

%%%%%%%%%%%%%%%%%%%% REFERENCES %%%%%%%%%%%%%%%%%%

% The best way to enter references is to use BibTeX:

\bibliographystyle{mnras}
\bibliography{cdl_mnras_2022} % if your bibtex file is called example.bib

% Alternatively you could enter them by hand, like this:
% This method is tedious and prone to error if you have lots of references
% \begin{thebibliography}{99}
% \bibitem[\protect\citeauthoryear{Author}{2012}]{Author2012}
% Author A.~N., 2013, Journal of Improbable Astronomy, 1, 1
% \bibitem[\protect\citeauthoryear{Others}{2013}]{Others2013}
% Others S., 2012, Journal of Interesting Stuff, 17, 198
% \end{thebibliography}
\clearpage
%%%%%%%%%%%%%%%%%%%%%%%%%%%%%%%%%%%%%%%%%%%%%%%%%%

%%%%%%%%%%%%%%%%% APPENDICES %%%%%%%%%%%%%%%%%%%%%

%% \appendix

\section*{Appendix A: Creating synthetic populations of asteroids with common proper elements.}
\label{sec:appendixa}
\renewcommand\thefigure{A\arabic{figure}}
For each test particle in our simulations, the semimajor axis $a$, eccentricity $e$, inclination $I$ as well as two angles $\varpi$ and $\Omega$ that define the orientation of the orbit {\sl osculate} due to the gravitational pull of the planets \citep{MurrayDermott1999}. Apart from the semimajor axis, the variation can be regarded as the vector sum of the respective {\sl forced} and {\sl proper} vectors, e.\,g. for the eccentricity
\begin{equation}
\label{eq:decomp}
\setlength\arraycolsep{0pt}
\left( \begin{array}{r}  e \cos{\varpi} \\ e \sin{\varpi} \end{array} \right) = \mbox{ }\left( \begin{array}{r}  e_{\rm forced} \cos{\varpi_{\rm forced}} \\ e_{\rm forced} \sin{\varpi_{\rm forced}}  \end{array}  \right)\mbox{}+\mbox{ }\left( \begin{array}{r} e_{P} \cos{\varpi_{P}} \\ e_{P} \sin{\varpi_{P}} \end{array}  \right)
\end{equation}
where the forced component depends primarily on the semimajor axis $a$ of the asteroid and of the planets and on the planetary masses, while the proper component has a constant modulus that is intrinsic to that asteroid. The proper semimajor axis $a_{P}$ - or, equivalently, the proper mean motion $n_{P}$ from Kepler's third law - is also constant and approximately the long-term average of $a$.

\begin{figure}
\hspace{-1mm}\includegraphics[angle=0,width=88mm]{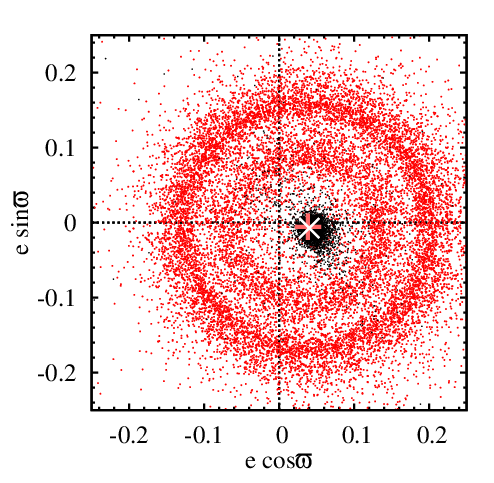}\\
\includegraphics[angle=0,width=87mm]{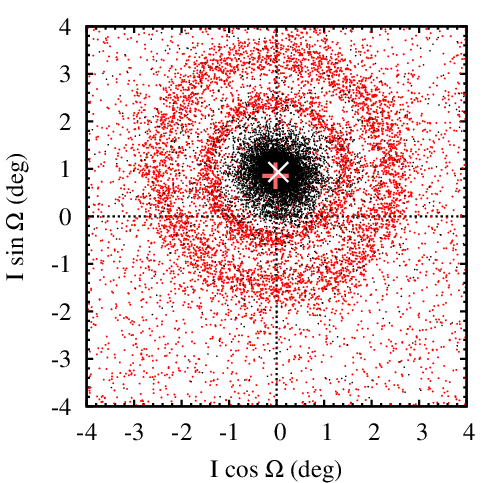}
\caption{Osculating (red) and forced (black) elements for asteroids near the 1M-2A resonance as described in the text. The large plus sign and cross symbols indicate the respective geometric averages.}
\label{fig:forced}
\end{figure}
To generate test particle orbits with a given proper eccentricity and/or inclination for the simulations we first determine the respective forced vector at or near the starting epoch. This vector is common to all asteroids with the same $a$ and is obtained by averaging the osculating element vectors for a large enough number of asteroids. We describe here the procedure for the eccentricity.

We searched through a catalogue of numbered asteroid proper elements available through the {\sl AstDys} information service \citep[][accessed in April 2014]{ASTDYS} and identified 14,653 asteroids with proper mean motion $n_{P}$ between 94.5 and 96.5 deg/yr, straddling the 1A-2M resonance. We then calculate the common forced vector of the asteroids in two different ways:
\begin{itemize}
\item[--]
\noindent
By processing asteroid osculating elements obtained from {\sl AstDys} for MJD 57000 = 2014 Dec 9.0 UT and using available tools within {\sl orbfit} \citep{ORBFIT} to calculate their proper elements as per \citet{KnezevicMilani2003}. From Eq.~\ref{eq:decomp} the forced vector for each asteroid is
\begin{equation}
\setlength\arraycolsep{0pt}
\left( \begin{array}{r}  e_{\rm forced} \cos{\varpi_{\rm forced}} \\ e_{\rm forced} \sin{\varpi_{\rm forced}} \end{array} \right) = \mbox{ }\left( \begin{array}{r}  e \cos{\varpi} \\ e \sin{\varpi}  \end{array}  \right)\mbox{}-\mbox{ }\left( \begin{array}{r} e_{P} \cos{\varpi_{P}} \\ e_{P} \sin{\varpi_{P}} \end{array}  \right)\mbox{.}
\end{equation}
The common force vector is then estimated as the geometric average $<$$\left( e_{\rm forced} \cos{\varpi_{\rm forced}},e_{\rm forced} \sin{\varpi_{\rm forced}} \right)$$>$ of these vectors for all the asteroids.
\item[--]
\noindent
By forming the osculating vector for each asteroid and taking the average $<$$\left( e \cos{\varpi},e \sin{\varpi} \right)$$>$. If the azimuths of proper vectors are assumed random, averaging estimates the common forced vector.
\end{itemize}
Fig.~\ref{fig:forced} shows good agreement between the two methods, the result differing by $2 \times 10^{-3}$ or $\sim$5\% for the eccentricity vector and $0^{\circ}$.1 or $\sim$10\% for the inclination vector.
% The rings apparent in the distribution of osculating elements are due to family asteroids with similar proper $e$ \& $I$.
 % [312 words].
Since the second method requires {\it a priori} knowledge of the osculating elements only, we have adopted it for the work reported here.
\clearpage

\renewcommand\thefigure{\arabic{figure}}
\setcounter{figure}{0}
\hspace{-5mm}
\begin{table*}
\caption[Summary of the numerical simulations in this work.]{Summary of the numerical simulations in this work.}
\begin{tabular}{l@{}c@{}cc@{  }cc@{}cc@{}}
\noalign{\smallskip}
\hline \hline
      & Duration  & \multicolumn{2}{c}{Number of orbits} &  \multicolumn{2}{c}{Proper Eccentricity} &  \multicolumn{2}{c}{Proper Inclination (deg)}   \\
Designation        & (Myr) &  ({\bf Res}onant) &   ({\bf Con}trol)  &   ({\bf Res}onant) &   ({\bf Con}trol)  &  ({\bf Res}onant) &   ({\bf Con}trol)   \\\hline \noalign{\smallskip}
{\tt 1M2A025100}    &   100   & 4 $\times$ 400 & 3 $\times$ 400  &   0.15, 0.19, 0.23, 0.27 & 0.15, 0.19, 0.20  & $2.5^{\circ}$ & $2.5^{\circ}$  \\
{\tt 1M2A065100}    &   100   & 5 $\times$ 400   & 4 $\times$ 400  &   0.11, 0.15, 0.19, 0.23, 0.27 & 0.11, 0.15, 0.19, 0.20, 0.23${}^{\dagger}$, 0.27${}^{\dagger}$ &   $6.5^{\circ}$ & $6.5^{\circ}$ \\
{\tt 1M2A100100}    &   100   & 5 $\times$ 400  & 4 $\times$ 400   &   0.11, 0.15, 0.19, 0.23, 0.27 & 0.11, 0.15, 0.19, 0.20 &   $10^{\circ}$ & $10^{\circ}$\\
\hline \hline
\multicolumn{8}{l}{\parbox{117mm}{${}^{\dagger}$Additional runs used to investigate protection from Mars encounters in Section~\ref{sec:close}.}}
\end{tabular}
\label{tab:sims}
\end{table*}
\clearpage

\begin{table*}
\centering
\caption[Diffusion parameters for the different groups of resonant test particles in the numerical runs. For each group we report the estimated exponent $b$ and, in brackets, the value of $\sigma_{8}$ from the power-law fit (Eq.~\ref{eq:power_law}) to the data.]{Diffusion parameters for the different groups of resonant test particles in the numerical runs. For each group we report the estimated exponent $b$ and, in brackets, the value of $\sigma_{8}$ from the power-law fit (Eq.~\ref{eq:power_law}) to the data.}
\begin{tabular}{lccccc}
\noalign{\smallskip}
\hline \hline
\noalign{\smallskip}
      $e_{P}(0)$          &  0.11  & 0.15 & 0.19 & 0.23 & 0.27  \\\hline
         &     \multicolumn{5}{c}{ Eccentricity   }   \\\hline
{\tt Res1M2A025100}        &  -- & 0.423 (0.0135) & 0.422 (0.0107) & 0.401 (0.0096)& --  \\
{\tt Res1M2A065100}      &   0.423 (0.0248) & 0.418 (0.0188) & 0.418 (0.0132) & 0.306 (0.0083) & --  \\
{\tt Res1M2A100100}      &   0.384 (0.0244) & 0.434 (0.0219) & 0.413 (0.0158) & 0.320 (0.0101)& --   \\  \hline
         &     \multicolumn{5}{c}{ Inclination${}^{\dagger}$ }   \\\hline \noalign{\smallskip}
{\tt Res1M2A025100}        &  -- & 0.429 (0.230) & 0.472 (0.194) & 0.350 (0.840)& --  \\
{\tt Res1M2A065100}      &   0.466 (0.233) & 0.435 (0.242)& 0.518 (0.235)& 0.387 (0.446) & --  \\
{\tt Res1M2A100100}      &   0.356 (0.325) & 0.447 (0.392)& 0.370 (0.377) & 0.436 (0.419)& -- \\
 \hline \hline
\multicolumn{6}{l}{\parbox{117mm}{
${}^{\dagger}$The value for $\sigma_{I,8}$ is given in degrees.}}
\end{tabular}
\label{tab:results100in}
\end{table*}
\clearpage

\begin{table*}
\caption[As Table~\ref{tab:results100in} but for the control set of particles started outside the resonance.]{As Table~\ref{tab:results100in} but for the control set of particles started outside the 1M-2A resonance.}
\begin{tabular}{lccccc@{}cc@{}c}
\noalign{\smallskip}
\hline \hline
\noalign{\smallskip}
      $e_{P}(0)$          &  0.11 &  0.15  &  0.19 &  0.20 &  \multicolumn{2}{c}{0.23} &  \multicolumn{2}{c}{0.27}   \\\hline
         &    \multicolumn{8}{c}{Eccentricity}   \\\hline
{\tt Con1M2A025100}  & -- & 0.044 (0.0006) & 0.064 (0.0012)  &  0.336 (0.0022)  &   \multicolumn{2}{c}{--}  &  \multicolumn{2}{c}{--} \\
{\tt Con1M2A065100}      & 0.238 (0.0013) & $-$0.005 (0.0010) & 0.150 (0.0014)   &  0.349 (0.0026)  &   &  &   & \\
{\tt Con1M2A100100}  &  0.028 (0.0022) &  0.055 (0.0020) & 0.194 (0.0022) &  0.316 (0.0028)    &   &  &   &  \\  \hline
         &     \multicolumn{8}{c}{Inclination}  \\\hline \noalign{\smallskip}
{\tt Con1M2A025100}      & -- & 0.303 (0.016) & 0.434 (0.030) & 0.534 (0.055)   & \multicolumn{2}{c}{--}   & \multicolumn{2}{c}{--}   \\
{\tt Con1M2A065100}      &  0.246 (0.021) & 0.130 (0.012) & 0.417 (0.042)  &  0.400 (0.129)   &   &  &   & \\
{\tt Con1M2A100100}  & 0.033 (0.047) & 0.033 (0.062) & 0.197 (0.101)   &    0.342 (0.149)   &   &  &   &  \\
\hline \hline  \noalign{\smallskip}
\multicolumn{9}{l}{\parbox{\columnwidth}{
% ${}^{\dagger}$The value for $\sigma_{I}$ is given in degrees.
% ${}^{\ast}$Extrapolated from  $t=t_{0}+74$ Myr.
}}
\end{tabular}
\label{tab:results100out}
\end{table*}
\clearpage

\begin{figure}
\hspace*{-4mm}\includegraphics[angle=0,width=8.9cm]{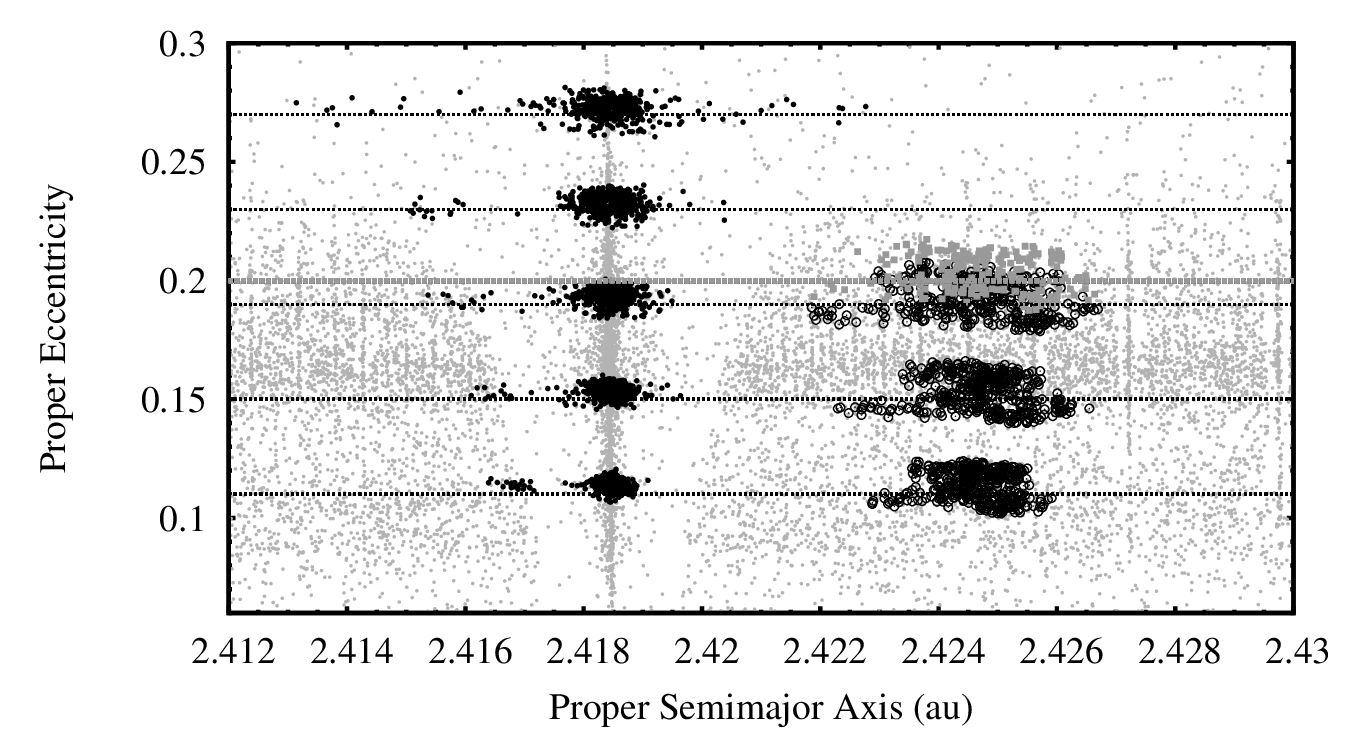}
\vspace{-2mm}
\hspace*{-0.5mm}\includegraphics[angle=0,width=9.1cm]{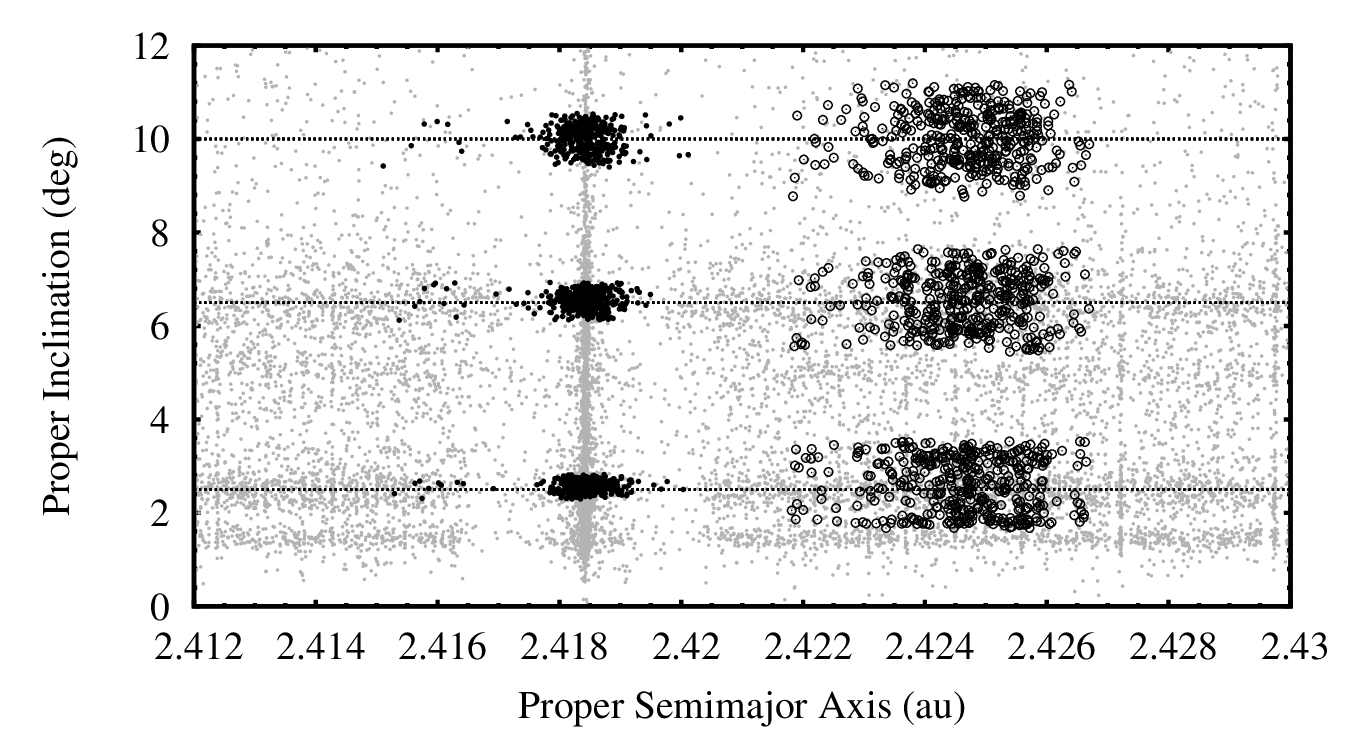}
\caption{Starting proper elements for test particles in the vicinity of the 1M-2A resonance. Small grey points represent numbered and multi-opposition asteroid orbits from {\sl AstDys} \citep{ASTDYS}. Horizontal lines indicate the target values for our procedure to generate starting orbits (Section~\ref{sec:methods}). {\it Top}: Proper eccentricity for particles with $I_{P}(0)$=$6.5^{\circ}$ where large grey points represent the group with $e_{P}(0)$=$0.20$. {\it Bottom}: Proper inclination for particles with $e_{P}(0)$=$0.19$.}
\label{fig:ic_100m}
\end{figure}
\clearpage

\begin{figure}
\centering
\hspace*{-4mm}\includegraphics[angle=0,width=87mm]{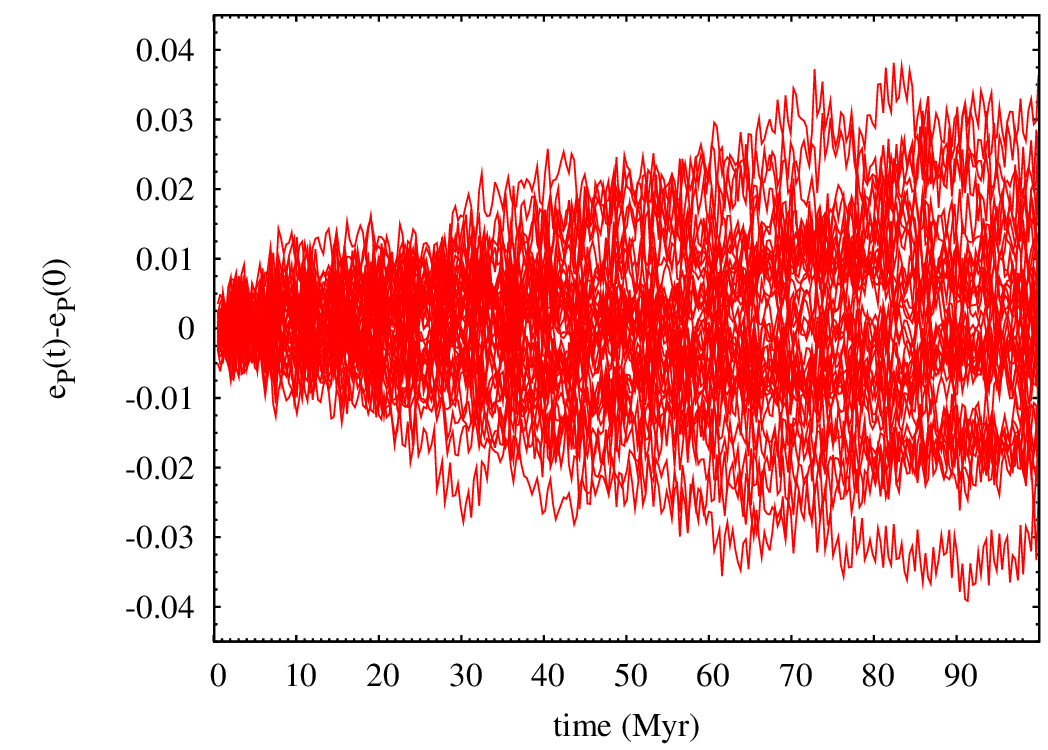}\\
\hspace*{-2mm}\includegraphics[angle=0,width=87mm]{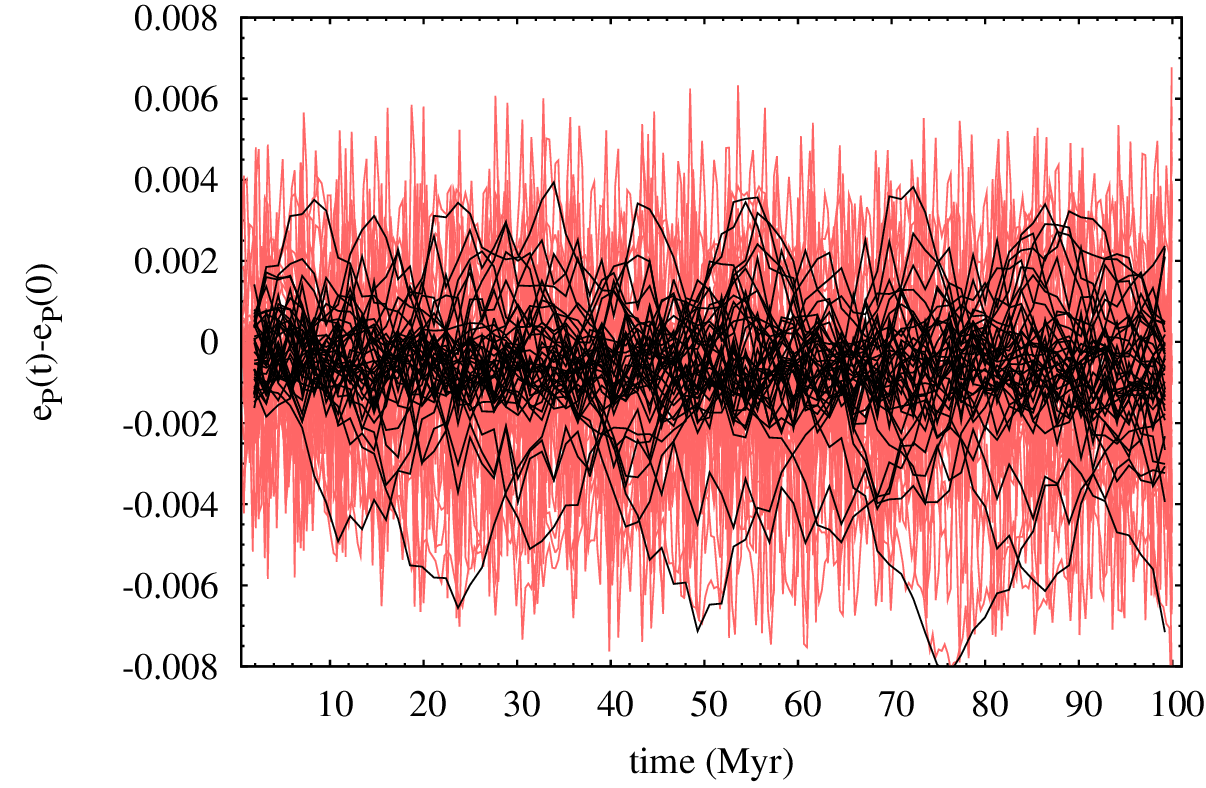}
\caption{Examples of proper eccentricity diffusion in the vicinity of the 1M-2A resonance. {\it Top}: Particles in resonance. {\it Bottom}: Particles outside the resonance. Only 1 for every 10 particles are shown to preserve plot clarity. Note the $\sim$$6\times$ finer vertical scale in the bottom panel.}
\label{fig:itime_100m}
\end{figure}
\clearpage

\begin{figure*}
\includegraphics[angle=0,width=5.5cm]{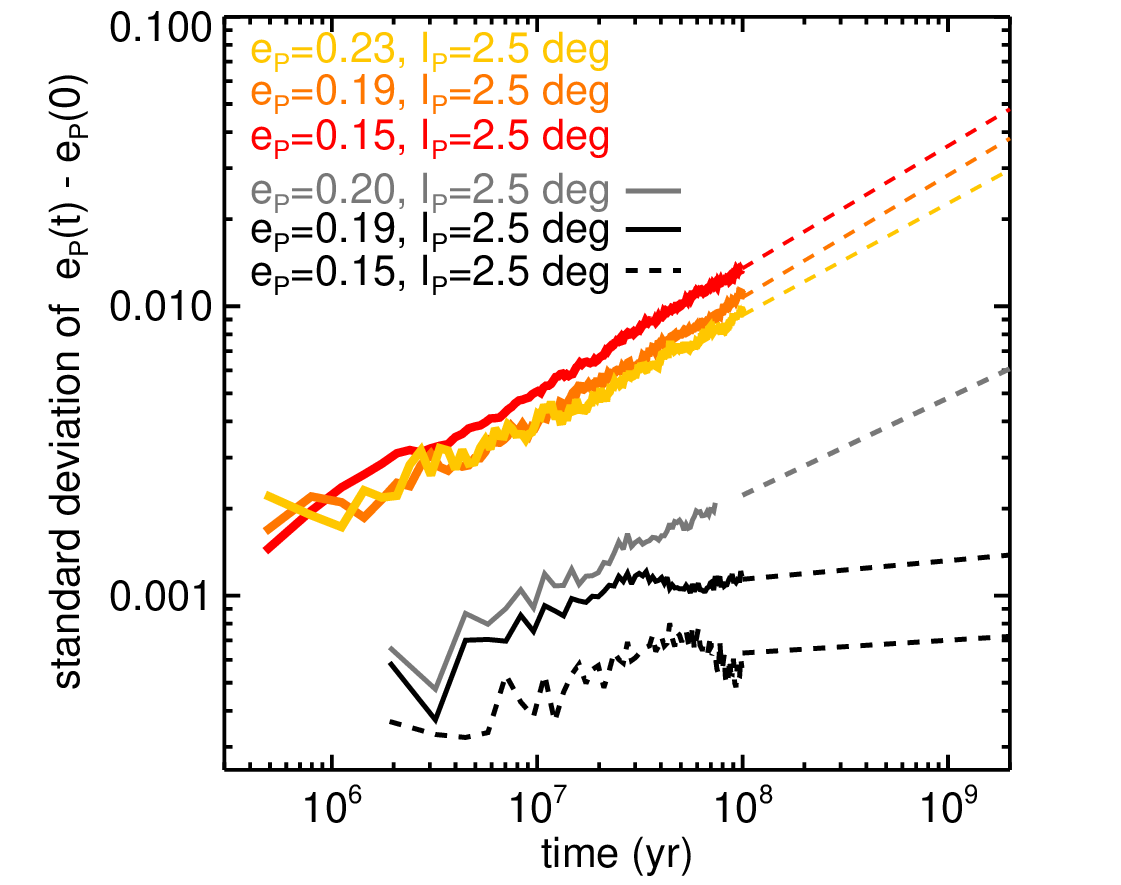}\includegraphics[angle=0,width=5.5cm]{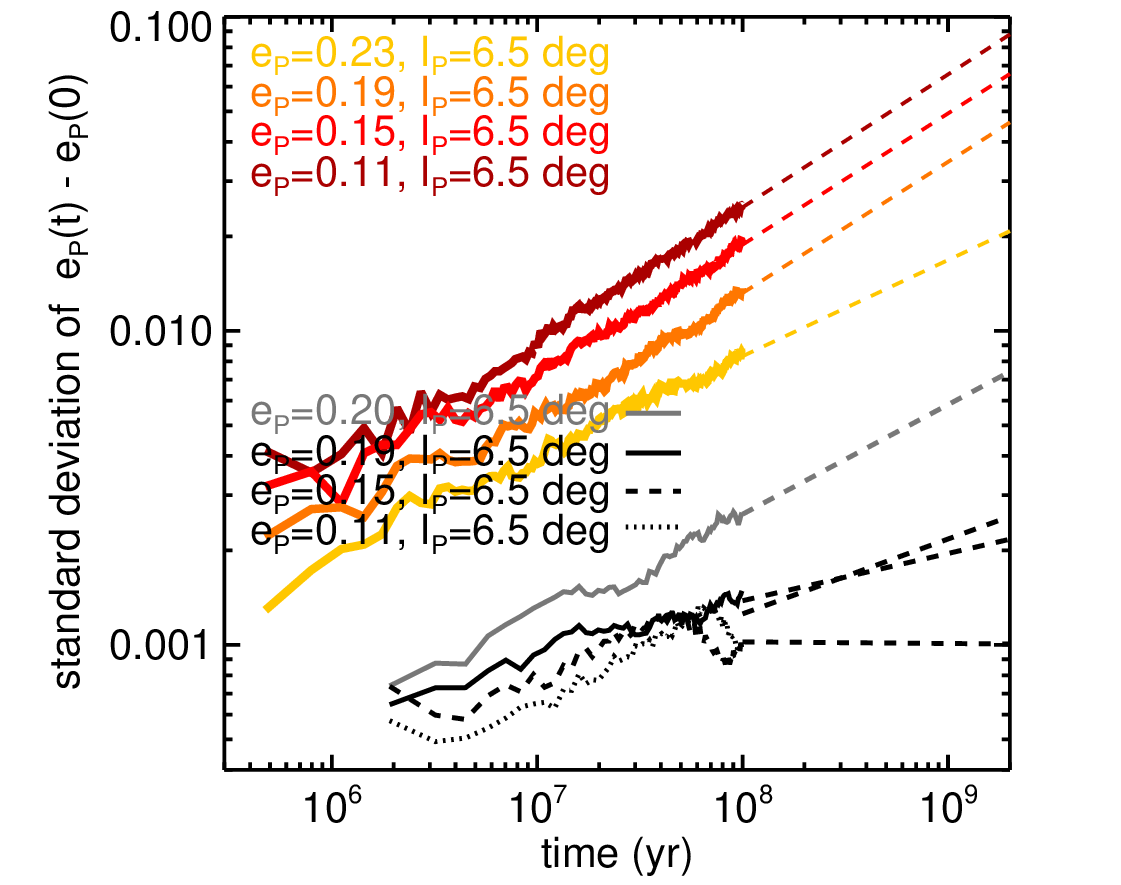}\includegraphics[angle=0,width=5.5cm]{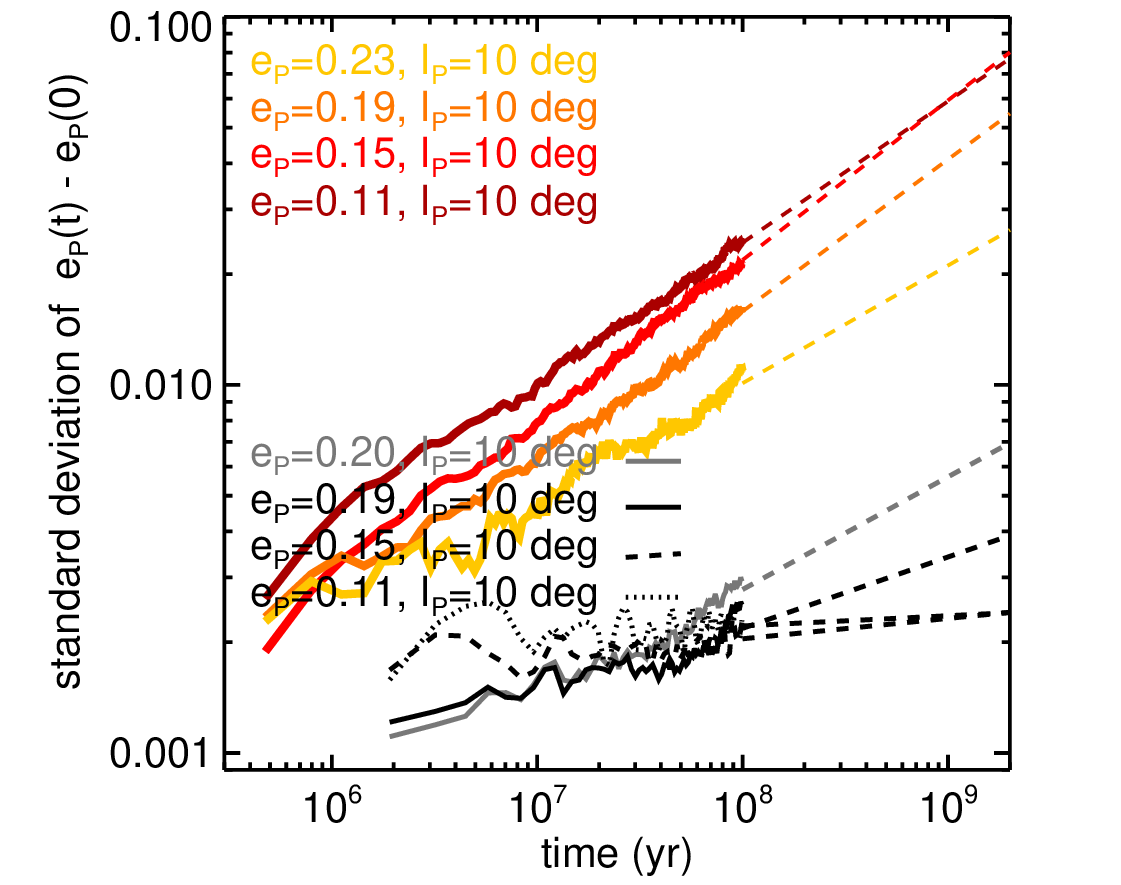}\\
\includegraphics[angle=0,width=5.5cm]{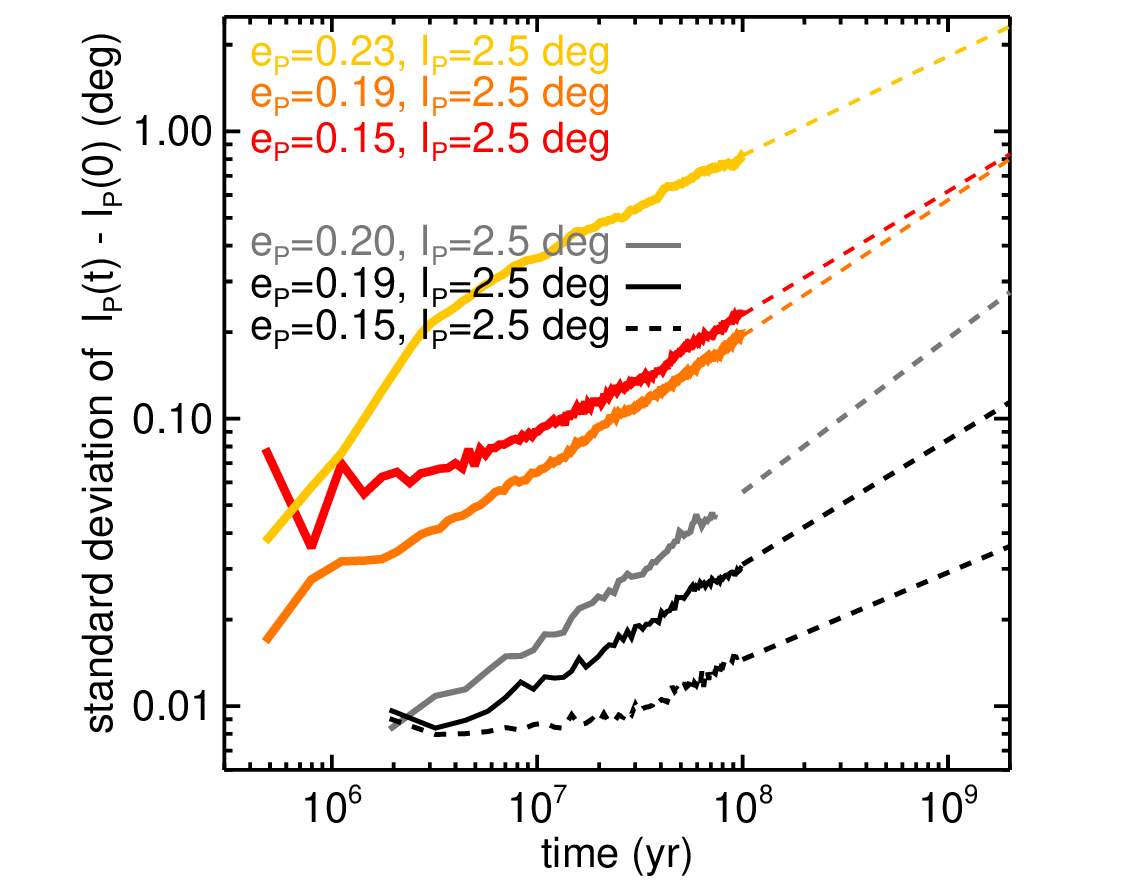}\includegraphics[angle=0,width=5.5cm]{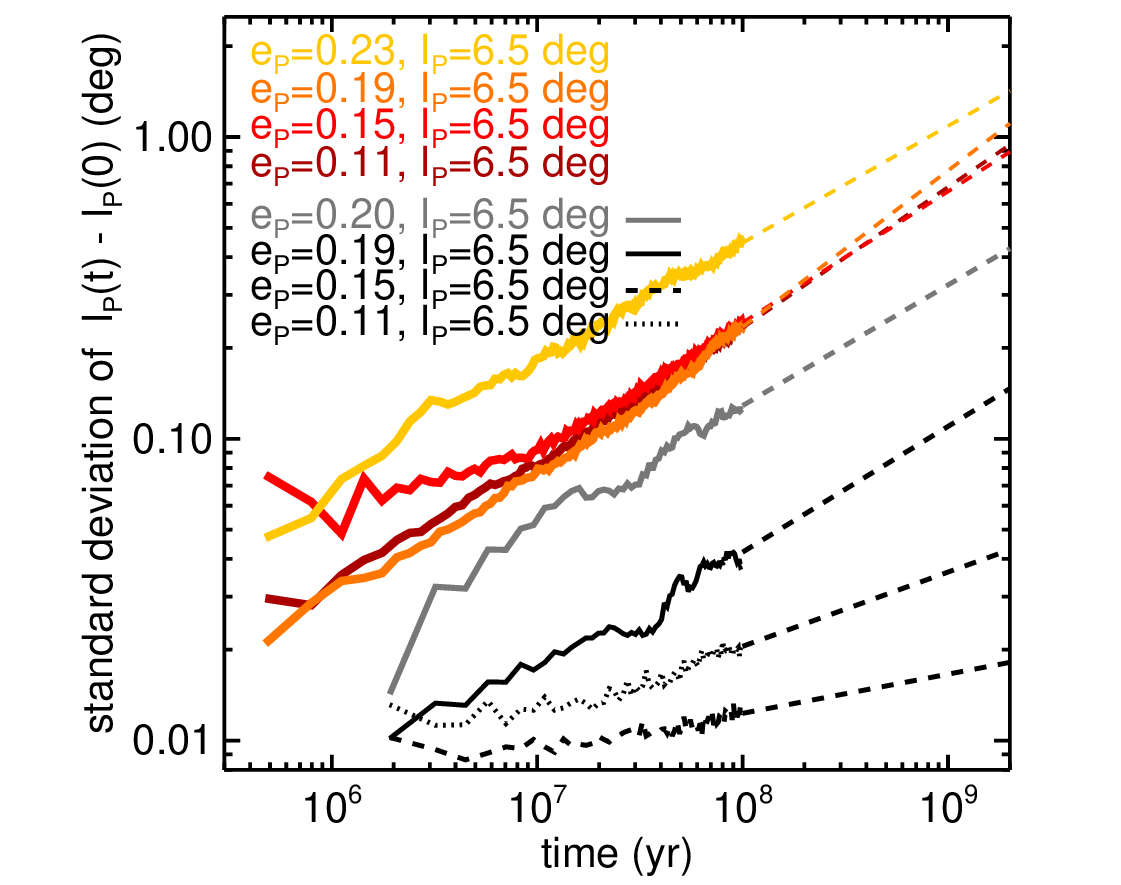}\includegraphics[angle=0,width=5.5cm]{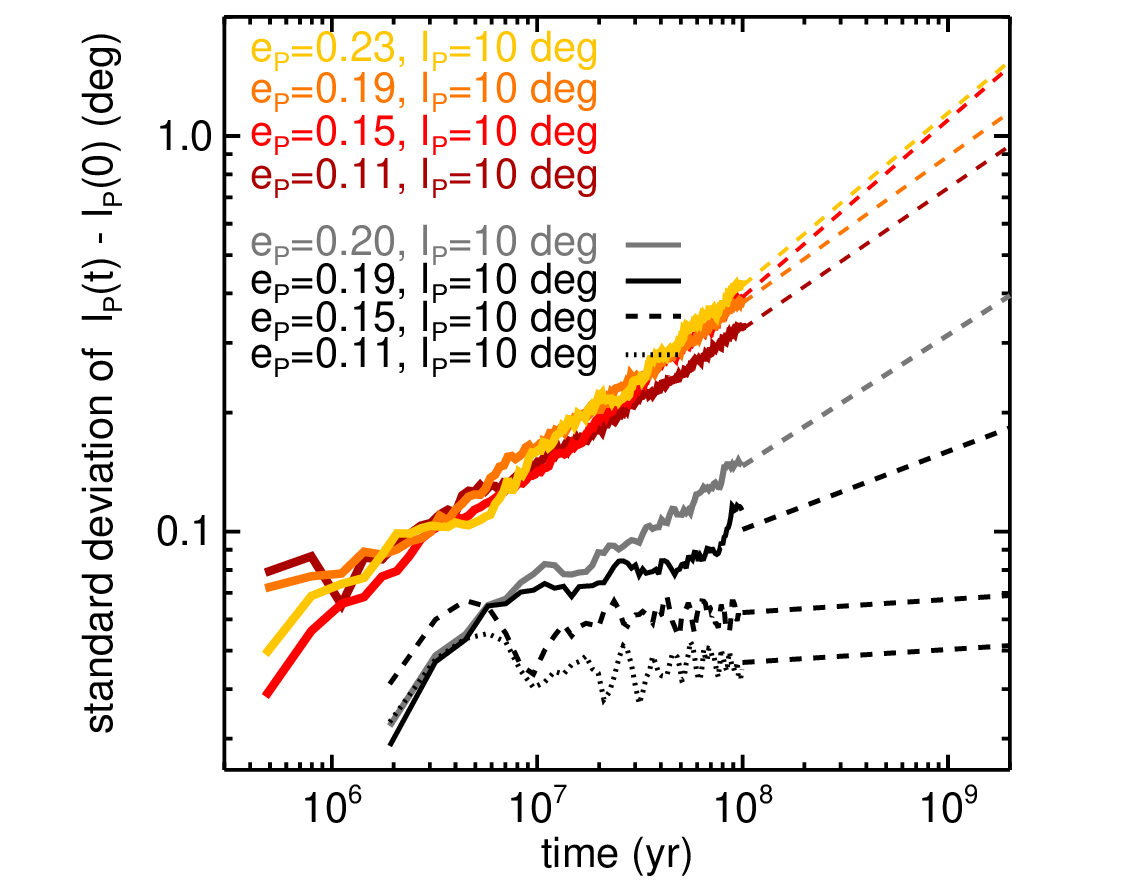}
\caption{Time evolution of $\sigma_{e}$ (top) and $\sigma_{I}$ (bottom) of resonant and control test particles from the 100 Myr simulations. Each panel corresponds to a different starting value of $I_{P}$ while a brighter colour indicates a higher starting value of $e_{P}$.}
\label{fig:sigma_100m_in2to1m}
\end{figure*}
\clearpage

\begin{figure*}
\hspace*{-2mm}\includegraphics[angle=0,width=5.8cm]{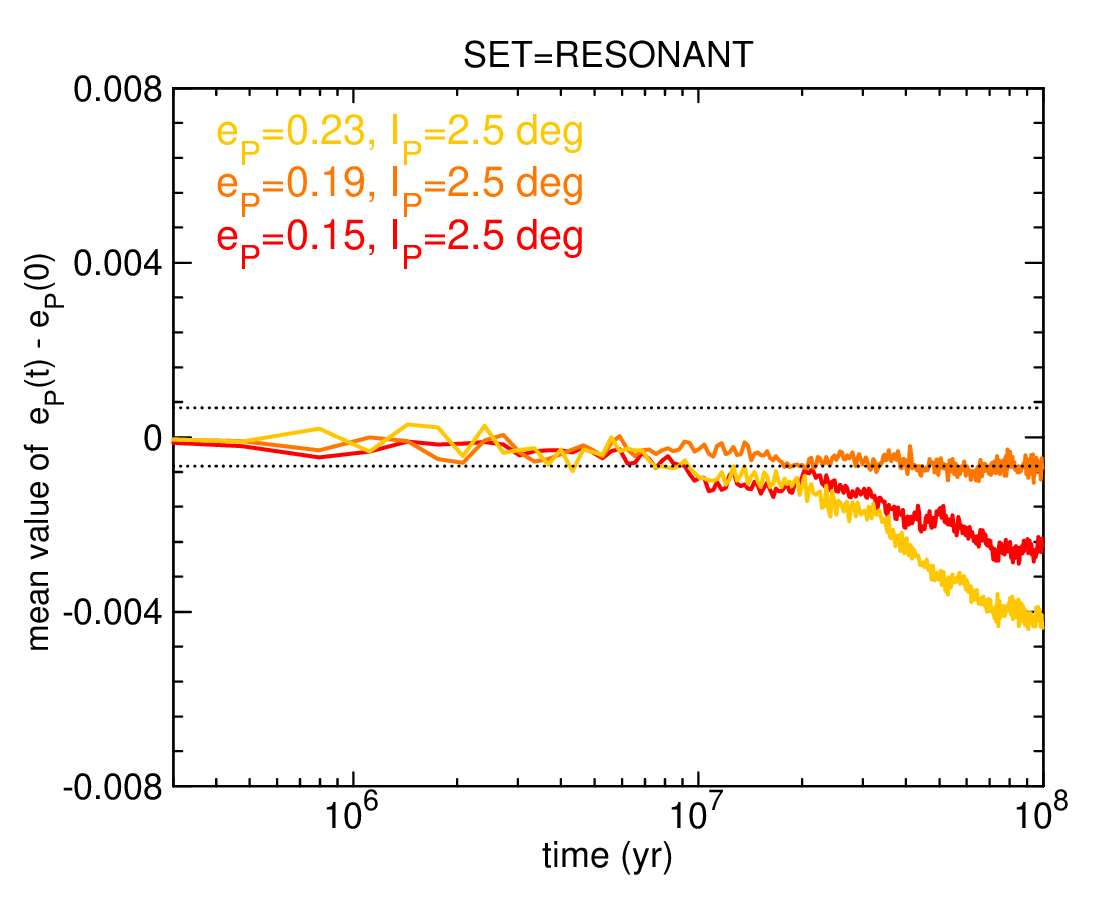}\hspace*{-2mm}\includegraphics[angle=0,width=5.8cm]{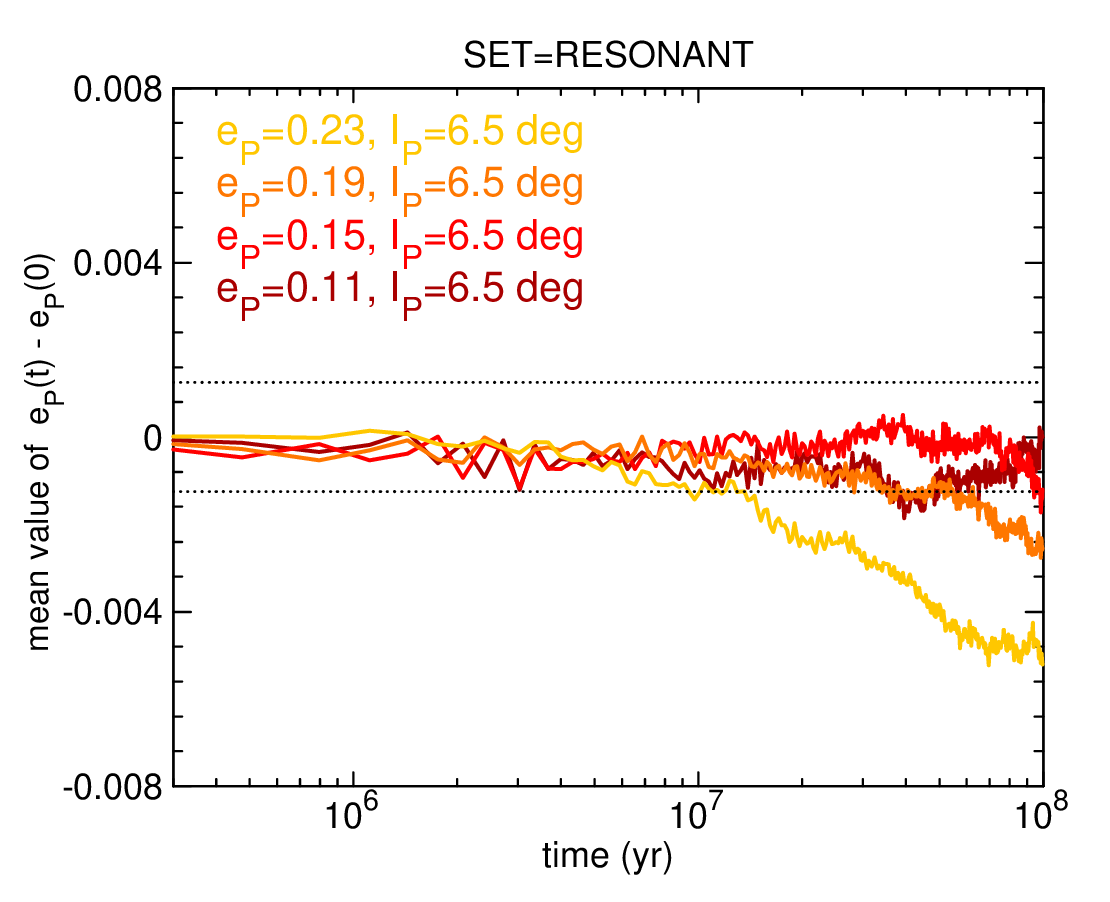}\hspace*{-1mm}\includegraphics[angle=0,width=5.8cm]{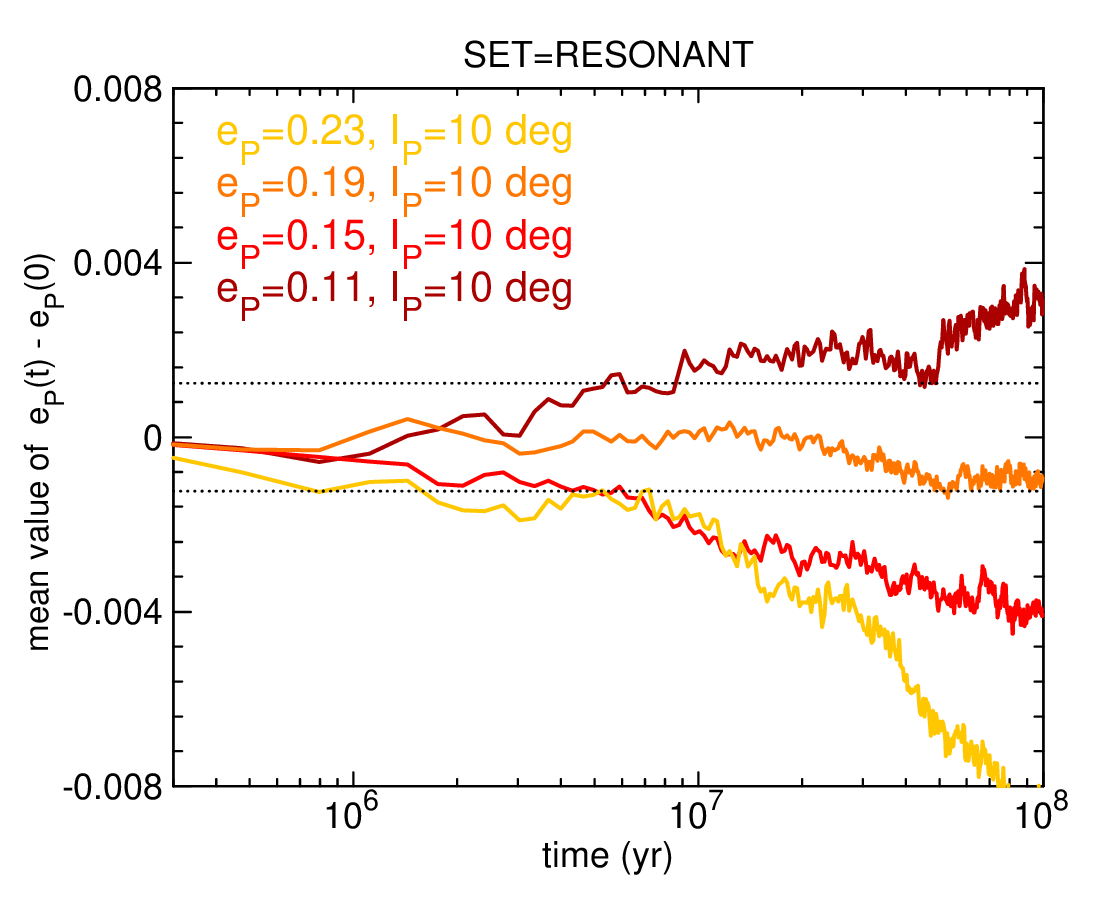}\\
\includegraphics[angle=0,width=5.6cm]{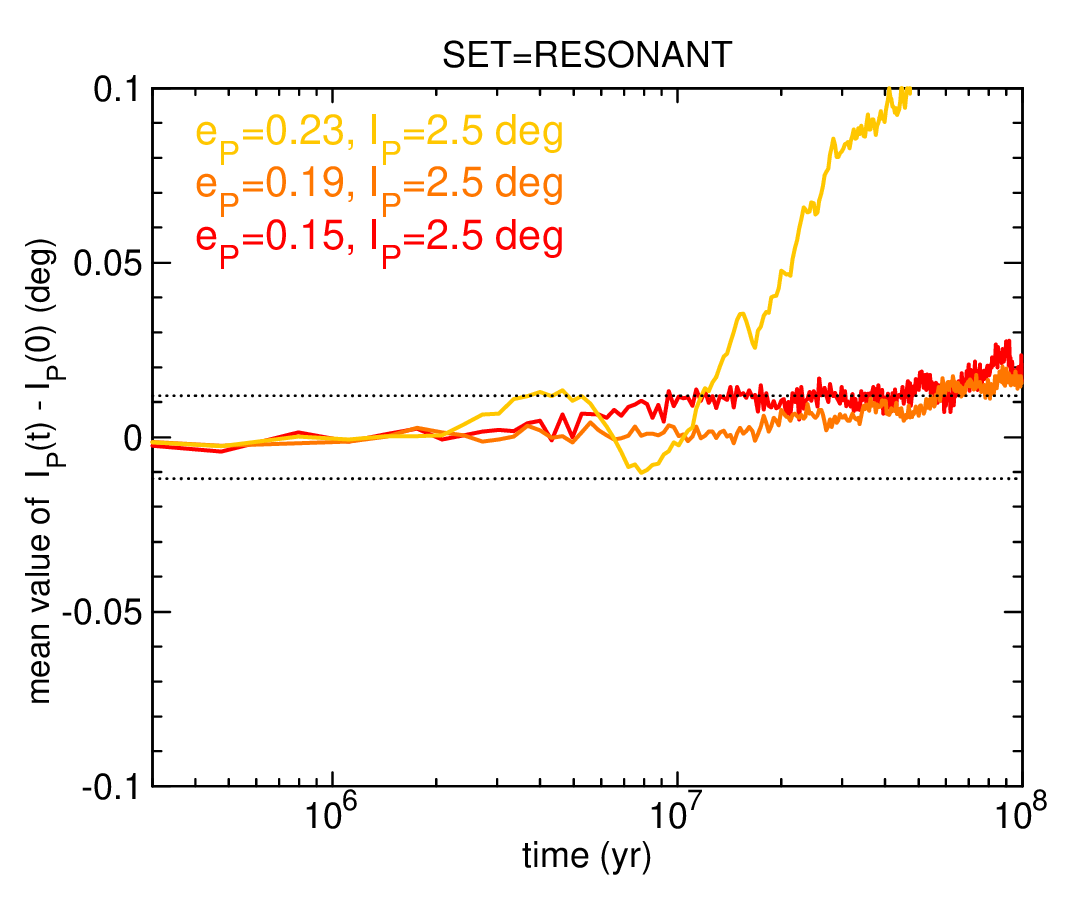}\includegraphics[angle=0,width=5.6cm]{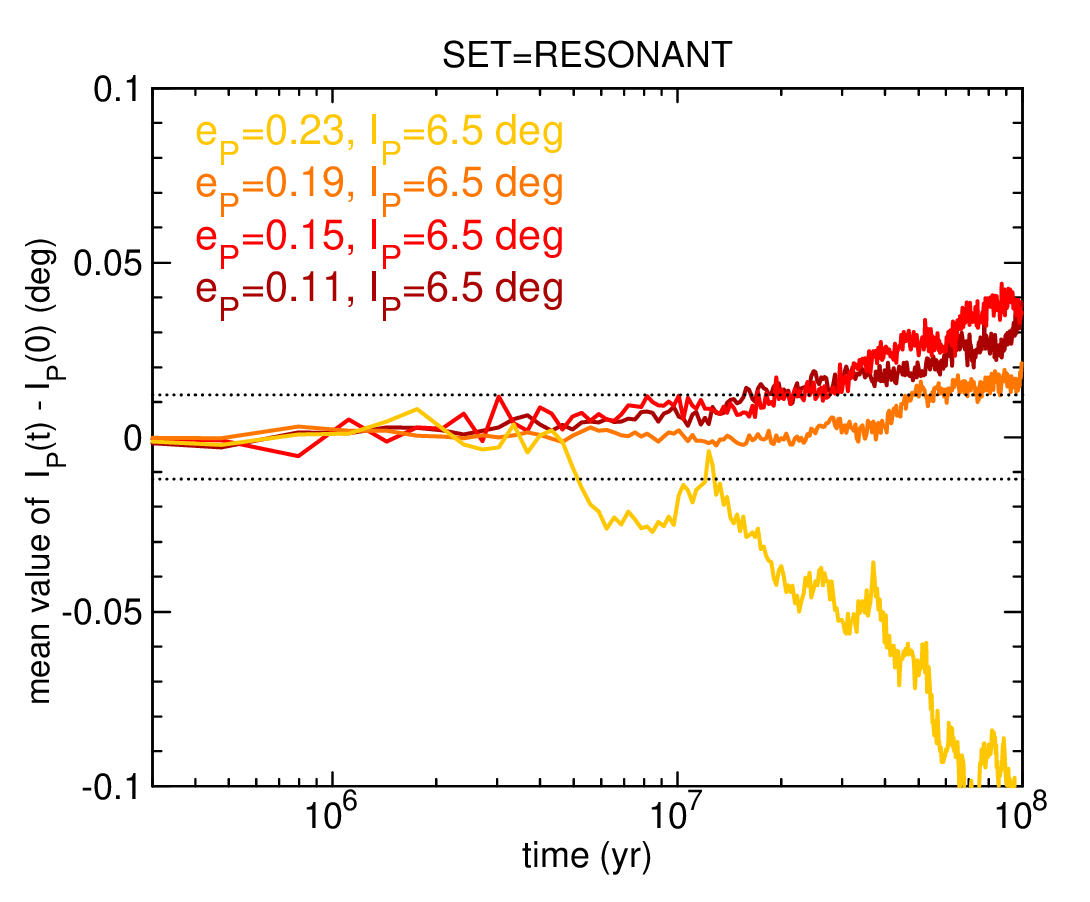}\includegraphics[angle=0,width=5.6cm]{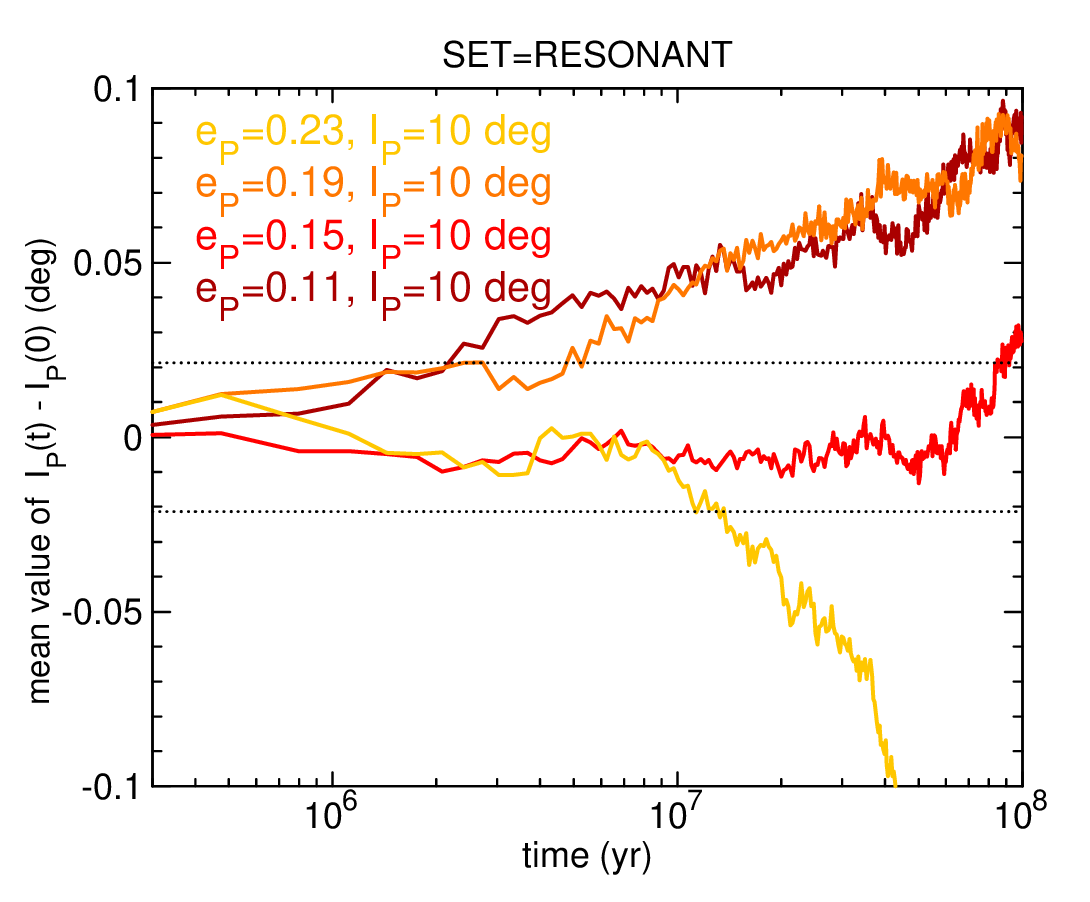}
\caption{Time evolution of $<$$e$$>$ (top) and $<$$I$$>$ (bottom) of test particles in the 1M-2A resonance from the 100 Myr simulations. Each panel corresponds to a different starting value of $I_{P}$ while a brighter colour indicates a higher starting value of $e_{P}$. The dotted horizontal lines indicate the formal $1$-$\sigma$ dispersion expected at $t$=$t_{0}$+$T$.}
\label{fig:mean_100m_in2to1m}
\end{figure*}
\clearpage

\begin{figure*}
\flushleft\hspace{-3mm}\includegraphics[angle=0,width=6.0cm]{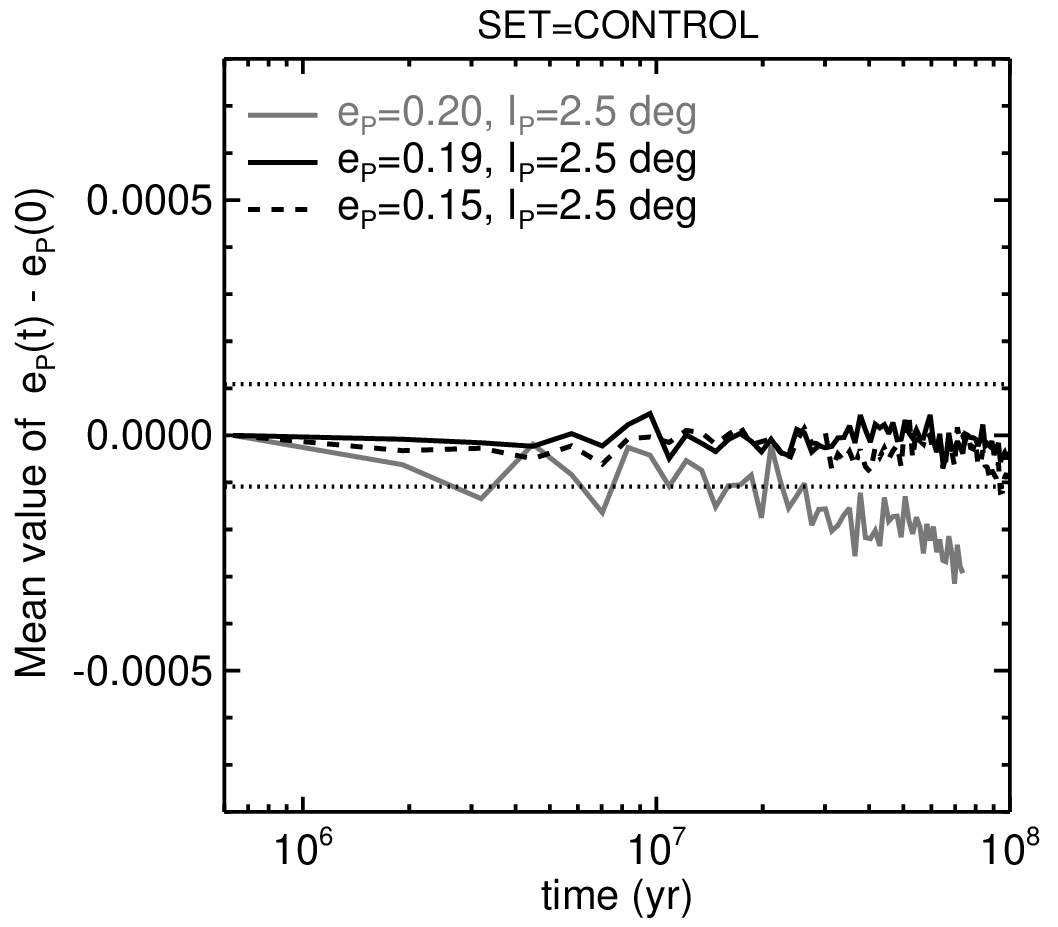}\mbox{}\includegraphics[angle=0,width=6.0cm]{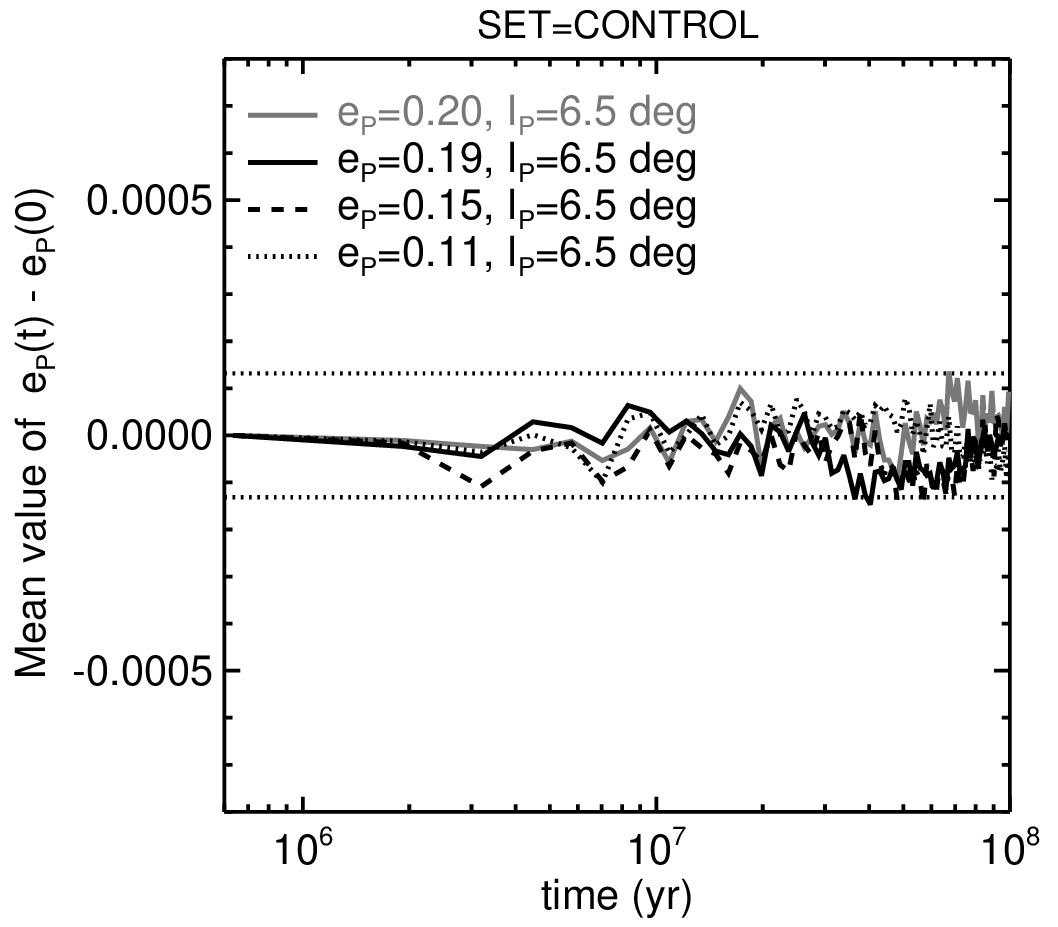}\mbox{}\includegraphics[angle=0,width=6.0cm]{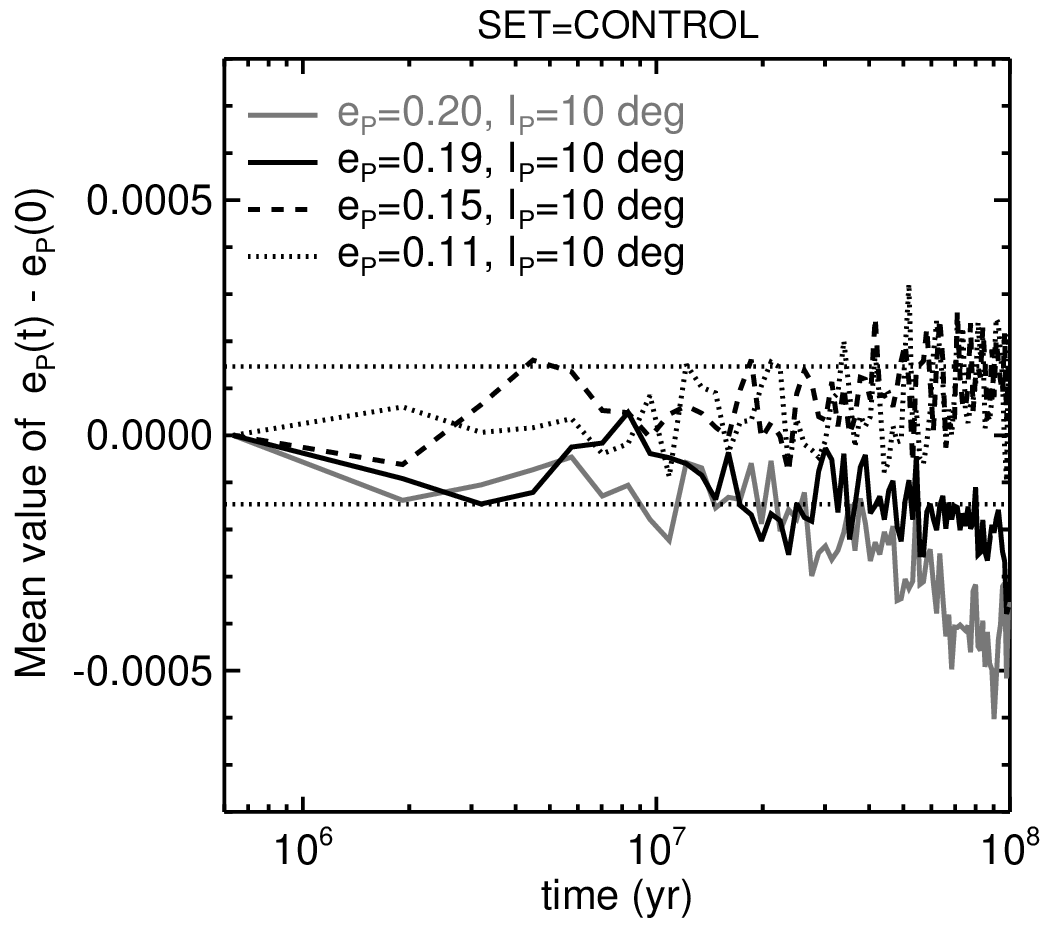}\\\flushleft\hspace{-3mm}\includegraphics[angle=0,width=6.0cm]{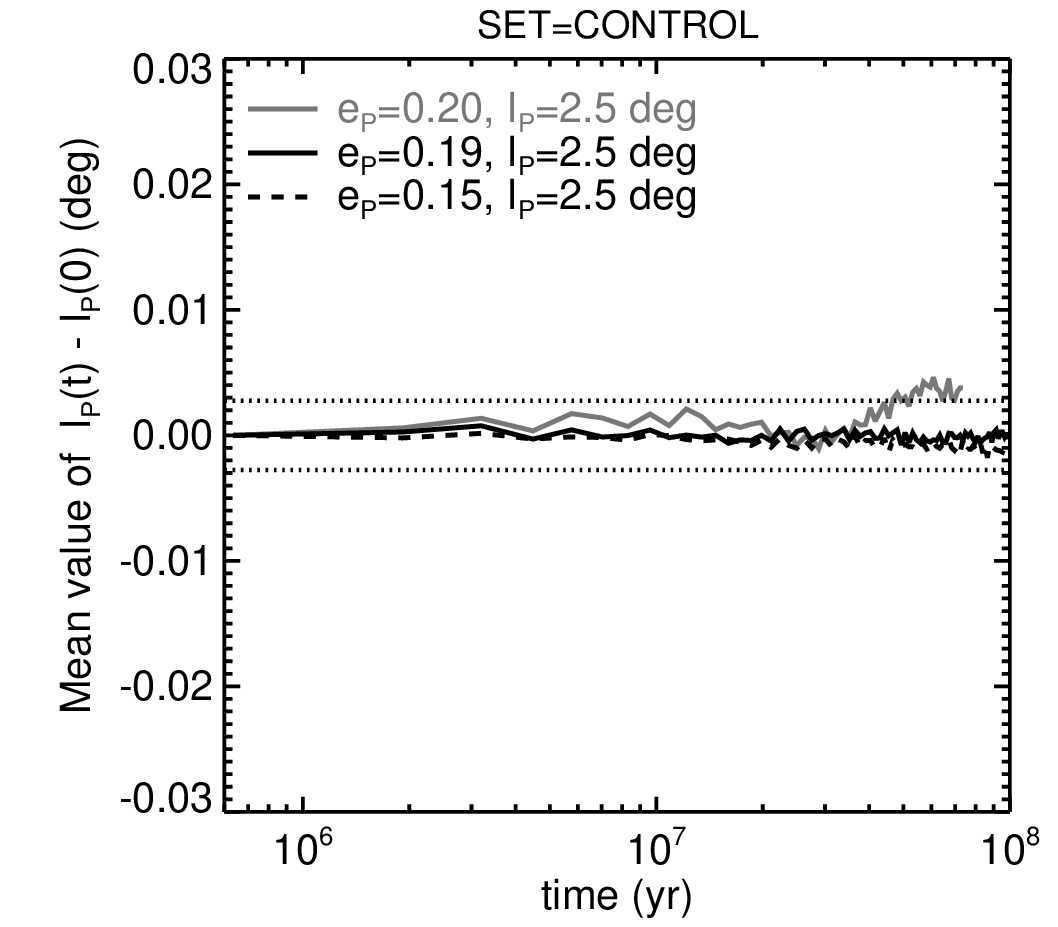}\mbox{}\includegraphics[angle=0,width=6.0cm]{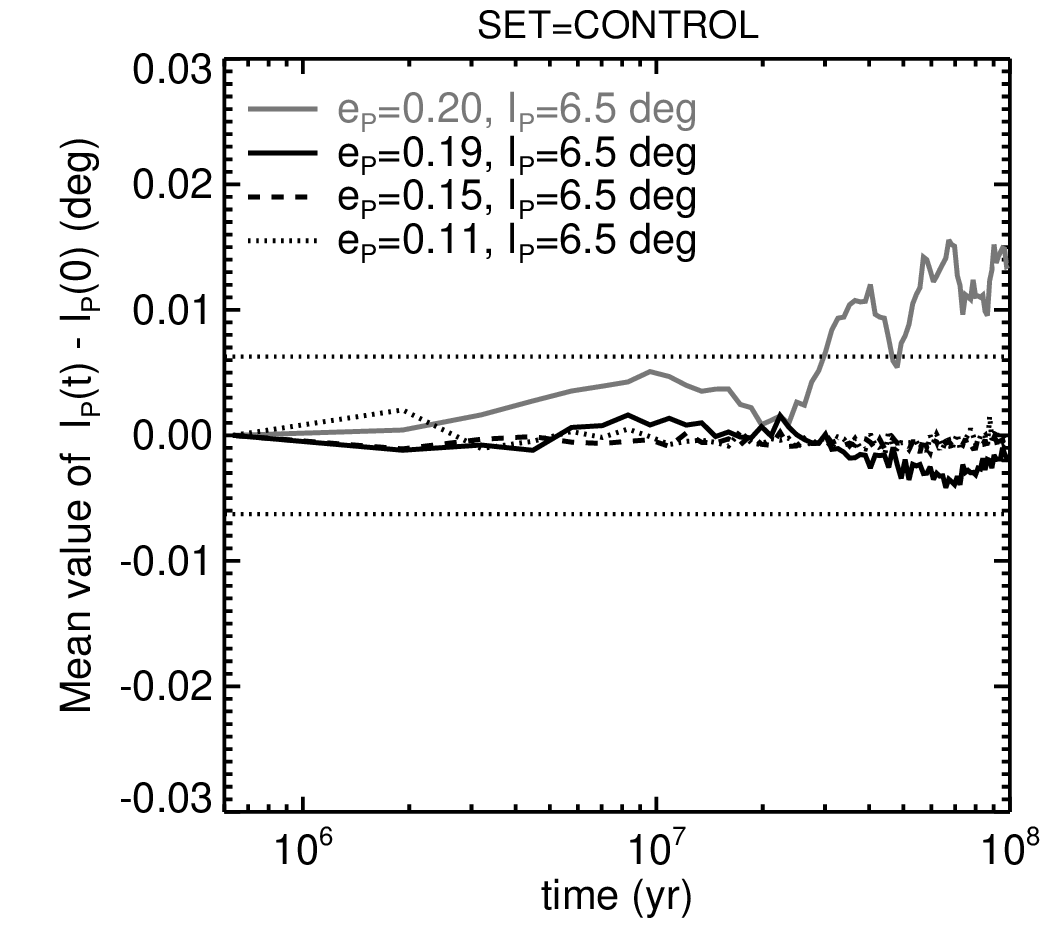}\mbox{}\includegraphics[angle=0,width=6.0cm]{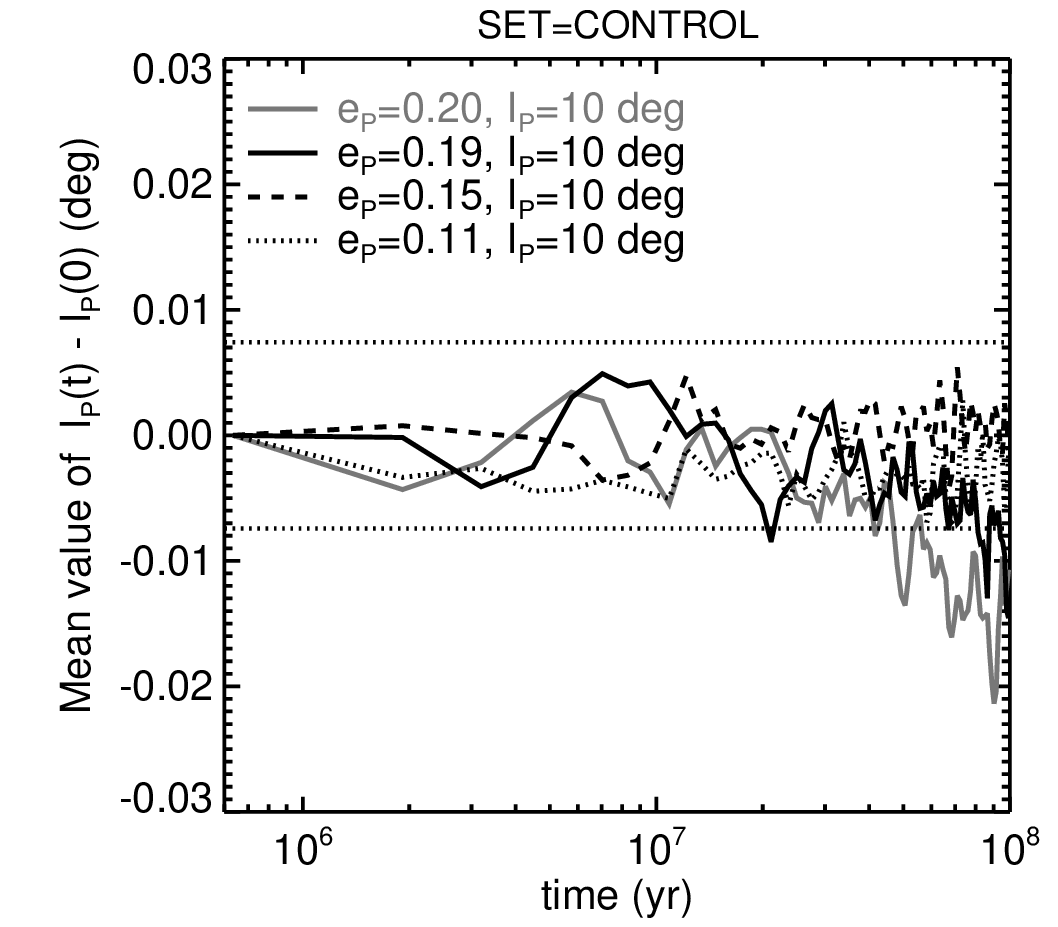}
\caption{As Fig.~\ref{fig:mean_100m_in2to1m} but for the respective groups of non-resonant particles.}
\label{fig:mean_100m_out2to1m}
\end{figure*}
\clearpage

\begin{figure}
\centering
\includegraphics[angle=0,width=\columnwidth]{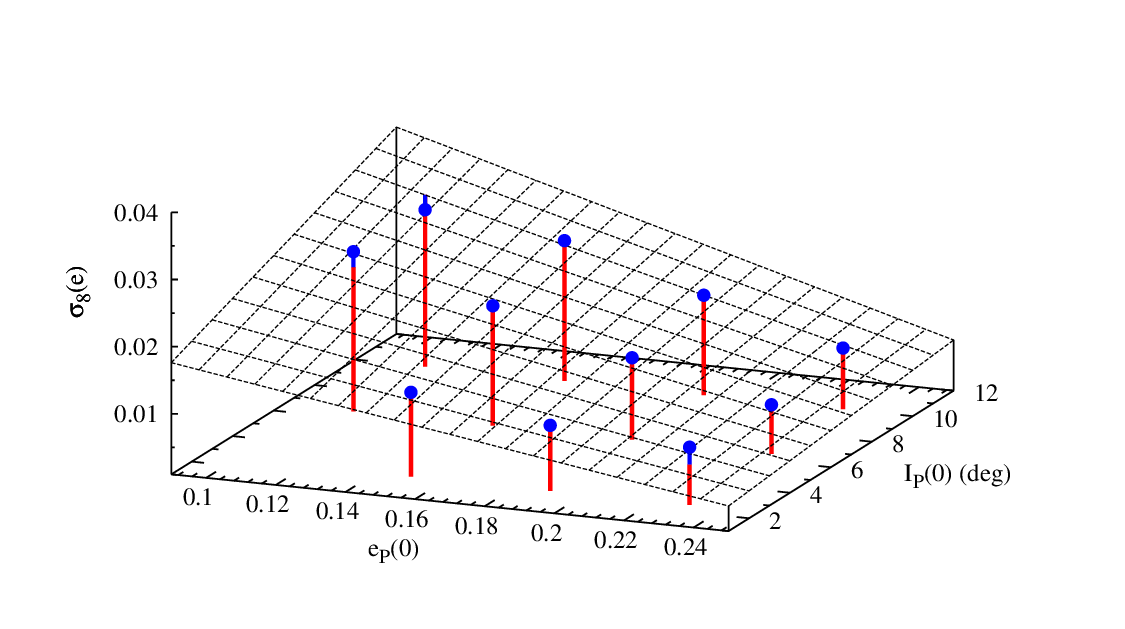}
\caption{Fit of Eq~\ref{eq:sigma_100m_fit} to the $\sigma_{8}$ eccentricity data (Table~\ref{tab:results100in}), shown here as blue points. Fit residuals are represented as blue segments superimposed on the fitted values (red segments). Data points with $e_{P}(0)=0.23$ were not considered in the fit.}
\label{fig:e_diffusion_fit}
\end{figure}
\clearpage

\begin{figure*}
\hspace*{-2mm}
\includegraphics[angle=0,width=78mm]{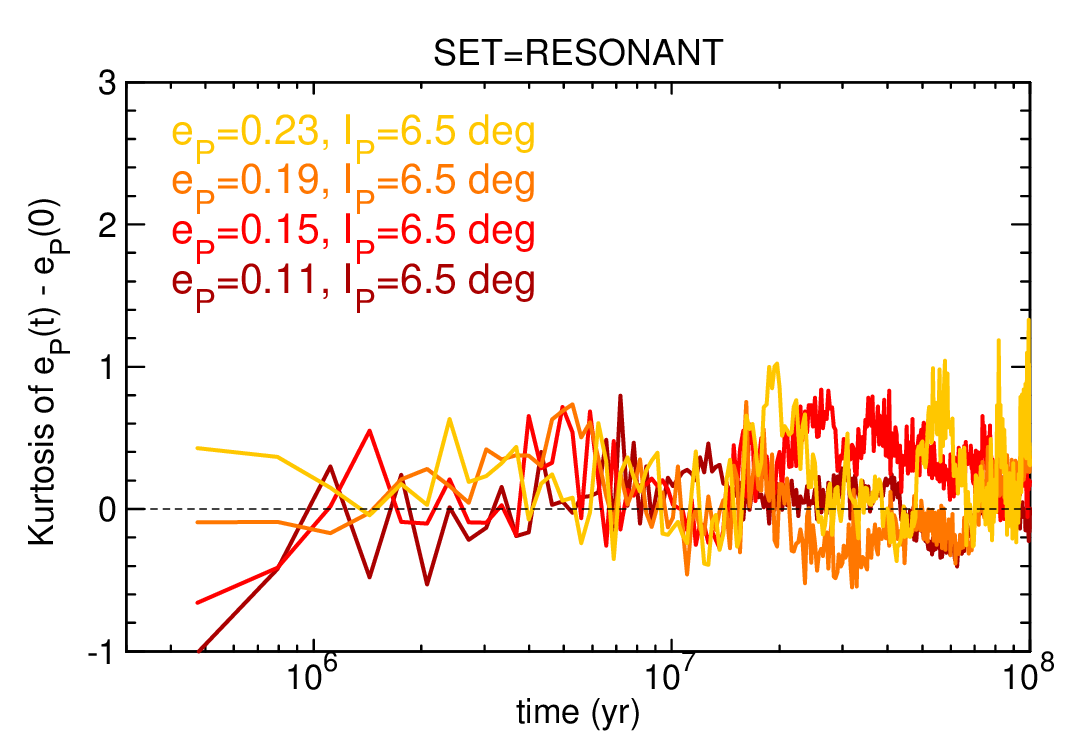}\includegraphics[angle=0,width=78mm]{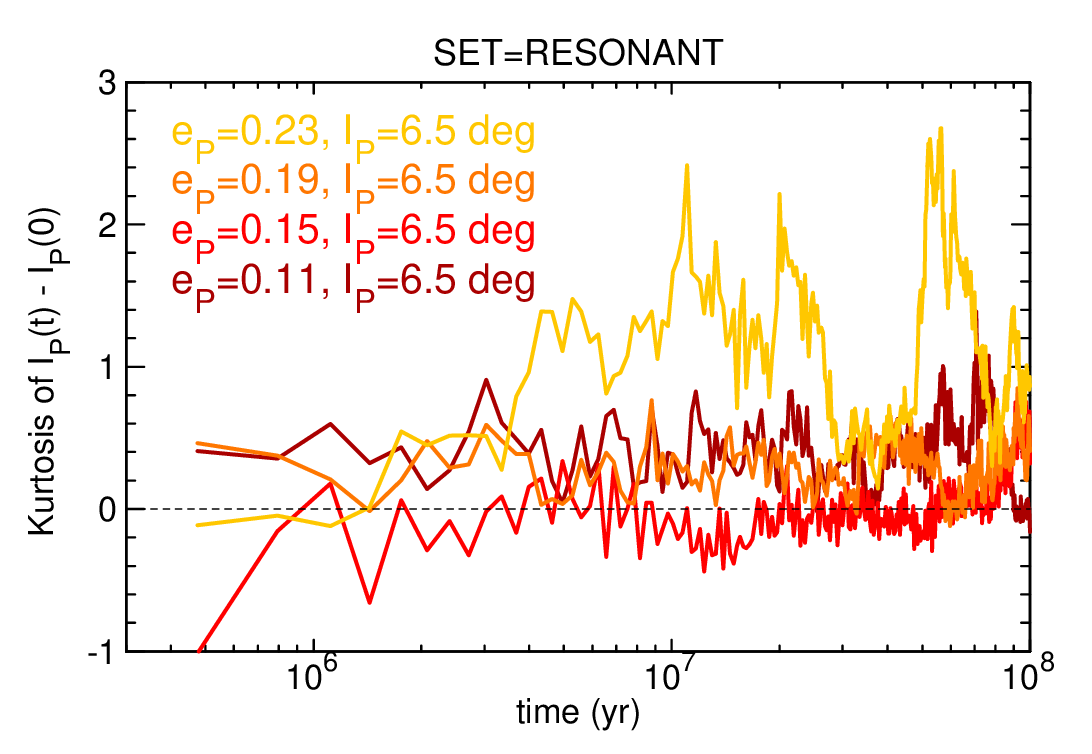}
\includegraphics[angle=0,width=78mm]{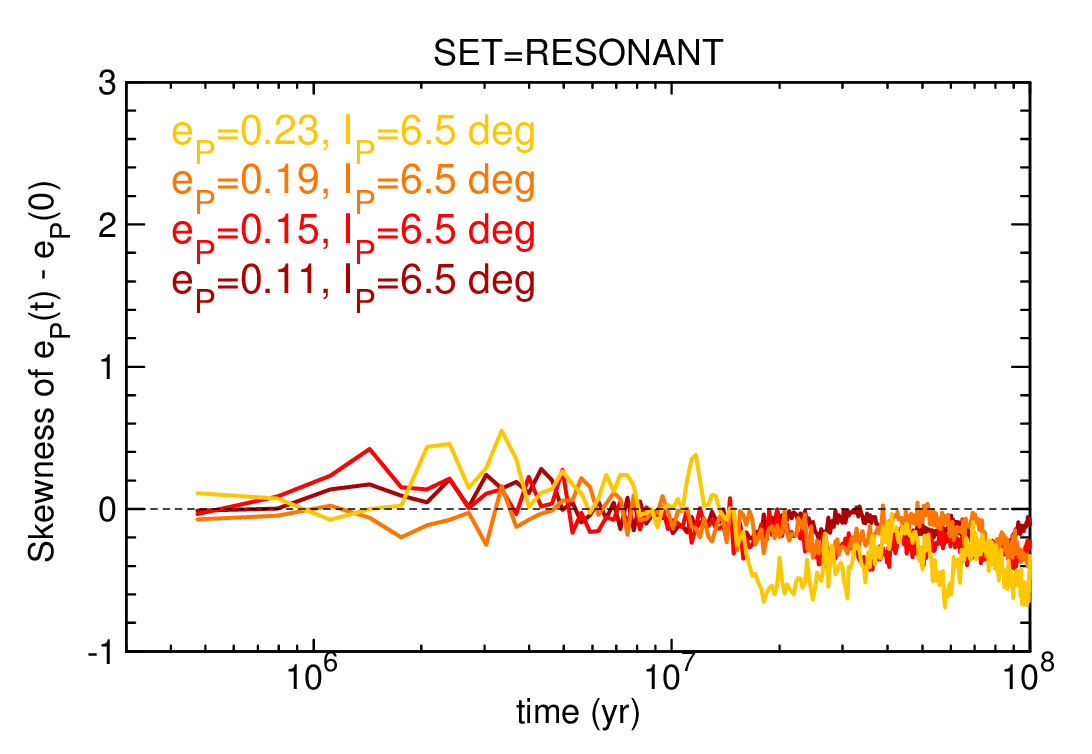}\includegraphics[angle=0,width=78mm]{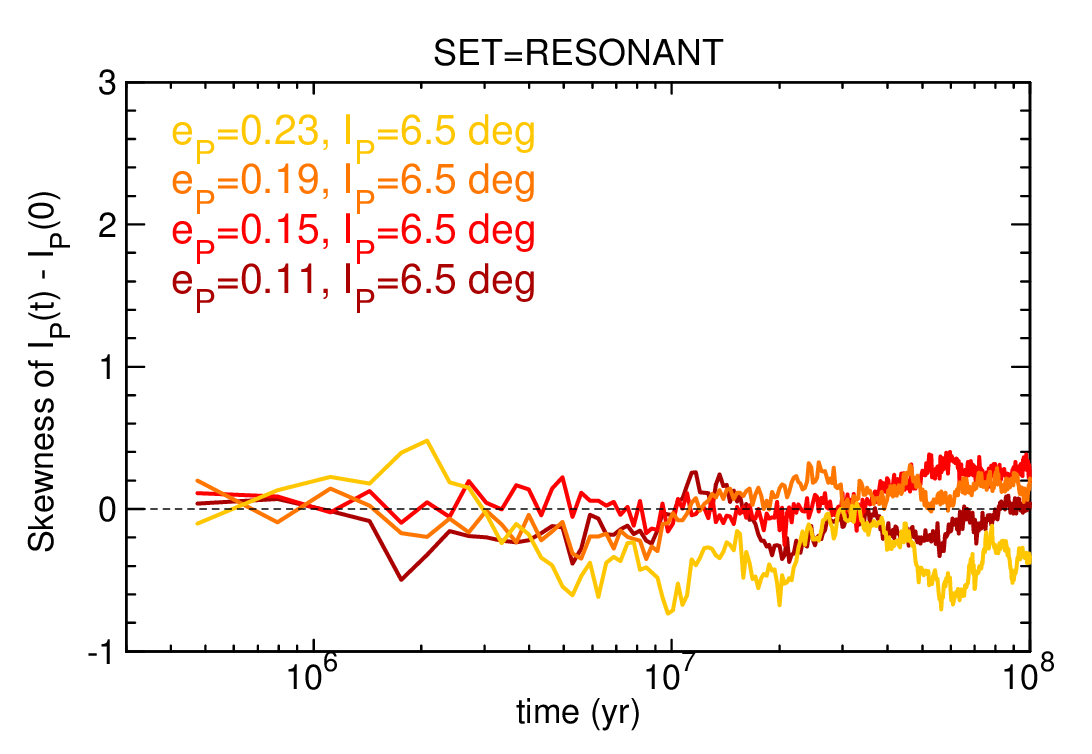}
\caption{{\it Top}: Time evolution of the kurtosis (fourth moment) of the distribution of $e_{P}(t)$ - $e_{P}(0)$ (left) and $I_{P}(t)$ - $I_{P}(0)$ (right) relative to that of a gaussian distribution for particles in the 1M-2A resonance with $I_{P}(0)$=$6.5^{\circ}$. Brighter colours correspond to higher starting values of $e_{P}$. The dashed horizontal line indicates a value equal to that of a gaussian. {\it Bottom}: Time evolution of the skewness (third moment) for the same group of particles.}
\label{fig:kurtosisskewness_100m_in2to1m}
\end{figure*}
\clearpage

\begin{figure*}
\includegraphics[angle=0,width=70mm]{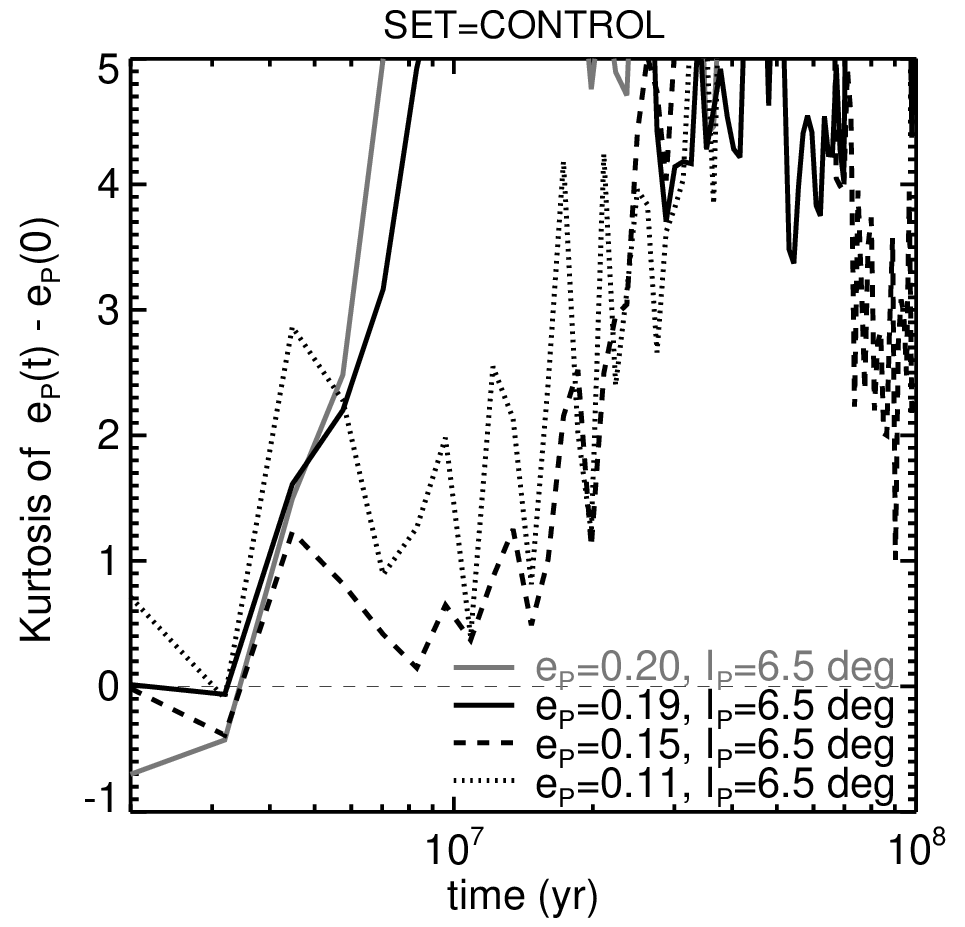}\includegraphics[angle=0,width=70mm]{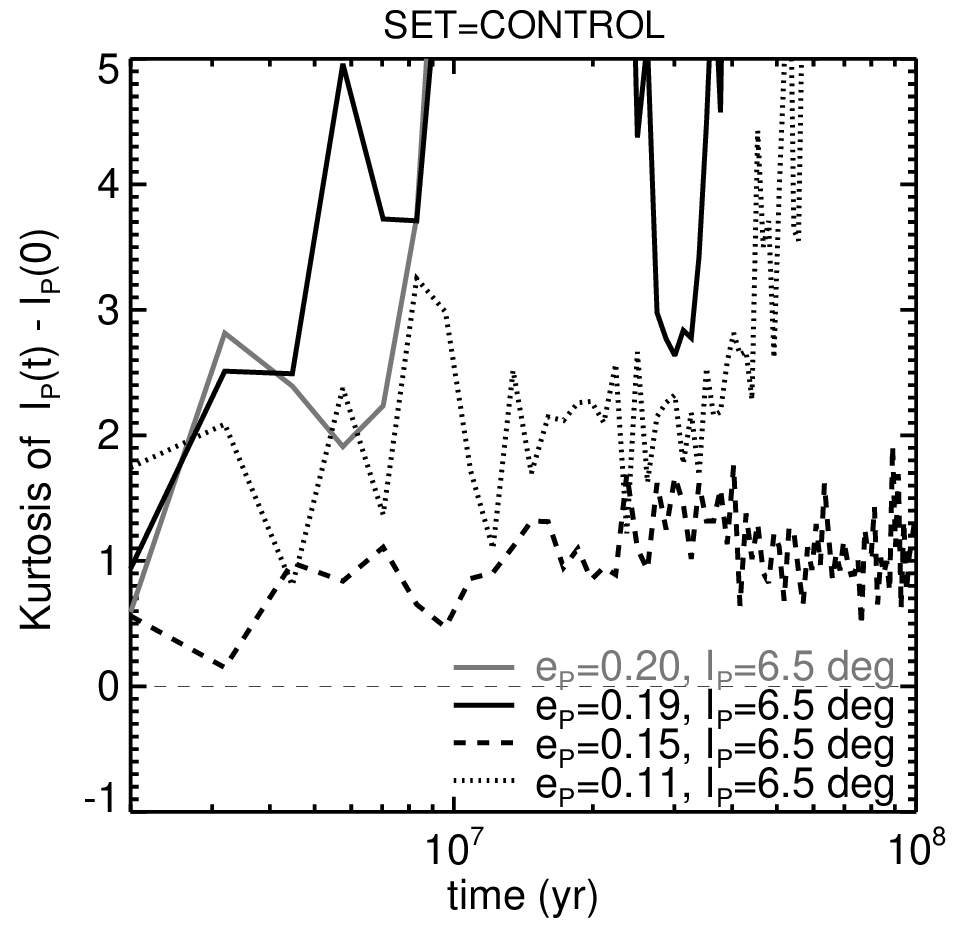}
\includegraphics[angle=0,width=70mm]{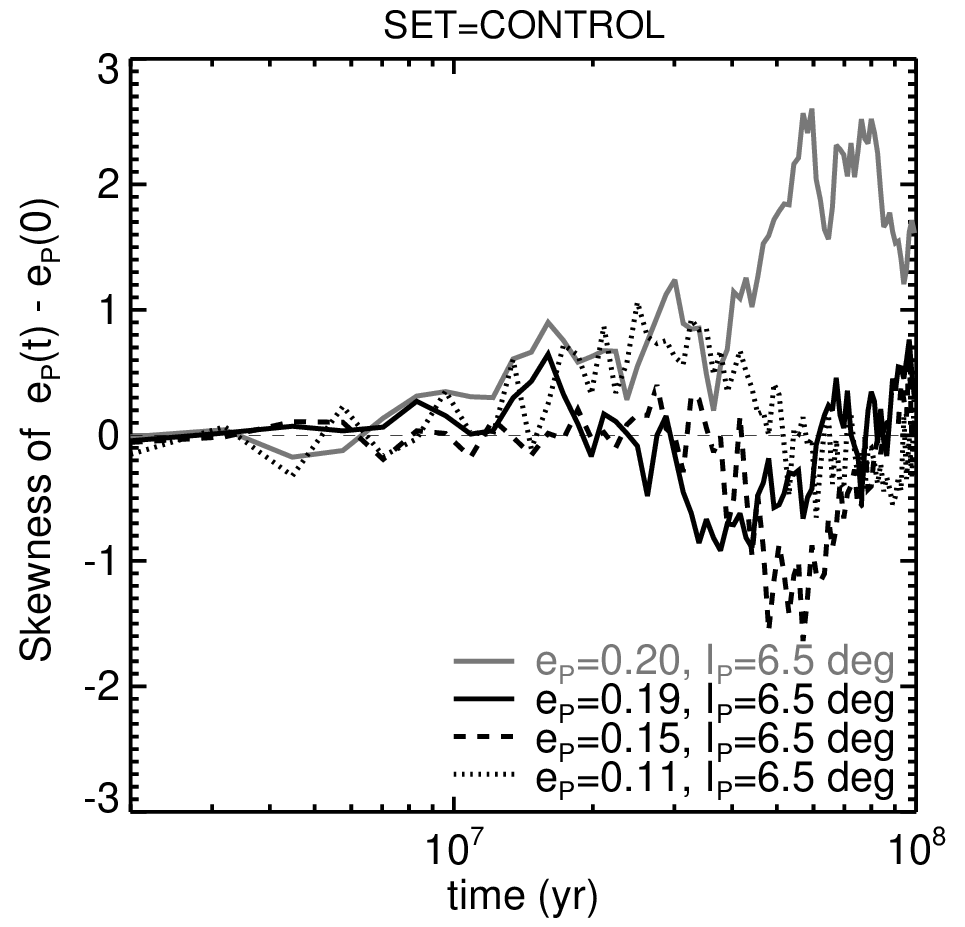}\includegraphics[angle=0,width=70mm]{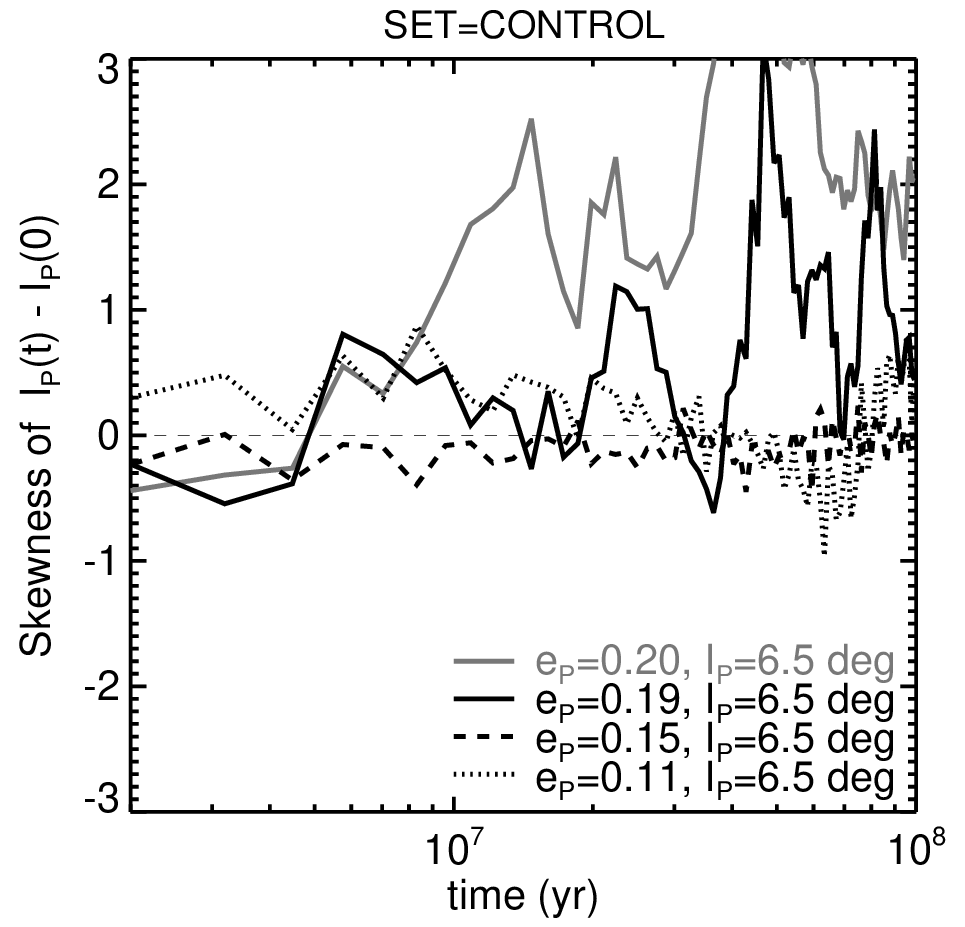}
\caption{As Fig.~\ref{fig:kurtosisskewness_100m_in2to1m} but for the control set of particles.}
\label{fig:kurtosisskewness_100m_out2to1m}
\end{figure*}
\clearpage

\begin{figure}
\centering
\includegraphics[angle=0,width=87mm]{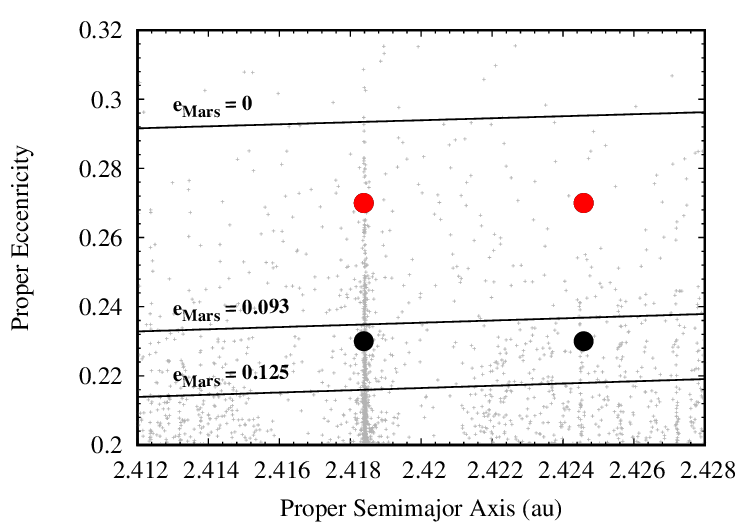}
\caption{Location of test particles placed on potentially Mars-crossing orbits in the proper $a-$proper $e$ plane against the background of real asteroids. Bold curves represent the minimum proper eccentricity that allows close approaches to Mars for different values of the planetary eccentricity: a circular orbit ($e_{\rm Mars}$$=$$0$), the present value ($e_{\rm Mars}$$=$$0.093$) and the maximum ($e_{\rm Mars}$$=$$0.125$) recorded during our simulations.}
\label{fig:qast_vs_adistmars}
\end{figure}
\clearpage

\begin{figure}
\centering
\includegraphics[angle=0,width=87mm]{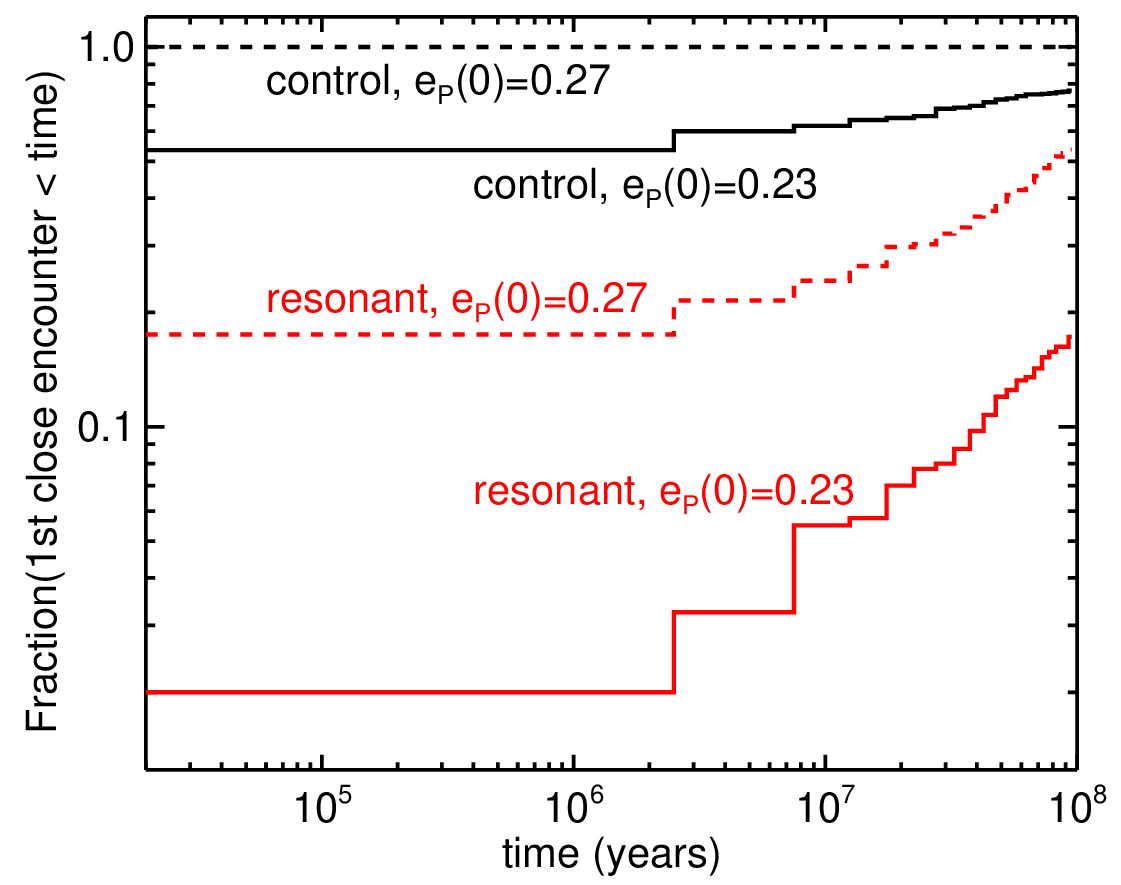}
\caption{Cumulative distribution of the 1st encounter between Mars and a test particle inside (red line) and outside (black line) the 1M-2A resonance with $e_{P}(0)$$=$$0.23$ (bold line) and $e_{P}(0)$$=$$0.27$ (dashed line) as recorded by {\sl orbit9} during the simulations.}
\label{fig:ec2327_mars}
\end{figure}
\clearpage

\begin{figure}
\centering
\includegraphics[angle=0,width=87mm]{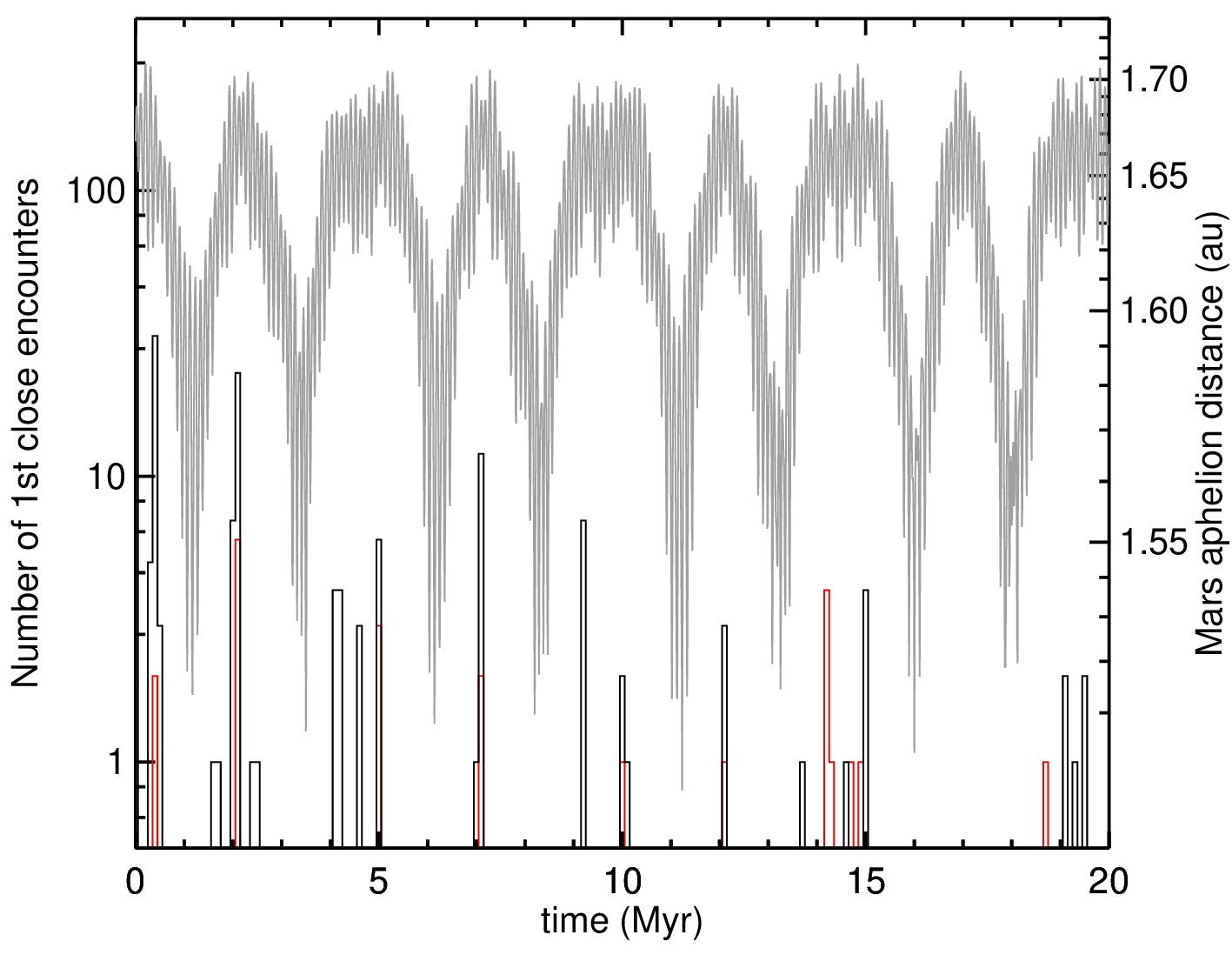}\\\vspace{-2mm}\hspace{-0.5mm}\includegraphics[angle=0,width=87mm]{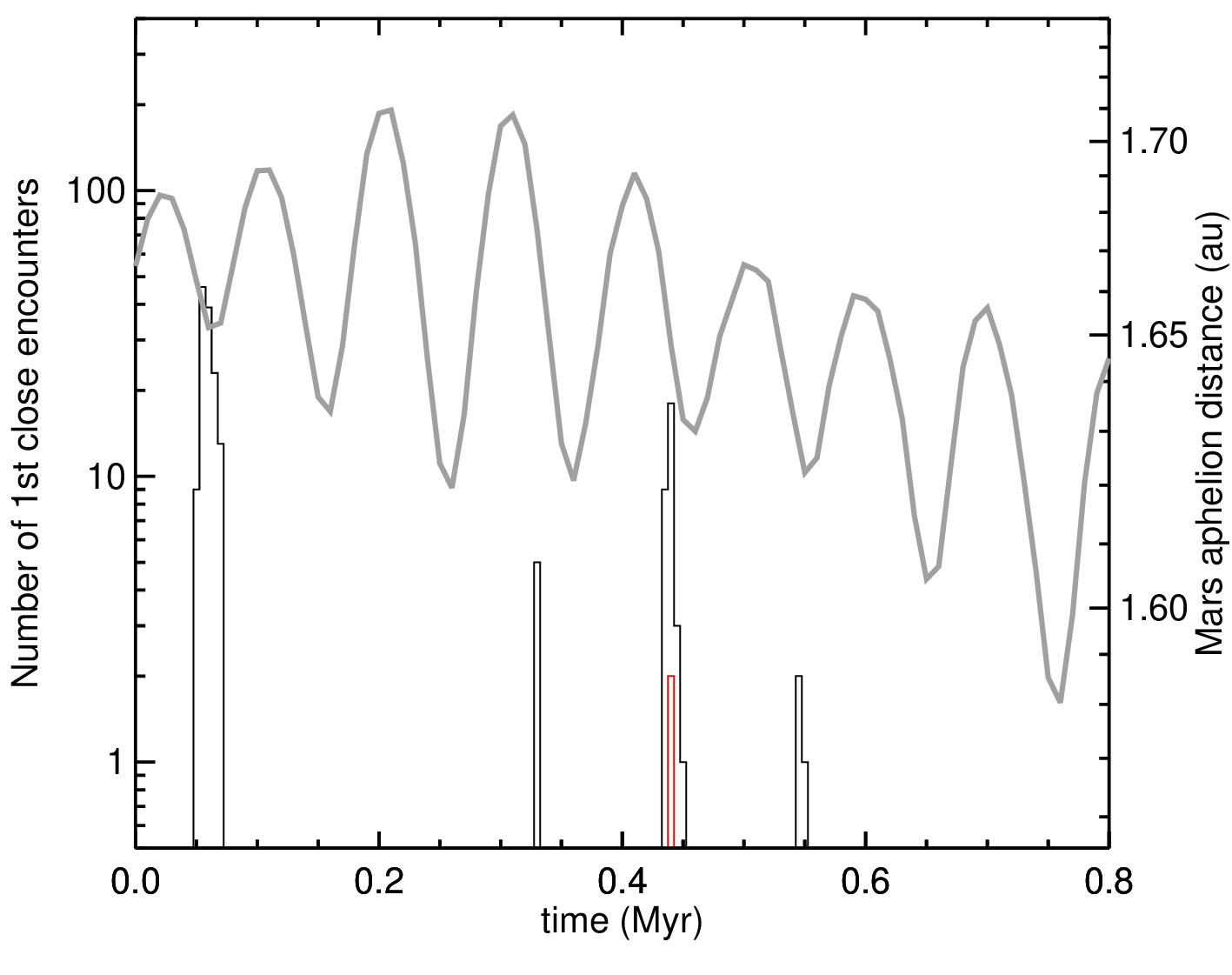}
\caption{{\it Top}: Incremental distribution of 1st encounters between control (black) or resonant (red) particles with $e_{P}(0)$$=$$0.23$, $I_{P}(0)$$=$$6.5^{\circ}$ and Mars, against the planet's aphelion distance (grey curve). Bin size is $10^{5}$ yr. {\it Bottom}: Detail of the first $0.8$ myr of data in the top panel. Bin size is 5$\times$$10^{3}$ yr.}
\label{fig:ec23_adistmars}
\end{figure}
\clearpage

\begin{figure*}
\hspace*{2mm}\includegraphics[angle=0,width=89mm]{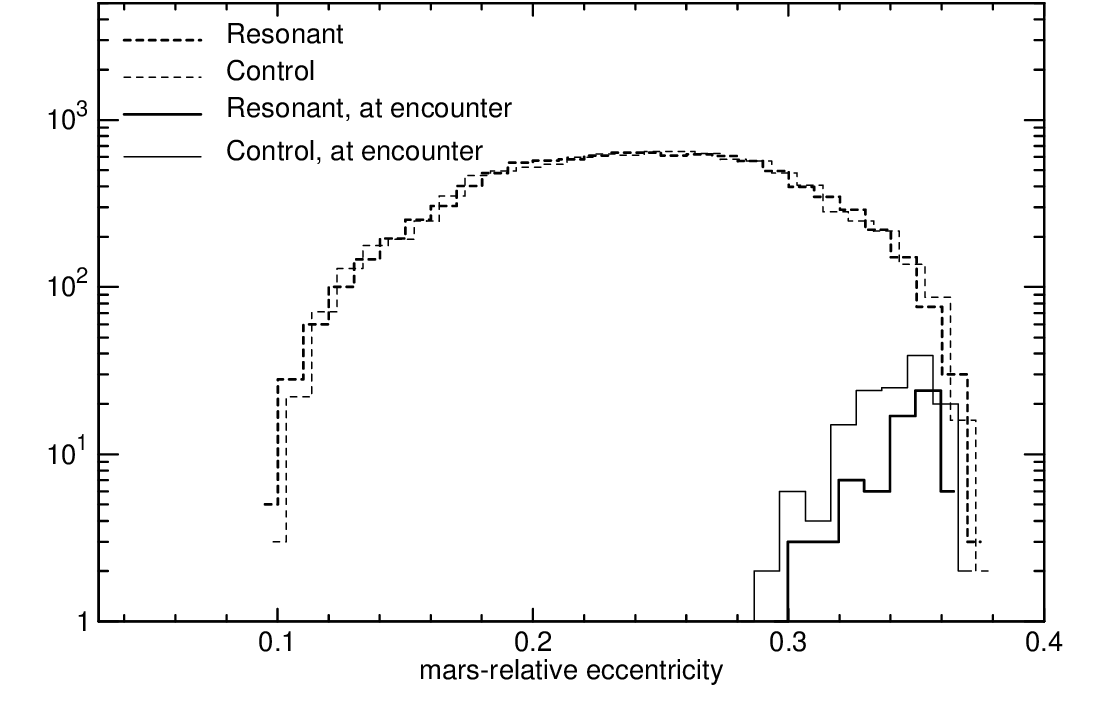}\includegraphics[angle=0,width=89mm]{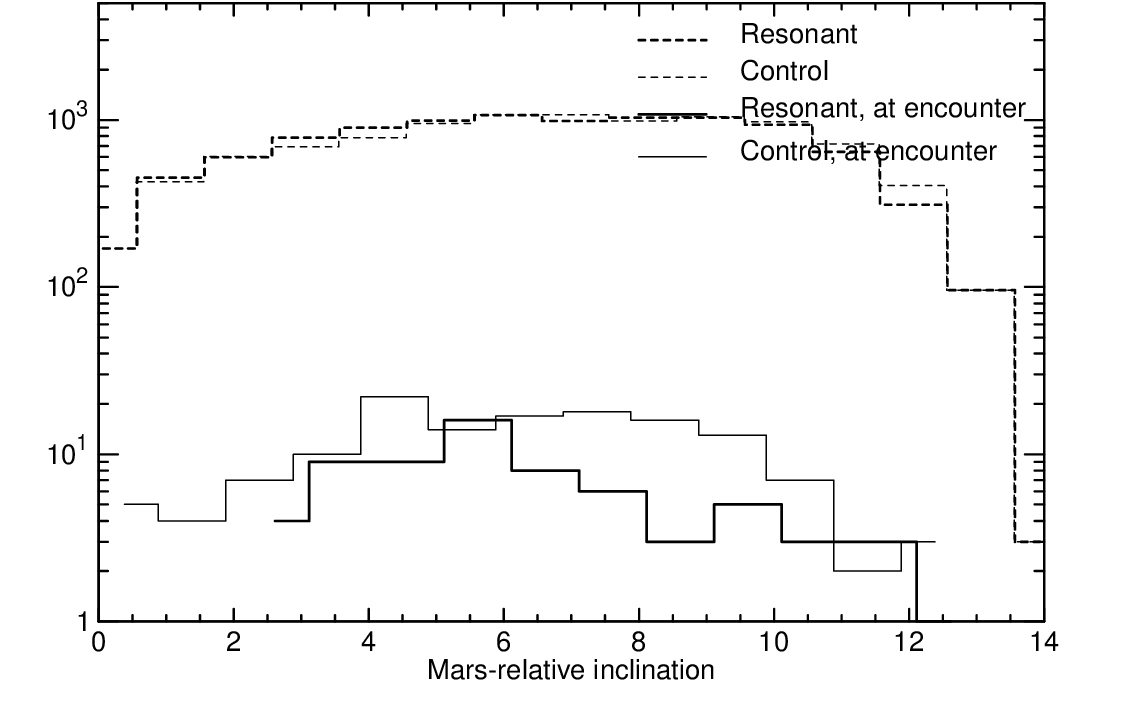}\\\hspace{-5mm}\includegraphics[angle=0,width=85mm]{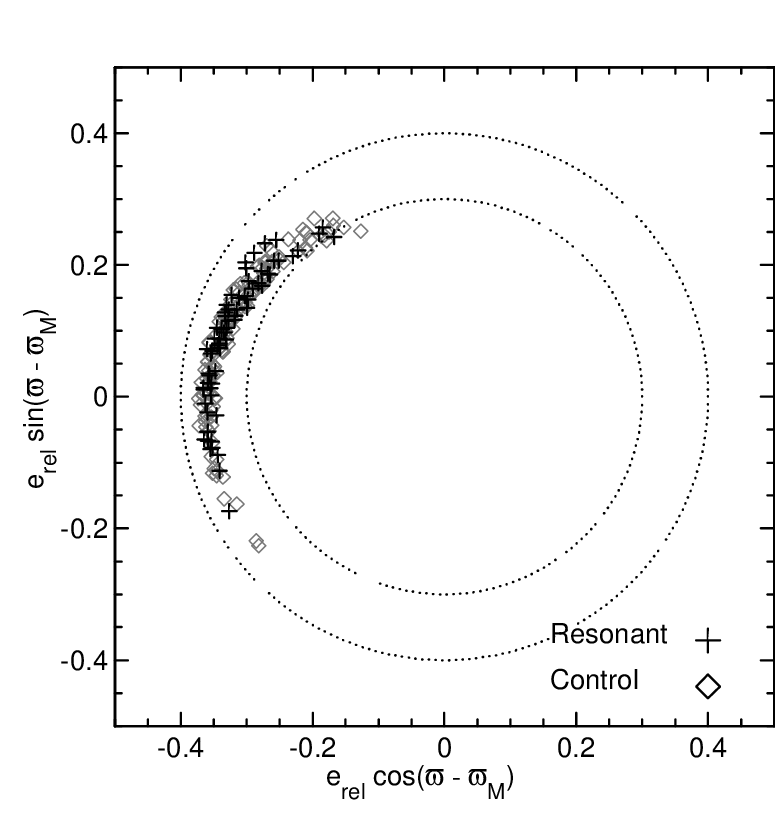}\hspace*{2mm}\includegraphics[angle=0,width=85mm]{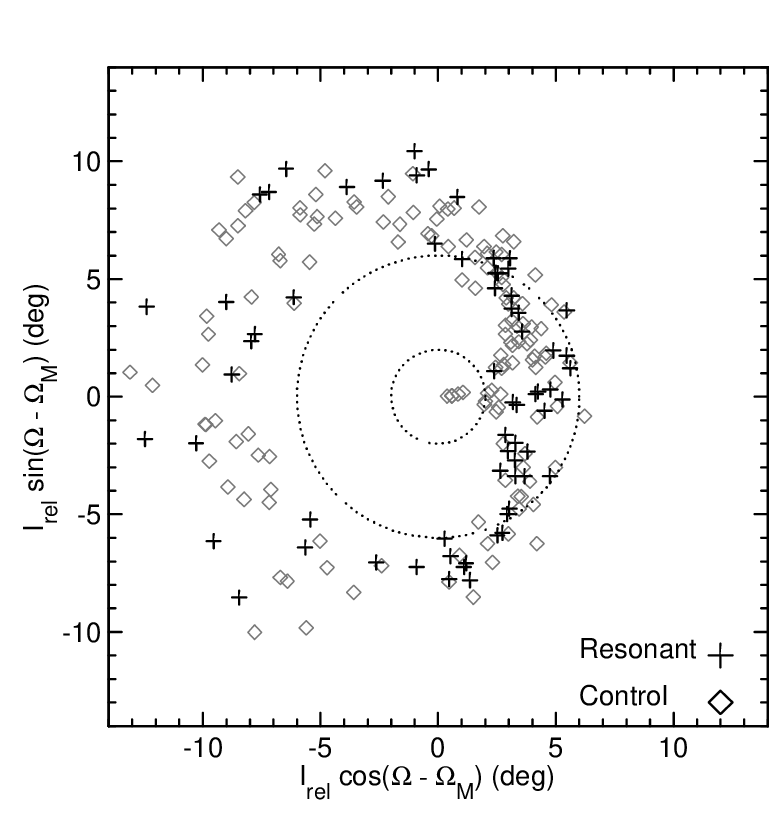}
\caption{{\it Top}: Distributions of relative eccentricity ${e_{\rm rel}}$ and inclination ${I_{\rm rel}}$ for particles with $I_{P}(0)$$=$$6.5^{\circ}$ upon 1st encounter with Mars. {\it Bottom}: Polar plots of $e_{\rm rel}$ and $I_{\rm rel}$ as functions of $\varpi - \varpi_{\rm M}$ and $\Omega - \Omega_{\rm M}$ respectively. Circles with $e_{\rm rel}=0.3$ \& $0.4$ and $I_{\rm rel}=2^{\circ}$ \& $6^{\circ}$ have been added to guide the eye.}
\label{fig:ereleireli_6p5d_100m_inout2to1m}
\end{figure*}
\clearpage

\begin{figure*}
\flushleft\includegraphics[angle=0,width=59mm]{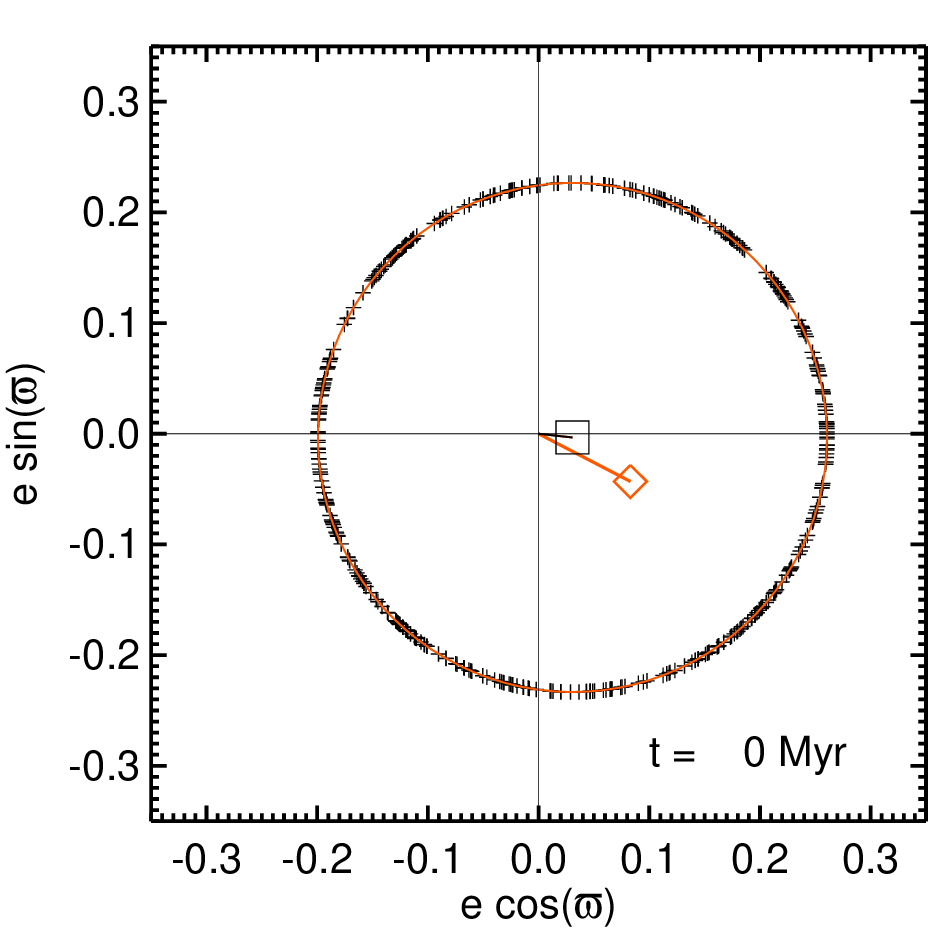}\includegraphics[angle=0,width=59mm]{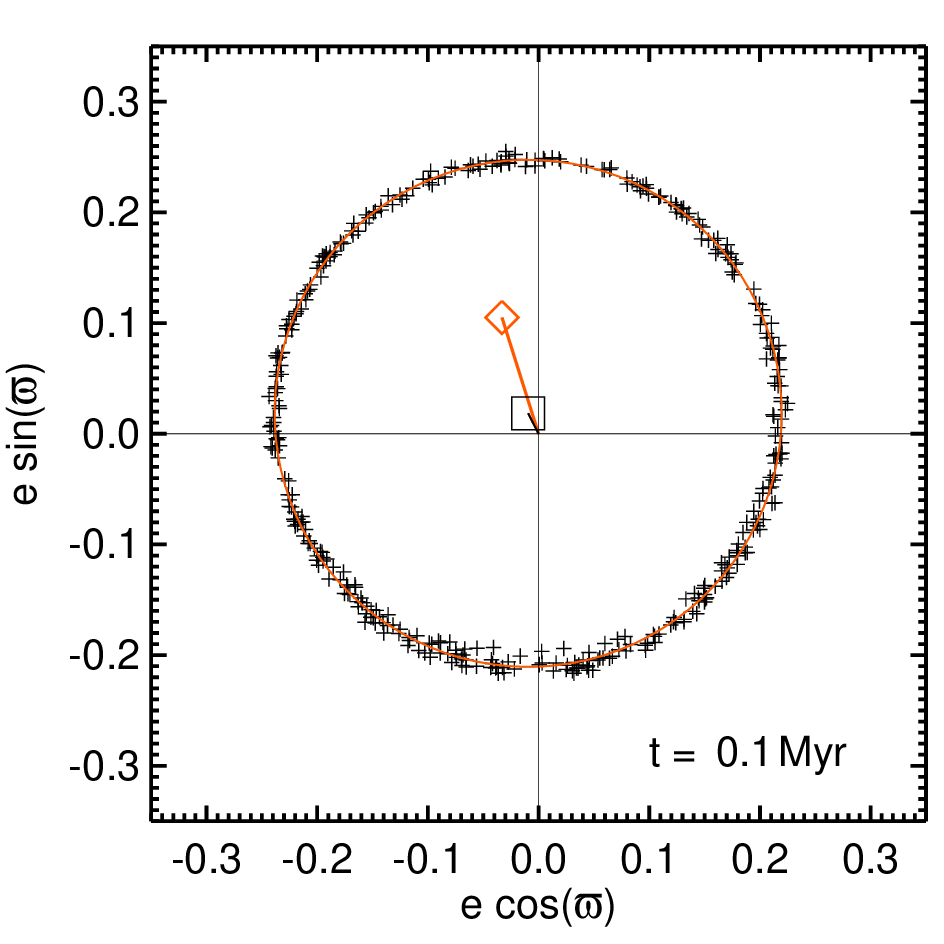}\includegraphics[angle=0,width=59mm]{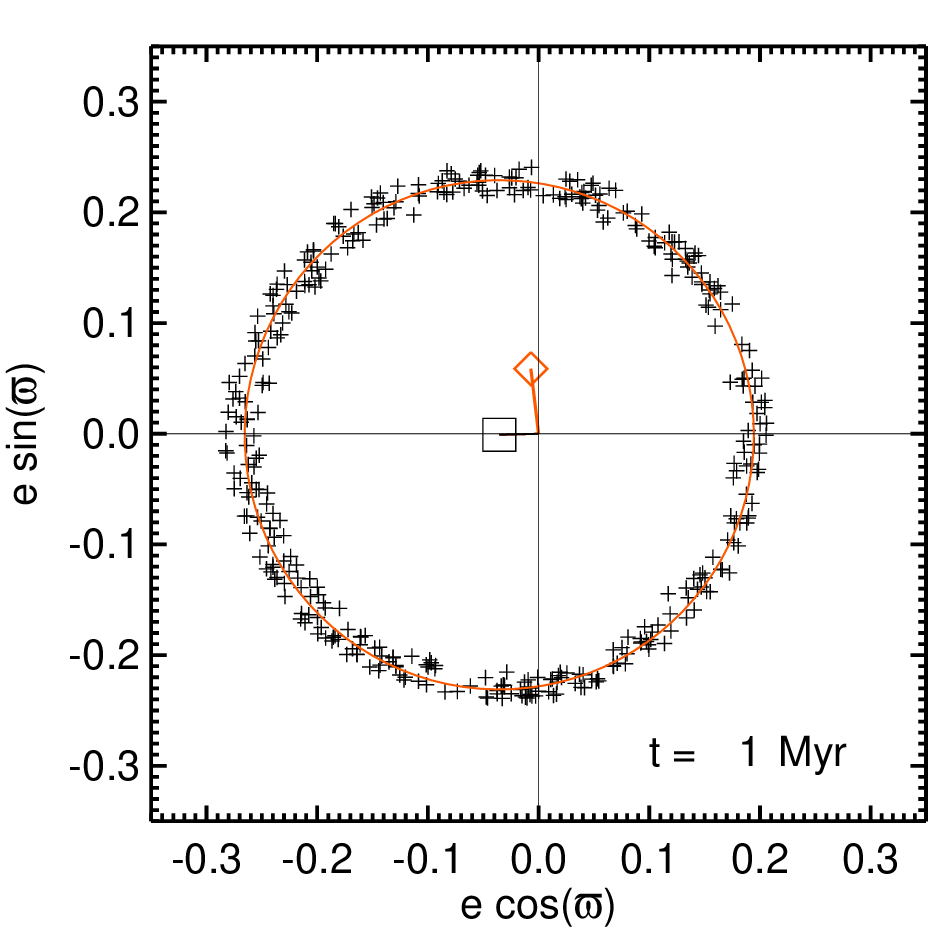}\caption{Eccentricity vectors of control particles with $e_{P}(0)=0.23$ and $I_{P}(0)=6.5^{\circ}$ at $t=0$, $10^{5}$ and $10^{6}$ yr compared to the common forced eccentricity (black square) as well as the osculating Martian eccentricity (red diamond). Encounters with Mars are more likely when these two vectors are anti-aligned. The red circle represents a least-squares fit to the eccentricity data.
%See also Fig.~\ref{fig:ec23_dispersion}.
}
\label{fig:ec23_ecc_vector}
\end{figure*}
\clearpage

\begin{figure}
\centering
\includegraphics[angle=0,width=87mm]{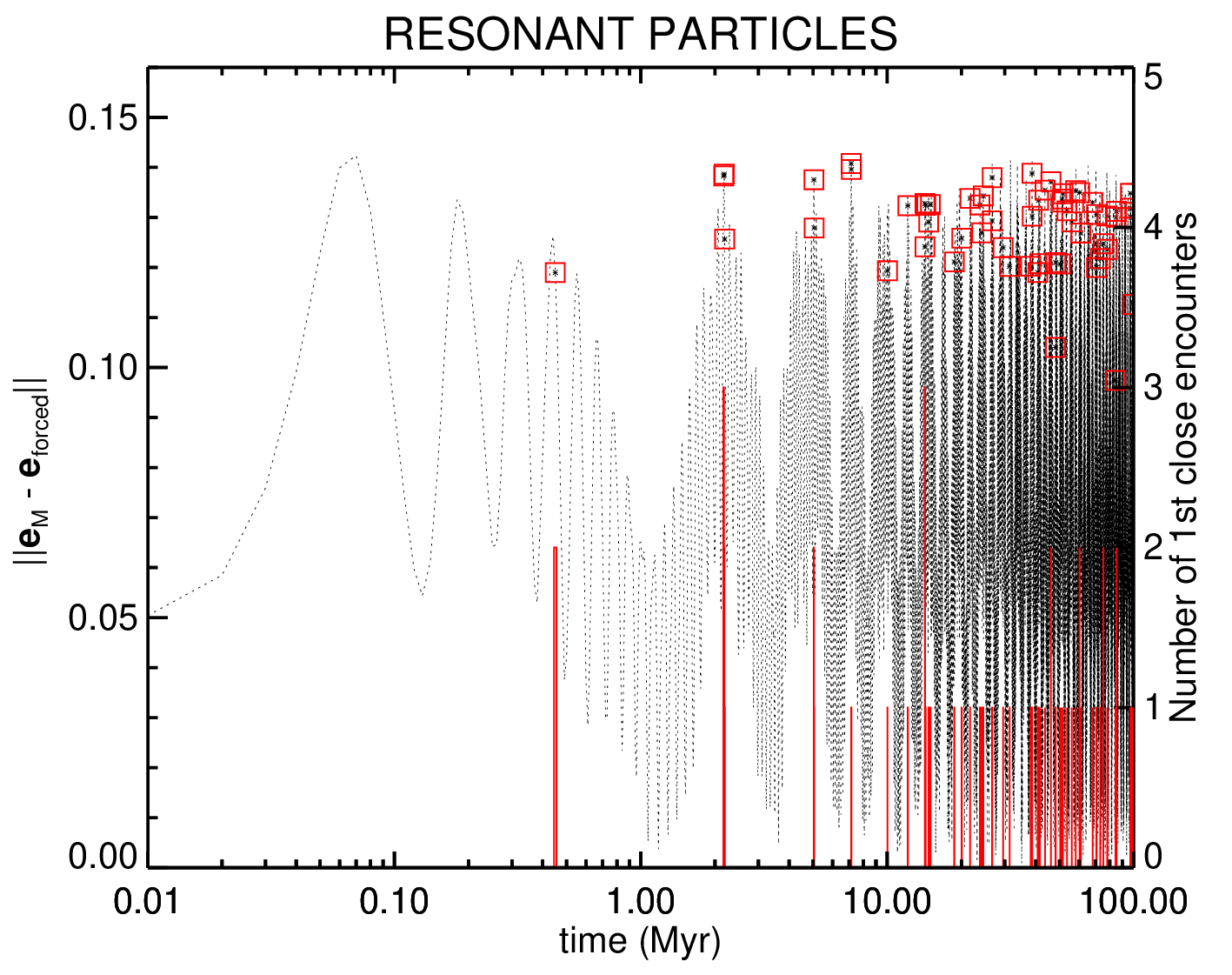}\\\vspace{-2mm}\includegraphics[angle=0,width=87mm]{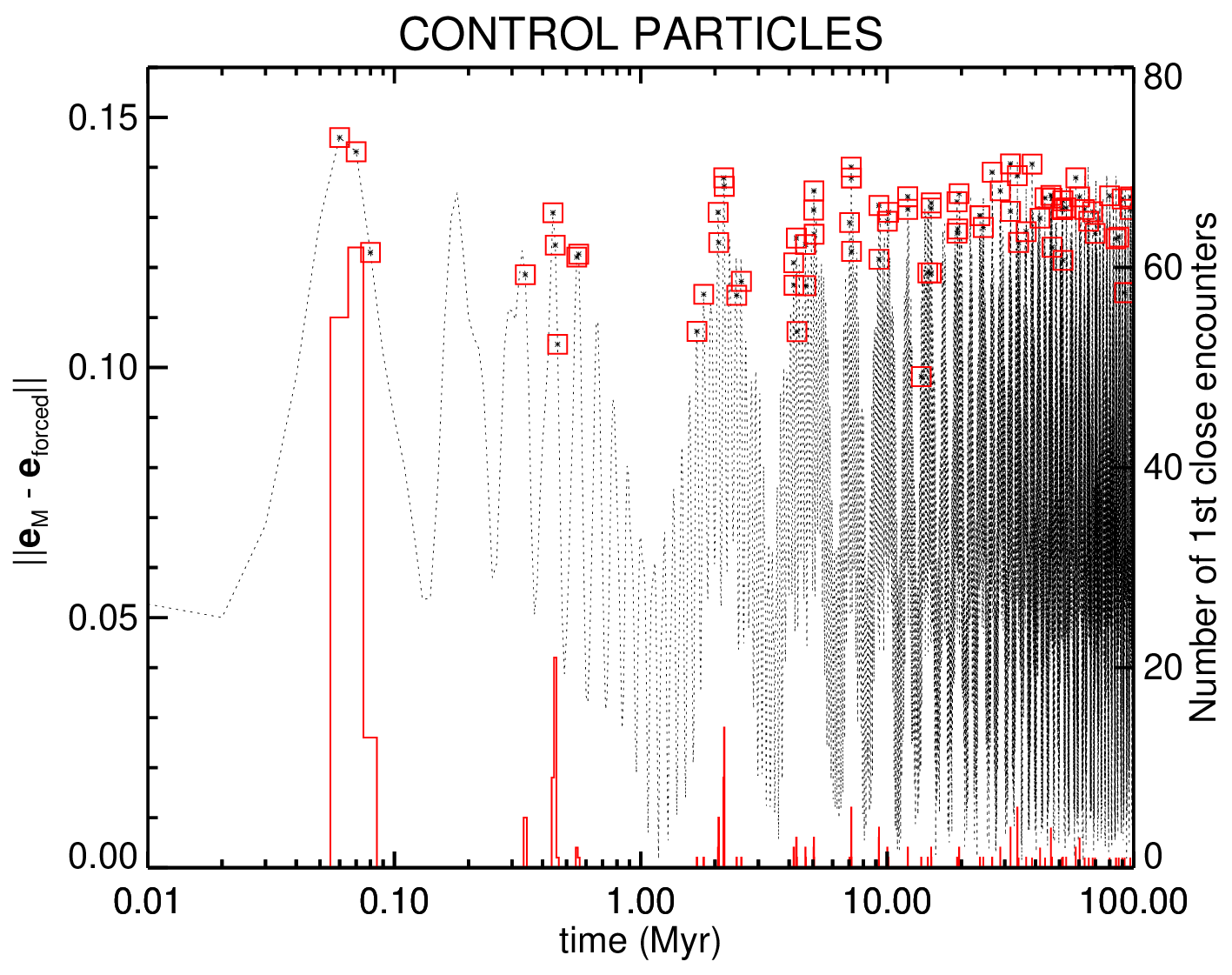}
\caption{Magnitude of $\mathbf{e}_{M} - \mathbf{e}_{\rm forced}$ (dotted line) shown against the time history of 1st Mars encounters from Fig.~\ref{fig:ec23_adistmars} (red histogram) for resonant (top) and for control (bottom) particles. Function values when encounters are recorded are highlighted as red squares. Histogram bin size is 1$\times$$10^{4}$ yr.}
\label{fig:ec23_eforced}
\end{figure}
\clearpage

\begin{figure}
\centering
\includegraphics[angle=0,width=78mm]{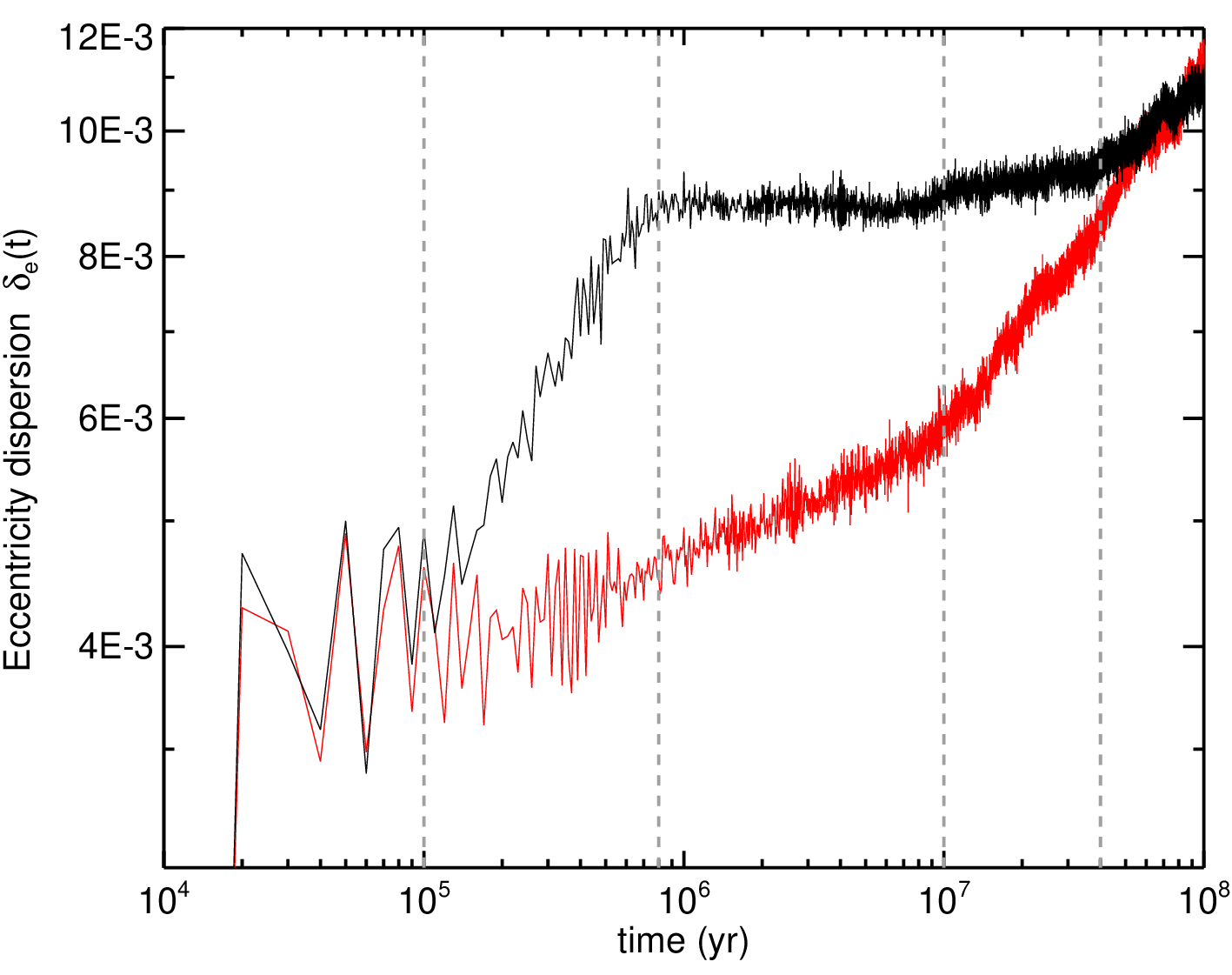}\\
\includegraphics[angle=0,width=78mm]{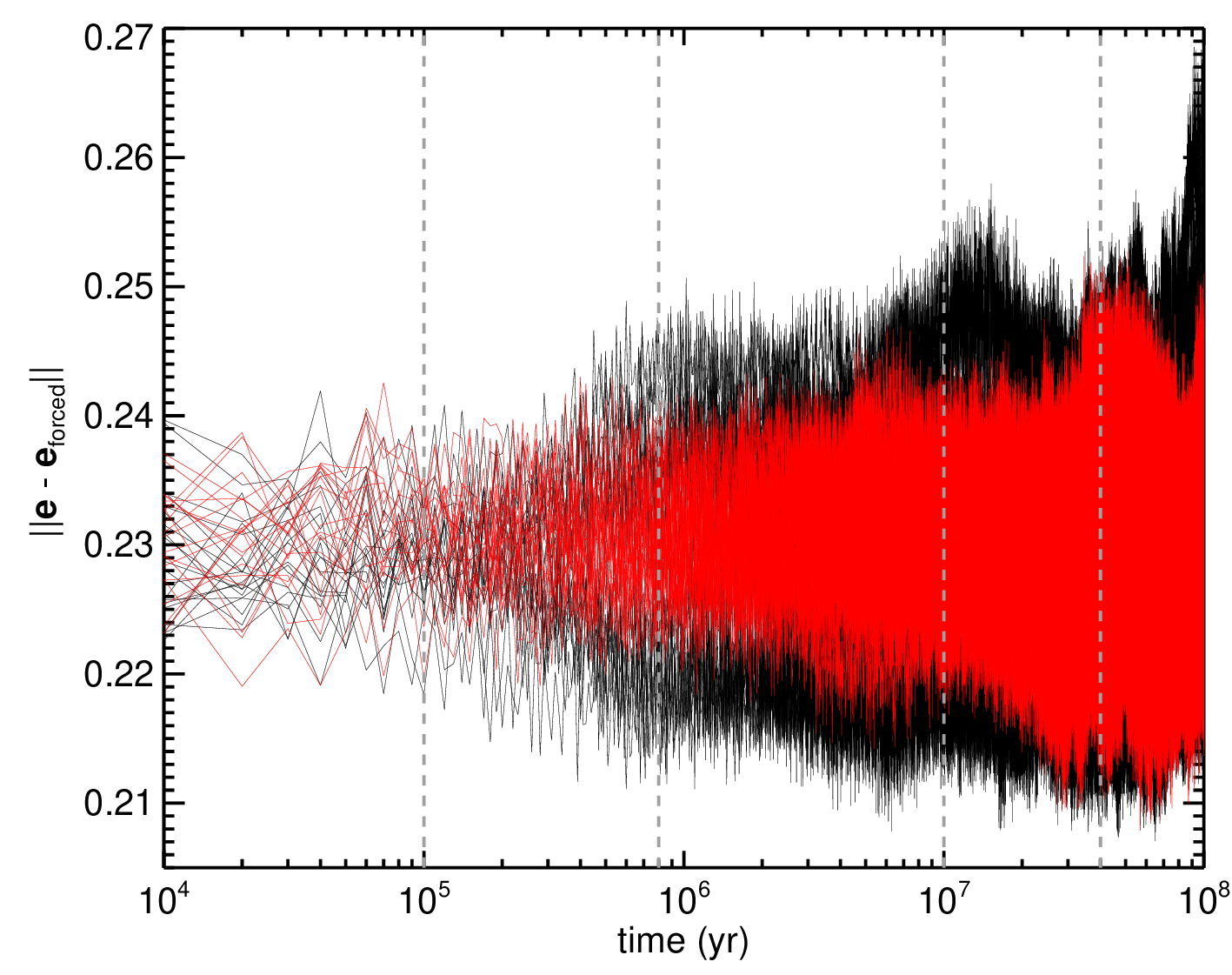}\\\includegraphics[angle=0,width=78mm]{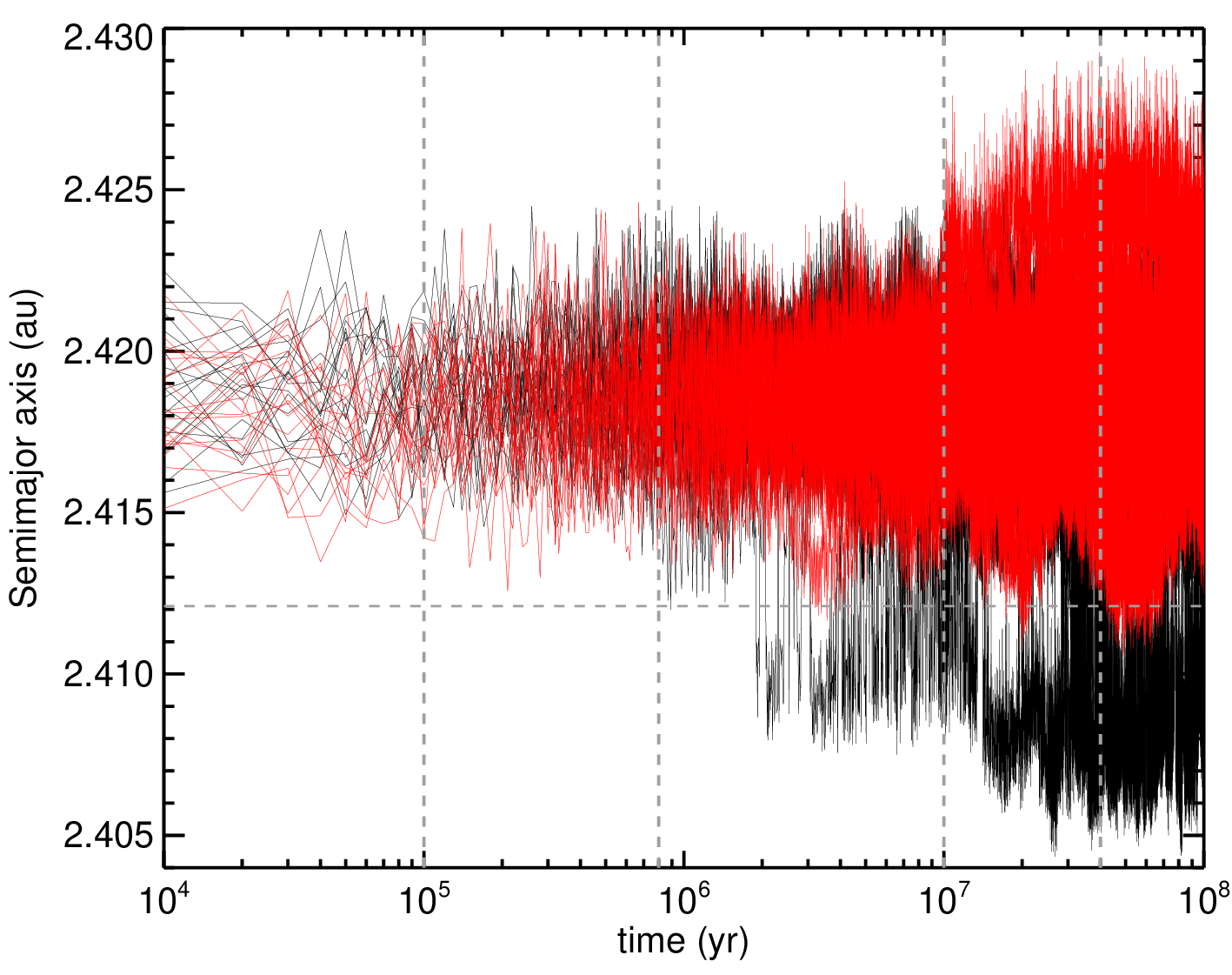}
\caption{{\it Top}: Eccentricity dispersion $\delta_{e}(t)$ (Eq.~\ref{eq:ecc_dispersion}) for resonant and control particles in Fig.~\ref{fig:ec23_eforced}, shown in red and black respectively. Vertical dashed lines highlight specific features in the time evolution of $\delta_{e}$. {\it Middle}: The magnitude of $\mathbf{e - e_{\rm forced}}$ for indidividual particles. Only 1 in 20 particles are shown for clarity. {\it Bottom}: The semimajor axis $a(t)$ for indidividual particles, where control particles have been displaced by $\delta a = -0.0063$ au to help compare to the resonant particles. The horizontal line indicates the 1M-2A resonance location for the displaced control particles.}
\label{fig:ec23_dispersion}
\end{figure}
%% \section{Some extra material}
%% If you want to present additional material which would interrupt the flow of the main paper, it can be placed in an Appendix which appears after the list of references.
%%%%%%%%%%%%%%%%%%%%%%%%%%%%%%%%%%%%%%%%%%%%%%%%%%
% Don't change these lines
\bsp	% typesetting comment
\label{lastpage}
\end{document}